\documentclass[10pt,preprint]{aastex}
\usepackage[usenames,dvipsnames]{color}
\usepackage{color}
\usepackage{lscape}
\usepackage{hyperref}
\usepackage[normalem]{ulem}

\slugcomment{To be submitted to ApJ}

\shorttitle{MT binaries}
\shortauthors{Bardalez Gagliuffi et al.}

\begin{document}

\title{SpeX Spectroscopy of Unresolved Very Low Mass Binaries. II. Identification of Fourteen Candidate Binaries with Late-M/Early-L and T Dwarf Components}

\author{Daniella C. Bardalez Gagliuffi\altaffilmark{1,11}, Adam J. Burgasser\altaffilmark{1,11}, Christopher R. Gelino\altaffilmark{2,12}, Dagny L. Looper\altaffilmark{3,11}, Christine P. Nicholls\altaffilmark{1,4}, Sarah J. Schmidt\altaffilmark{5,11}, Kelle Cruz\altaffilmark{6,7}, Andrew A. West\altaffilmark{8}, John E. Gizis\altaffilmark{9}, Stanimir Metchev\altaffilmark{10}.}
\altaffiltext{1}{Center for Astrophysics and Space Sciences, University of California, San Diego, 9500 Gilman Dr., Mail Code 0424, La Jolla, CA 92093, USA.~\texttt{daniella@physics.ucsd.edu}}
\altaffiltext{2}{NASA Exoplanet Science Institute, Mail Code 100-22, California Institute of Technology, 770 South Wilson Ave, Pasadena, CA 91125, USA.}
\altaffiltext{3}{Tisch School of the Arts, New York University, 721 Broadway, New York, NY 10003, USA.}
\altaffiltext{4}{School of Mathematics and Physics, The University of Queensland, Brisbane, QLD 4072, Australia.}
\altaffiltext{5}{Department of Astronomy, Ohio State University, 140 West 18th Ave., Columbus, OH 43210, USA.}
\altaffiltext{6}{Department of Physics and Astronomy, Hunter College, City University of New York, 695 Park Avenue, New York, NY 10065, USA.}
\altaffiltext{7}{Department of Astrophysics, American Museum of Natural History, Central Park West at 79th St., New York, NY 10024, USA.}
\altaffiltext{8}{Department of Astronomy, Boston University, CAS 422A, 725 Commonwealth Ave., Boston, MA 02215, USA.}
\altaffiltext{9}{Department of Physics and Astronomy, University of Delaware, 104 The Green, Newark, DE 19716, USA.}
\altaffiltext{10}{Department of Physics and Astronomy, Western University, London, Ontario, Canada, N6A 3K7.}
\altaffiltext{11}{Visiting Astronomer at the Infrared Telescope Facility, which is operated by the University of Hawaii under Cooperative Agreement no. NNX-08AE38A with the National Aeronautics and Space Administration, Science Mission Directorate, Planetary Astronomy Program.}
\altaffiltext{12}{Infrared Processing and Analysis Center, Mail Code 100-22, California Institute of Technology, 1200 E California Blvd., Pasadena, CA 91125, USA.}

\begin{abstract}
Multiplicity is a key statistic for understanding the formation of very low mass (VLM) stars and brown dwarfs. Currently, the separation distribution of VLM binaries remains poorly constrained at small separations ($\leq$ 1 AU), leading to uncertainty in the overall binary fraction.  We approach this problem by searching for late-M/early-L plus T dwarf spectral binaries whose combined light spectra exhibit distinct peculiarities, allowing for separation-independent identification. We define a set of spectral indices designed to identify these systems, and use a spectral template fitting method to confirm and characterize spectral binary (SB) candidates from a library of 815 spectra from the SpeX Prism Spectral Libraries.  We present eleven new binary candidates, confirm three previously reported candidates and rule out two previously identified candidates, all with primary and secondary spectral types between M7-L7 and T1-T8, respectively. We find that subdwarfs and blue L dwarfs are the primary contaminants in our sample and propose a method for segregating these sources. If confirmed by follow-up observations, these systems may add to the growing list of tight separation binaries, whose orbital properties may yield further insight into brown dwarf formation scenarios.
\end{abstract}


\section{Introduction}\label{sec:intro}

Brown dwarfs are self-gravitating objects with physical and atmospheric properties intermediate between stars and planets. With masses below $\simeq$0.075~M$_{\bigodot}$\footnote{Minimum mass for Hydrogen fusion may vary between 0.072-0.078~M$_{\bigodot}$ depending on age and metallicity. See~\citet{1997ApJ...491..856B} for an extensive discussion of evolutionary models.}~\citep{1963ApJ...137.1121K, 1963PThPh..30..460H}, these objects cannot sustain hydrogen fusion, and hence cool and dim as they age, radiating primarily at infrared wavelengths. The evolution of their spectra spans the spectral classes M, L, T, and Y, with transitions demarcated by the appearance and disappearance of absorption lines and bands as molecules form and condense out of their atmospheres at different temperatures and pressures~(\citealt{2005ARA&A..43..195K} and references therein).  
%with M being the hottest (3500 $> \Teff >$ 2100~K) and Y the coldest ($\Teff <$ 500~K;~\citet{2004AJ....127.3516G};~\citet{2009ApJ...702..154S};~\citet{2013Sci...341.1492D}),

Despite having a basic understanding of their evolution, brown dwarf formation remains an open question.  Standard Jeans collapse of molecular clouds requires high densities so that gravity can overcome thermal pressure. Once the collapse has begun, halting the accretion becomes problematic~\citep{1987ARA&A..25...23S}.  Several mechanisms have been proposed to resolve this issue, including turbulent fragmentation of protostellar clouds~\citep{2002ApJ...576..870P}, fragmentation of pre stellar disks~\citep{2009AIPC.1094..557S}, ejection by dynamical interactions with other protostars~\citep{2001AJ....122..432R}, and photoerosion of prestellar cores~\citep{2004A&A...427..299W}. In principle, these formation mechanisms should leave traces on the statistical properties of brown dwarfs, including the occurrence of multiple systems and distributions of their separation, relative masses and eccentricity. 
%For example, if x is the dominant mechanism, we would expect a strong dependent of the binary separation on mass, while if y is the dominant mechanisms, something else.

Observationally, it has been shown that multiplicity increases with primary mass, even at the lower mass end of the main sequence, with the G dwarf binary fraction being 57$\%$ higher than that for M dwarfs~\citep{1992ApJ...396..178F, 2004MNRAS.351..617D}.  Current estimates of the binary fraction of very low mass (VLM) late-M to T dwarfs (VLM $\mathrm{M_{total}} < 0.1~ \mathrm{M_{\bigodot}}$) are $20-25\%$, with a peak in separation at $\sim$ 4~AU and a mass ratio distribution peaking at nearly equal masses~\citep{2003AJ....126.1526B, 2003IAUS..211..249C, 2006ApJS..166..585B, 2007ApJ...668..492A, 2012ApJ...757..141K}. 
%Briefly, how does this compare to model predictions? or what models are favored by these stats? - Reread Bate et al. 2012 and all of those listed at the end.
However, these multiplicity statistics have been largely determined from resolved imaging programs, sampling separations greater than 3~AU.~\citet{2007prpl.conf..427B} pointed out that the current peak in the binary angular separation distribution is coincident with the resolution limit of HST and ground-based adaptive optics (AO) imaging, indicating that tight ($< 1$AU) VLM binaries could be undercounted. Likewise, ~\citet{2003MNRAS.342.1241P} and~\citet{2005MNRAS.361.1323C} report a higher unresolved binary fraction ($30-50\%$) based on overluminous binary candidates in color-magnitude plots. Conversely, spectroscopic radial velocity (RV) studies find binary fractions of 2.5$\%$ in systems separated by $< 1$ AU~\citep{2010ApJ...723..684B} and $2-28\%$ up to 3~AU~\citep{2008A&A...492..545J}. For the 0-6 AU range,~\citet{2006AJ....132..663B} estimate a binary fraction of 26$\%\pm10\%$. However, the difficulty of obtaining high resolution spectra of faint VLM dwarfs results in small sample size. Since total binary fractions for VLM stars and brown dwarfs could range between $2-50\%$, it is imperative to constrain this statistic to make conclusions about brown dwarf formation.

An alternative method for detecting tight unresolved binaries, developed by~\citet{2007ApJ...659..655B}, involves identifying blended light pairs, or spectral binaries.  We will refer as \emph{spectral} binaries to those objects whose combined-light spectrum shows distinct peculiarities that come from the highly structured spectra of single M, L and T dwarfs when blended together, as opposed to \emph{spectroscopic} binaries which are binaries that show RV variations. The first brown dwarf spectral binary, 2MASS J05185995$-$2828372, was serendipitously identified by~\citet{2004ApJ...604L..61C} based on its hybrid characteristics containing features of both L and T dwarfs.  The superposition of L plus T dwarf spectra proved to be the simplest model of its peculiar spectrum and it was later resolved as a binary using the Hubble Space Telescope~\citep{2006ApJS..166..585B}. The unusually blue L dwarf SDSS J080531.84+481233.0 was next identified as a spectral binary with L4.5 and T5 components by~\citet{2007AJ....134.1330B}, based on a peculiar methane absorption band starting at $1.60~\mu$m, and was later confirmed as an astrometric variable by~\citet{2012ApJS..201...19D}. A third system, 2MASS J03202839$-$0446358, was concurrently identified as an unresolved M9+T5 spectral binary~\citep{2008ApJ...681..579B} and a RV variable with an orbital period of eight months~\citep{2008ApJ...678L.125B}. These examples serve to illustrate how spectral binaries can encompass a broad range of system architectures. To date, 34 VLM spectral binaries and candidates have been reported~(see Table~\ref{tab:SB}), and ten have been confirmed by direct imaging, RV or astrometric variability~(\citealt{2011AJ....141...70B, 2011A&A...525A.123S, 2012ApJ...757..110B, 2012ApJS..201...19D, 2012ApJ...752...56F, 2013arXiv1310.2191M}; Gelino et al. in prep).

Detecting binaries using the spectral binary method is particularly useful for multiplicity statistics, as the method is independent of separation within 0$\farcs$5, which translates to $<$10-20 AU for field brown dwarfs at distances of 20-40 pc. The closest separation pairs can be followed-up to measure orbits and component masses, as well as infer ages by comparison to evolutionary models~\citep{2009AJ....137.4621B}.  Systems with independent age constraints can also be used to test the evolutionary models directly~\citep{2009arXiv0912.0738D, 2010ApJ...722..311L, 2011AJ....141...70B}. Additionally, unresolved binaries are strong contaminants in luminosity functions that later lead to uncertainties in mass functions and studies of formation history through stellar populations~\citep{2013MNRAS.430.1171D}, so their identification is extremely important. Finally, spectral binaries with late-M/early-L primaries and T dwarf secondaries can straddle the hydrogen burning limit, thus giving additional insight into brown dwarf evolution.

%The greatest advantage of utilizing this technique for confirming and selecting candidate binaries is that it does not depend on the resolution of the instrument used for observing. Other methods for companion detection like direct imaging and radial velocity variation rely heavily on the precision of the instrument to determine a significant signature of the secondary object, and thus the binary must have an angular separation greater than the limiting resolution of the telescope.  The spectral fitting method is completely separation-independent, since it is based on the relative addition of flux. Therefore, it allows us to probe angular separations that are beyond telescope detection limits.  However, it is important to keep in mind that this method is not sensitive to pairs of objects with equal spectral types, and that it does not provide certain confirmation of binarity, but only a strong candidacy probability. 

In this paper we adapt the technique of~\citet{2010ApJ...710.1142B} to search for spectral binaries composed of late-M or early-L dwarf primaries with T dwarf secondaries.  M dwarfs are the most common stars in the galaxy \citep{2010AJ....139.2679B}, and are the brightest VLM objects, enabling better statistics through larger magnitude-limited search volumes and sample sizes.  M-dwarf spectra are also intrinsically distinct from T-dwarf spectra, but differ in brightness by several magnitudes, rendering peculiar features extremely subtle. In \S~\ref{sec:obs} we describe our spectral sample used to find spectral binaries, drawn from the SpeX Prism Libraries and new observations. In~\S~\ref{sec:MTbin} we explain our two methods to identify spectral binary candidates: by visual examination (\S~\ref{sec:visual}) and through spectral indices (\S~\ref{sec:specind}).  In \S~\ref{sec:specfit}, we perform single and binary template fitting to identify fourteen binary candidates.  In \S~\ref{sec:candidates}, we describe the properties of the candidates. In \S~\ref{sec:blue} and \S~\ref{sec:separation}, we discuss our major contaminant, blue L dwarfs, and show preliminary evidence that the separations of spectral binaries are tighter than the resolved population. Our results are summarized in \S~\ref{sec:summary}.

\section{SpeX Spectral Sample}\label{sec:obs}

The SpeX Prism Library is composed of low resolution ($\lambda/\Delta\lambda = 75-120$) spectra acquired with the SpeX $0.8-2.5~\mu$m spectrograph, mounted on the 3.0 m NASA Infrared Telescope Facility (IRTF), located in Mauna Kea, Hawaii~\citep{2003PASP..115..362R}. All spectra were obtained using the prism-dispersed SpeX mode, which continuously samples wavelengths between $0.75-2.5~\mu$m at a dispersion of $20-30$~\AA~pixel$^{-1}$.  The library includes close to 2,000 sources, both previously published data\footnote{e.g.~\citet{2010ApJ...710.1142B,2006AJ....131.2722C,2003AJ....126.2421C}.} and 530 new spectra acquired between November 2000 and December 2013~(Table~\ref{tab:newobs}). The new observations were obtained with the 0$\farcs$5 or 0$\farcs$7 slit, generally aligned with the parallactic angle. Total integration times ranged between 360 s and 1200 s, depending on source brightness and atmospheric conditions, and were obtained in an ABBA dither patter along the slit. Spectra of nearby A0 V stars were used to flux calibrate the raw spectra and correct for telluric absorption.  Internal flat fields and argon arc lamps were observed with each flux standard for pixel response and wavelength calibration.  All data were reduced with the SpeXtool package~\citep{2004PASP..116..362C, 2003PASP..115..389V} using standard settings.  A detailed description of our reduction procedures is given in~\citet{2007AJ....134.1330B}. 

The sources observed have optical and/or near-infrared spectral classifications reported in the literature.  To obtain a self-consistent set of spectral types, we computed SpeX spectral types based on spectral indices, following the method described in~\citet{2007ApJ...659..655B}. From these, we selected two samples: the ``candidate'' sample which has been purged of spectral types outside the M7-L7 range, optical subdwarfs, giants, and poor quality spectra, but keeping binaries, objects suspected of being binaries from previous studies, young objects and unusually red and blue dwarfs, and the ``template'' sample which has been purged of binaries, candidate binaries, giants and poor quality spectra (as determined by visual inspection) only. The ``candidate'' sample contains 815 spectra of 738 objects with SpeX spectral types between M7-L7, as those would be the potential primaries for late-M/early-L plus T binaries. The ``template'' sample comprise 1110 spectra of 992 single sources whose spectral types range between M7-L7 for primaries and T1-T8 for secondaries used in spectral fitting.  
%The only difference between the two samples, besides from the spectral type cuts, is the number of sources, since the spectral index sample includes new SpeX observations (table~\ref{tab:newobs}). While the index sample includes only late-M/early-L spectra to find the telltale signs of a hidden T secondary, the fitting sample covers the entire spectral type range from late-M to late-T because we search for template spectra for both primary and secondary.
%db_singles: 915 sources from M5-T9 observed between 20001106 and 20130814 - minmax(Resolution) =  120,75
%source_db: 1010 sources from M5-T1.2 observed between 20001106 and 20130814
%candidate sample includes candidate binaries from before
%source_db - bad spectra = 
%binaries - candidate binaries - bad spectra = singles_db
%singles_db - spt

The distribution of spectral types for both samples is shown in Figure~\ref{fig:sample}. In both samples, the number of spectra decreases toward later spectral types due to declining space densities for L dwarfs~\citep{2003AJ....126.2421C} and sensitivity limits for late L and T dwarfs. Since there are significantly more sources with late-M spectral types in our samples, it is more likely to find binaries with a late-M primary. The sources included were observed as part of several different programs, including our ongoing program to compile a magnitude-limited sample of L dwarfs (Burgasser et al. in prep.).  As such, we do not claim the sample to be complete or unbiased.

%some sources had several spectra and were selected as 

\section{Identification of spectral binaries}\label{sec:MTbin}

\subsection{Visual inspection}\label{sec:visual}

The spectral morphology of unresolved late-M/early-L plus T dwarf binary systems gives rise to a distinctive feature in blended-light spectra: a small ``dip'' centered at 1.63~$\mu$m, which is the combination of CH$_4$ absorption from the secondary and FeH from the primary~\citep{2003ApJ...582.1066C,2007AJ....134.1330B}. Methane does not exist in the spectra of late-M/early-L dwarfs, so its presence indicates a T dwarf companion.  However, this feature is very weak in blended-light spectra since a T dwarf is significantly fainter than the M/L primary (e.g. $\Delta J\sim$ 3.5 mag between an M8 and a T5, which is the case for 2MASS J03202839$-$0446358). Moreover, variations in the spectral slope for a blue or red L dwarf, can make this feature ambiguous, as can poor correction of Hydrogen lines in the A0V calibrators. Alternative indicators such as a relatively higher flux around the 1.25~$\mu$m peak and \textbf{an} inflated bump short ward of 2.2~$\mu$m, may also reveal the presence of a T dwarf companion, or that the spectrum of the source is unusually blue. 

To facilitate our visual inspection, we fit the candidate sample to templates of single objects, following the same chi-squared minimization routine as in Section~\ref{sec:specfit}, and then subtracted the median combination of the ten best fitting single sources from each spectrum. The objects with residuals consistent with a T dwarf spectrum were selected as visual candidates. To validate this procedure, we also performed the same template subtraction on four confirmed spectral binaries: SDSS J000649.16$-$085246.3, 2MASS J03202839$-$0446358, SDSS J080531.89+481233.0, and 2MASS J13153094$-$2649513~(see Table~\ref{tab:SB}). The residuals from these subtractions clearly exhibited T dwarf-like morphologies. Twelve sources were selected as visual candidates.

\subsection{Spectral indices}\label{sec:specind}

In addition to visual inspection, we also used spectral indices to identify additional candidate binaries due to the subtlety of T dwarf features in combined-light spectra~\citep{2010ApJ...710.1142B}. We initially examined standard classification indices from~\citet{2006ApJ...637.1067B}, as well as the ``\emph{H}-dip'' index from~\citet{2010ApJ...710.1142B}, and further defined five new indices.  The new indices were designed by comparing the residuals of the four known binary spectra after subtracting their best single template fits. As a control sample, we also examined single templates subtracted from each other, which showed no evidence for a T dwarf companion.

The new spectral indices specifically designed in this paper are:
\begin{itemize}
\item \emph{H}-bump: measures the peak in the continuum from the T dwarf in the \emph{H} band relative to the dip centered around 1.63~$\mu$m seen in M+T binaries, making this index complimentary to \emph{H}-dip. A higher value of \emph{H}-bump implies a larger flux at 1.55~$\mu$m, possibly caused by the presence of a T dwarf.
\item \emph{J}-curve: designed to detect the flux coming from both the 1.05~$\mu$m and 1.27~$\mu$m peaks of a T dwarf, as compared to the deep methane absorption at 1.12~$\mu$m.
\item \emph{J}-slope and \emph{K$_s$}-slope: measure the slope of the peaks in the \emph{J} and \emph{K$_s$} bands at 1.27~$\mu$m and 2.10~$\mu$m. In both cases, the peaks in a single late-M/early-L should look somewhat flat, giving values close to one, whereas in a late-M/early-L plus T dwarf binary the slope of the \emph{J} and \emph{K$_s$} band peaks are slightly negative and positive, respectively.
\item H$_2$O-\emph{Y}: measures the prominence of the \emph{Y}-band peak of the T dwarf at $\sim 1.05~\mu$m compared to the water and methane absorption around $\sim 1.15~\mu$m. M and L dwarfs do not present peaks in the \emph{Y}-band.
\end{itemize}

The thirteen indices examined are described in Table~\ref{tab:index}. We also used \emph{J-K$_s$}, \emph{J-H} and \emph{H-K$_s$} colors synthesized from the spectra themselves, and the source spectral type, for a total of seventeen parameters. 

Comparing all seventeen parameters against each other yielded 136 pairings.  After visual examination to determine which pairings best segregated the four known M/L+T binaries, twelve combinations were selected (Figure~\ref{fig:paramspace}). We then used two techniques to define regions of interest in each combination for candidate selection.  If a trend among all sources was clear, we fit the points to a second order polynomial and defined a region demarcated in the y-axis by the +1$\sigma$ or -1$\sigma$ curves from the fit function, and in the x-axis by the horizontal spread of the binary benchmarks. Conversely, if the points did not indicate any trends, then the region was demarcated such that it included the four binary benchmarks. The limits to these regions are described in Table~\ref{tab:selectreg}.

Objects falling in eight or more selection regions were considered strong index candidates, those falling in four to eight regions were considered weak index candidates (Figure~\ref{fig:histogram}). The number of selected sources rises steadily below four or five combinations, suggesting that sources selected fewer than four times could be spurious.  Three of our benchmarks were selected by all twelve combinations, while SDSS J0006$-$0852 missed only the SpT/CH$_4$-H cut, since it falls within one standard deviation from the fitting curve. 

In total, eight strong and twenty-two weak candidates were selected, including the previously identified spectral binaries 2MASS J20261584$-$2943124~\citep{2010AJ....140..110G} and 2MASS J13114227+3629235~\citep{2011ApJS..197...19K}. Seven visual candidates overlapped with the index candidates: five as strong and two as weak.

\subsection{Spectral template fitting}\label{sec:specfit}

To statistically test the binary hypothesis for our visual and index-selected candidates, we compared each spectrum to templates of both single sources and binary systems, using the method described in~\citet{2010ApJ...710.1142B}. The candidates determined by visual inspection or spectral index selection were first rejected from the template pool.  Then, all spectra were interpolated onto a common wavelength scale from $0.8$ to $2.4~\mu$m and normalized to the peak flux between $1.2-1.3~\mu$m. Each candidate spectrum $C[\lambda]$ was directly compared to all single templates $T[\lambda]$ and ranked by a weighted chi-squared statistic.

\begin{equation}\label{eq:chi2}
\chi^2 \equiv \sum_{\lambda} w[\lambda]\left[\frac{C[\lambda]-\alpha T [\lambda]}{\sigma_c [\lambda]}\right]^2
\end{equation}

%check ~\citet{2008ApJ...678.1372C} for this paragraph

\noindent where $w[\lambda]$ is a vector of weights proportional to the wave band size of each pixel (see~\citealt{2008ApJ...678.1372C}), $\alpha$ is a scaling factor minimizing $\chi^2$ and $\sigma_c [\lambda]$ is the noise spectrum for each candidate.  The statistic was computed over the wavelength range: $\{\lambda\} = 0.95-1.35~\mu$m,  $1.45-1.80~\mu$m and $2.00-2.35~\mu$m, avoiding regions of strong telluric absorption.  

Binary templates were constructed by first scaling each template spectrum to absolute fluxes using the 2MASS M$_\mathrm{Ks}$ versus spectral type relation of~\citet{2008ApJ...685.1183L}, and then combining all pairs of single templates, such that the spectral type of the primary was earlier than that of the secondary resulting in a total of $638,686$ binary templates~\footnote{We do not explicitly include uncertainties for the absolute magnitude to spectral type relation, but the extensive number of binary templates we use already models the intrinsic scatter in the population.}. More specifically, the primary spectral type was fixed to lie between M7-L7 while the secondary spectral type ranged between T1-T8, since types earlier than T1 do not evidence strong methane features yet. The best binary fits were ranked using a chi-squared minimization routine.  We determined the true significance that a binary template is superior to a single template  by comparing the $\chi^2$ distributions of the binary and single fits using the one-sided F-test statistic $\eta_{\mathrm{SB}}$:

\begin{equation}
\eta_{\mathrm{SB}}\equiv\frac{\mathrm{min}(\{\chi^2_{\mathrm{single}}\})}{\mathrm{min}(\{\chi^2_{\mathrm{binary}}\})}\frac{\mathrm{dof_{binary}}}{\mathrm{dof_{single}}} .
\end{equation}

\noindent Here, dof is the degrees of freedom for each fit~(Equation 2 in \citet{2010ApJ...710.1142B}). Candidates with an F-statistic falling under the $90\%$ confidence level were rejected, including five visual candidates. In particular, 2MASS J14493784+2355378 and 2MASS J14232186+6154005 (also a weak index-selected candidate), two previously identified spectral binary candidates from~\citet{2003AJ....125.3302G} and \citet{2011ApJ...732...56G} were rejected due to their low confidence level that the binary fit was statistically better than the single fit. Since our template sample includes a wide range of objects such as young and unusually blue and red dwarfs, the peculiarities of these candidates may be better explained by factors other than unresolved binarity. One exception to the index selection was 2MASSI J1711457+223204 whose SpeX spectral type was too late to be included in the candidate sample, yet it was a visual candidate and passed the binary fit F-test. Figures~\ref{fig:strongfit} and~\ref{fig:weakfit} show the best single (left) and binary (right) template fits to our strong and weak candidates. Table~\ref{tab:cand} summarizes the results of these fits.

Upon further visual examination, some binary fits still proved unsatisfactory. This was the case for the blue L dwarfs 2MASS J11181292$-$0856106, 2MASS J14162409+1348267, 2MASS J15150083+4847416, 2MASS J17114558+40285779, and the subdwarfs 2MASS J03303847$-$2348463, 2MASS J03301720+3505001, 2MASS J04024315+1730136, 2MASS J15412408+5425598 and 2MASS J23311807+4607310. Section~\ref{sec:blue} discusses these issues in more detail.  As a result, fourteen candidates have been recognized, of which eleven are newly identified.

In an effort to balance the tradeoff between fidelity of binary candidates and completeness, we are leaning towards the former. Our binary selection criteria are conservative and it is likely that other spectral binaries may be identified with slightly looser constraints.
% Your binary selection criteria are necessarily conservative, and likely other spectral binaries can be identified if the selection regions are made somewhat larger.  You do mitigate the limitations of your choice of any one given region by using multiple criteria.  However, it may be worth considering that you can not exclude (at say, the 99% confidence level) that objects that do not meet at least 5 of the criteria are not spectral binaries, too.

\section{Individual candidates}\label{sec:candidates}

In summary, from the $\sim$800 sources compiled in the candidate sample, twelve were selected by visual inspection and thirty were selected by spectral indices. Seven sources overlapped the results of these selection methods. After fitting all thirty-five candidates, seventeen were rejected due to their confidence level being lower than $90\%$, and four more due to their unusually blue colors (See Section~\ref{sec:blue}), leading to a final count of fourteen: six strong, seven weak and one visual candidate not overlapping with the index-selected. Labels of strong and weak candidates come from index selection.

\subsection{Strong candidates}

%Strong candidates are those sources selected by eight or more spectral indices (out of twelve) and subsequently having a confidence $>99\%$ that a binary template fits better than a single template.  It is worth noting that all four M+T benchmarks (0006-0852~\citep{2012ApJ...757..110B}, 0320-0446~\citep{2008ApJ...681..579B, 2008ApJ...678L.125B}, 0805+4812~\citep{2012ApJS..201...19D}, and 1315-2649~\citep{2011ApJ...739...49B}), as well as recently identified binaries like 2252-1730~\citep{2006ApJ...639.1114R} were also selected as strong candidates, reinforcing the effectiveness of our selection method.  From the visual candidates subsample, those with a confidence $>99\%$ also overlapped with strong and weak index candidates, while the rest were rejected.

\subsubsection{2MASS J02361794+0048548}
Originally discovered by~\citet{2002ApJ...564..466G}, 2MASS J0236+0048 was classified as an L6 in the optical and L6.5 in the infrared by~\citet{2008MNRAS.390.1517C}.  In their study,~\citet{2008MNRAS.390.1517C} comment that this object may belong to the Pleiades moving group, given its proper motion of $[\mu_{\alpha}cos\delta, \mu_{\delta}] = [161.33\pm10.10, 176.33\pm19.16]$~mas yr$^{-1}$ and agreement between photometric and moving group distances at d = 26~pc.  However, ~\citet{2009AIPC.1094..955S} reclassified this object as an L9, reducing its spectroscopic distance to 18~pc while its strong FeH band at 0.99$\mu$m argues against low surface gravity~\citep{2007ApJ...657..511A}. Nevertheless, the spectrum of this source does not show any signs of youth~\citep{2007ApJ...657..511A}. 2MASS J0236+0048 is selected by eleven out of twelve spectral index combinations, and its binary fit is significantly better than its single fit, making this a strong binary candidate with L5.0$\pm$0.6 and T1.9$\pm$1.1 components.

\subsubsection{SDSS J093113.09+280228.9}
~\citet{2010AJ....139.1808S} discovered SDSS J0931+2802 in the SDSS catalog and classified it as an L3 at a mean distance of 29$\pm$9~pc.  Its spectrum shows excess flux in the \emph{J}-band at $\sim1.27\mu$m and a noticeable dip in the \emph{H}-band in the vicinity of $1.63\mu$m, as expected for a T dwarf component.  This source was selected as a visual candidate, and by eleven out of twelve spectral index combinations, and our spectral fitting predicts component types of L1.4$\pm$0.1 and T2.6$\pm$0.9.

%\subsubsection{2MASS J11181292-0856106}
%This source was originally classified as an L6 in the optical and L6 peculiar in the near infrared by~\citet{2010ApJS..190..100K}.  Its J-K$_s$ color $= 1.56\pm0.02$ is $\sim$0.35 magnitudes bluer than the median for L6 dwarfs J-K$_s$ =1.82$\pm$0.07~\citep{2010AJ....139.1808S}. Selected by nine out of twelve spectral index combinations, the spectrum of this strong candidate is best fit by a combination of L5.0$\pm$1.2 and T1.6$\pm$2.2 components with a confidence $>95\%$.

\subsubsection{2MASS J13114227+3629235}
Identified as a brown dwarf candidate by~\citet{2009A&A...497..619Z}, 2MASS J1311+3629 is a peculiar L5. While also classified as unusually blue in wavelengths longer than \emph{J} band~\citep{2013ApJS..205....6M}, it lacks evidence of low metallicity or H$_2$ collision-induced absorption (CIA) in \emph{H} and \emph{K} bands. \citet{2011ApJS..197...19K} identified the methane feature in the \emph{H} band centered around 1.63~$\mu$m suggesting unresolved binarity. In this study, it was selected by eleven spectral index combinations and also as a visual candidate due to its methane absorption band starting at 1.60~$\mu$m. Template fitting gives spectral types of L4.8$\pm$0.6 and T4.1$\pm$2.7.

\subsubsection{2MASS J13411160$-$30525049}
2MASS J1341$-$3052 was discovered by~\citet{2008AJ....136.1290R} and classified as an L3 in the optical by~\citet{2009AJ....137....1F}, who also measured its parallax and distance ($24\pm2$~pc).  2MASS J1341$-$3052 was selected by eight spectral indices, and its spectral fitting suggests component spectral types of L1.2$\pm$0.3 and T6.3$\pm$1.0.

%\subsubsection{SDSS J141624.09+134826.7}
%~\citet{2010ApJ...710...45B} and~\citet{2010AJ....139.1045S} independently identified this object from the SDSS DR7~\citep{2009ApJS..182..543A} and classified it as a blue L6 and L5 at a distance of $8.4\pm1.9$~pc and $8.0\pm1.6$~pc, respectively, among the ten closest L dwarfs to the Sun. No Li absorption or H$\alpha$ emission was found in its spectrum.  Moreover,~\citet{2010MNRAS.404.1952B}, ~\citet{2010A&A...510L...8S} and~\citet{2010AJ....139.2448B} showed that this source has a common proper motion companion 9 arc seconds away, the unusually blue T7.5 ULAS J141623.94+134836.3.~\citet{2010ApJ...710...45B} rejected unresolved binarity on the primary based on a qualitative comparison to the unusually blue L dwarf 2MASS J11263991-5003550.  The most likely cause for these colors is the presence of either thin cloud coverage or large-grain clouds, thus allowing for reduced condensate opacity and a stronger emission in shorter wavelengths.  It is selected by ten out of twelve spectral index combinations. Its best fitting components are L2.0$\pm$0.9 and T3.4$\pm$1.2.  

\subsubsection{2MASS J14532589+1420418}
2MASS J1453+1420 was classified as an L1 in both the infrared~\citep{2010ApJS..190..100K} and the optical~\citep{2010AJ....139.1808S}, where it clearly shows excess flux in the \emph{J} band and a dip in the \emph{H} band. It is selected by eleven out of twelve spectral index combinations, and it is slightly blue with a \emph{J-K$_s$} color of 1.18$\pm$0.05 as compared to the median for \textbf{the} L1 spectral type 1.34$\pm$0.19~\citep{2010AJ....139.1808S}. It is best fit by L1.1$\pm$0.0 and T6.0$\pm$1.1 components.

\subsubsection{2MASS J20261584$-$2943124}
2MASS J2026$-$2943 had already been identified as a spectral binary candidate by~\citet{2010AJ....140..110G}, but it failed to be resolved by Keck AO, thus setting an upper limit in separation of 0$\farcs$25 or a projected separation of 9 AU at a distance of 36$\pm$5 pc~\citep{2010AJ....140..110G}. This source clearly shows a dip in its spectrum centered at 1.63~$\mu$m, and it is best fit by a combination of L1.0$\pm$0.5 and T5.8$\pm$1.0 components. 

\subsection{Weak Candidates}

\subsubsection{2MASS J02060879+22355930}
2MASS J0206+2235 was discovered and classified as an L5.5 by~\citet{2006AJ....131.2722C}, and characterized as a blue L dwarf by~\citet{2014AJ....147...34S}. It was selected by seven spectral index combinations and fit to L5.1$\pm$0.5 and T3.2$\pm$2.3 components.

\subsubsection{DENIS-P J04272708$-$1127143}
2MASS J0427$-$1127 was discovered and classified as an M7 by~\citet{2010A&A...517A..53M}. It was selected by five spectral index combinations and best fit by M7.4$\pm$0.2 and T5.1$\pm$1.5 components.

%\subsubsection{2MASS J04430581$-$3202090}
%2MASS J0443$-$3202 was deemed anomalously blue for an L5 by~\citet{2003A&A...403..929K}. It was selected by five spectral index combinations and best fit by a combination of L5.0$\pm$0.9 and L9.8$\pm$1.1 components. Its spectrum does not show a pronounced methane absorption feature centered at 1.63~$\mu$m because the secondary is a late L with very weak methane absorption.

\subsubsection{2MASS J10365305$-$3441380}
2MASS J1036$-$3441 was classified as an L6~\citep{2002ApJ...575..484G} at a distance of 21$\pm$3~pc. It almost made the cut for a strong candidate, since it was selected by seven spectral index combinations. This source was best fit by components with L5.2$\pm$0.4 and T1.4$\pm$0.4 spectral types. Despite not having a pronounced methane absorption feature centered at 1.63~$\mu$m, the binary fit is significantly better than the single fit, especially at the \emph{J} band peak.

\subsubsection{2MASS J10595138$-$2113082}
2MASS J1059$-$2113 is an L1~\citep{2003AJ....126.2421C} at a distance of 32.1 $\pm$2.2~pc. This source was selected by four spectral index combinations and its best binary fit yields components with L0.6$\pm$0.4 and T3.4$\pm$1.3 spectral types. Its spectrum shows a strong absorption feature centered at 1.63~$\mu$m, as well as a flux excess at 1.23$\mu$m and 2.20$\mu$m.

\subsubsection{SDSS J142227.20+221557.5}
SDSS J1422+2215 was identified and classified as an L6 in the NIR by~\citet{2006AJ....131.2722C} and also as an unusually blue L dwarf, showing strong H$_2$O and FeH absorption bands, which may be due to subsolar metallicity and/or thinner condensate cloud decks. It was selected by four out of twelve spectral index combinations with most likely component spectral types of L4.2$\pm$0.6 and T4.1$\pm$2.3.

\subsubsection{WISE J16235970$-$0508114}
WISE J1623$-$0508 was classified as an L1 in the NIR~\citep{2013PASP..125..809T}. This source was selected by four spectral index combinations and best fit by L0.6$\pm$0.3 and T6.0$\pm$0.3 components.

%\subsubsection{2MASS J15150083+4847416}
%2MASS J1515+4847 is an L6 with old disk kinematics~\citep{2010ApJ...723..684B}. It is also a slow-rotating object with $V\sin i <$ 12.5 km/s, at a distance of 10.2$\pm$1.4~pc~\citep{2008AJ....135..580R}. This source was selected by seven spectral index combinations, and was best modeled by a combination of L6.0$\pm$1.1 and L9.2$\pm$1.1 components. Its spectrum shows no signs of methane absorption in the \emph{H} band given the late L type of the secondary, but it does present a prominence in the \emph{K} band at 2.2$\mu$m.

\subsubsection{2MASS J17072529$-$0138093}
2MASS J1707$-$0138 was discovered and classified as an L2 by~\citet{2010A&A...517A..53M}. Selected by five spectral index combinations, its spectrum is best fit by components with L0.7$\pm$0.5 and T4.3$\pm$2.0 spectral types. Its spectrum shows a strong absorption feature centered at 1.63$\mu$m.

\subsection{Visual Candidates}

\subsubsection{2MASSI J1711457+223204}
2MASS J1711+2232 was first identified and classified as an L6.5 in the optical by~\citet{2000AJ....120..447K}. Due to its FeH and CH$_4$ absorption features in the \emph{H} band,~\citet{2010ApJ...710.1142B} suggested it could be a spectral binary with L5.0 and T5.5 components. We find slightly different component spectral types of L1.5$\pm$0.6 and T2.5$\pm$1.0, yet this source was not selected by spectral indices because of its late SpeX spectral type of L8.8. Despite having been imaged with HST/WFPC, it remains unresolved~\citep{2003AJ....125.3302G}.
%no photometric variability~\citet{2014ApJ...782...77B}, khandrika 2013 says marginally yes~\citet{2013AJ....145...71K}

\section{Discussion}\label{sec:discussion}

\subsection{Blue L dwarfs as contaminants}\label{sec:blue}

%Scattered among most of our selection regions, we found unusually blue L dwarfs.  Unusually blue L dwarfs are a subclass of the L dwarfs, whose J-K$_s$ NIR colors are outliers compared to the median color for their spectral type~\citep{2009AJ....137....1F}. Unusually blue spectra is a result of low metallicity, high surface gravity, thin atmospheric clouds or large grain condensates, or unresolved binarity~\citep{2010AJ....139.1045S}, or all of the above.  Both low metallicity and high surface gravity objects show a flattened K band, as a result of collision induced H$_2$ absorption, but this effect is less pronounced than in a subdwarf~\citep{2003ApJ...592.1186B}. A thin or patchy cloud coverage, i.e. reduced cloud opacity, allows more light to emerge in the J band regime resulting in brighter and bluer spectra~\citep{2001ApJ...556..872A,2008ApJ...674..451B}. 

%Median and average colors for very low mass dwarfs have been calculated by~\citet{2000AJ....120..447K},~\citet{2010AJ....139.1808S} and~\citet{2011AJ....141...97W} for the spectral range M0-T9.

Four of the candidates selected by spectral indices were rejected after spectral fitting due to their poor binary fits. When we investigated these sources in detail, we found they were classified as blue objects in the literature and/or showed an unusually blue spectrum. 2MASS J11181292$-$0856106 was classified as a metal-poor subdwarf by~\citet{2010ApJS..190..100K}.  SDSS J141624.09+134826.7 is part of a resolved binary system with a T7.5 companion~\citep{2010MNRAS.404.1952B,2010AJ....139.2448B,2010A&A...510L...8S} that is itself a blue outlier.~\citet{2010ApJ...710...45B} rejected unresolved binarity for the primary based on a qualitative comparison to the unusually blue L dwarf 2MASS J11263991$-$5003550. The L6 2MASS J15150083+4847416 shows a stable RV of -29.97$\pm$0.14~km s$^-1$~\citep{2003IAUS..211..197W} and no signs of binarity from its spectrum. Finally, 2MASS J17114558+40285779 was discovered by~\citet{2008ApJ...689..471R} as an unusually blue wide companion to the K star G203-50. They discuss the possibility that the object may be unusual due to unresolved binarity, but argue in favor of low metallicity. For all of these sources, the lack of single templates akin to blue objects resulted in statistically better binary fits, yet the match is still relatively poor around the 1.63~$\mu$m methane absorption feature.

A few more previously unidentified NIR subdwarfs were also selected as weak candidates and subsequently rejected due to their poor binary fits. The best binary fits for 2MASS J03303847$-$2348463, 2MASS J03301720+3505001, 2MASS J04024315+1730136, 2MASS J15412408+5425598, and 2MASS J23311807+4607310 use another subdwarf as a primary, which again indicates that they are part of a rare blue population that has a short supply of examples in this sample.

M+T binaries have slightly bluer spectra caused by the extra flux in the \emph{J} band corresponding to the peak in the T dwarfs. Particularly, some sources originally classified as unusually blue have been later identified as spectral binaries (e.g. SDSS J0805+4812~\citealt{2007AJ....134.1330B, 2012ApJS..201...19D}).  In contrast, intrinsically blue L dwarfs have low metallicity, thin cloud coverage, large-grain clouds or a combination of these, causing a blue tilt to the NIR spectrum~\citep{2010AJ....139.1808S,2008ApJ...674..451B,2007AJ....133..439C}.~\citet{2009AJ....137....1F} have defined red and blue photometric outliers as the objects whose \emph{J-K$_s$} color placed them 2$\sigma$ or 0.4~mag away from the average for their spectral type, while pointing out the difficulty to distinguish outliers beyond a spectral type of L9 due to the small sample of objects. Figure~\ref{fig:J-K} shows the \emph{J-K} colors for our sample as compared to their spectral types, including the median and $\pm2\sigma$ lines as calculated from the sample (solid lines) and reported in the literature (dashed lines) by~\citet{2011AJ....141...97W} and~\citet{2010AJ....139.1808S} for samples of M and L dwarfs, respectively. Figure~\ref{fig:J-K} suggests that blue L dwarfs are a major contaminant in our sample since a significant fraction of both known binaries and candidates have similar colors and thus lie in the same region as blue sources. We conclude that the blue L dwarf contaminants can be recognized if rejected due to their poor fits to binary template spectra.

%important to expand spex library to include more blue examples. only 1422 and 1453 are blue and candidates.

% x number of the not previously identified are blue or not according to faherty

%Other definitions.
%the calculated median and $\pm 2\sigma$ limits are shown, compared against those reported in the literature. Clearly the spread is much more pronounced in the later L spectral types, as mentioned by~\citet{2009AJ....137....1F}.  Despite the distribution of J-K$_s$ color points per spectral type being notoriously non-Gaussian, the use of standard deviation as a delimiter was the best option considered.
%While blue dwarfs typically have an overall ``tilt'' in their spectrum, unresolved binary spectra show absorption features associated with a hidden T dwarf companion. 

\subsection{Separation distribution of binary systems}\label{sec:separation}

True confirmation of our candidates requires observational follow-up to either resolve the systems or measure RV or astrometric variability. As noted in the introduction, spectral binaries can be used to devise an unbiased method to measure the VLM binary separation distribution. Therefore, it is worth examining the separation distribution of VLM and brown dwarf spectral binaries confirmed to date, to see if there are any differences compared to the resolved population.

Figure~\ref{fig:separation} shows the distribution of projected separations from 122 confirmed VLM binaries\footnote{Based on the compilation at the Very-Low-Mass Binaries Archive, http://www.vlmbinaries.org, and more recent discoveries by~\citet{2013ApJ...768..129C, 2013A&A...555A.137D, 2013ApJ...767L...1L, 2013ApJ...778...36R, 2013A&A...556A.133S, 2012ApJ...757..110B, 2012ApJ...758...57L, 2011ApJ...739...48A, 2011ApJ...739...49B, 2011AJ....141....7D, 2011AJ....142...57G, 2011ApJ...732...56G, 2011ApJ...740..108L, 2010ApJ...715..561A, 2010ApJ...710.1142B, 2010ApJ...723..797H, 2010A&A...516A..37S, 2009ApJ...697..824A, 2009ApJ...691.1265L}.}. Among the observational methods for detecting binaries, such as direct imaging, radial velocity variations, astrometric variations, and microlensing, direct imaging has proven to be the most successful so far (73$\%$ of confirmed VLM binaries), but its biggest drawback is its limit in resolution.  At minimum angular scales of 0$\farcs$1-0$\farcs$2 for AO and HST programs, and typical distances of field brown dwarfs of 20-30 pc, telescope sensitivity reaches its limit at separations of around 2-6 AU. At 2.90 AU, the mean projected separation of eight independently-confirmed spectral binaries plotted in Figure~\ref{fig:separation} falls at the lower end of this sensitivity limit, at less than the mean of known VLM binaries excluding the spectral binaries (3.75 AU), raising the possibility that there may be significantly more tightly bound systems. 

To assess whether this is a significant difference, we performed a two-sample Kolmogorov-Smirnov test comparing the projected separation distributions of all binary systems to the confirmed spectral binaries.  Specifically, the distributions were constrained in angular separation to 50-500 mas, where the lower limit corresponds to the smallest possible imaging resolution in good seeing, while the upper limit restricts the maximum size of the slit. In addition, the distance was constrained to less than 30 pc, since objects that are farther away would be more difficult to confirm as binaries. In this way, we intend to fairly compare the spectral binary method to the other available methods for binary detection. These constraints reduced the number of spectral binaries to six. The result was a D statistic of 0.41 and a probability of $25\%$. While the low probability is indicative of a difference between the samples, the small sample size makes this statistic inconclusive. Many more of the existing spectral binaries need to be characterized before a significant difference can be confirmed or ruled out. 

%You can also add a statement in the separation distribution discussion that this sample isn't complete and will require simulations to assess the selection function, work that will appear in a future paper.
 
\section{Summary}\label{sec:summary}

We have identified fourteen brown dwarf binary candidates with late-M/early-L plus T dwarf components based on visual inspection of low resolution data, and analysis with spectral indices and template fitting.  We combined five new spectral indices, with previously defined ones, spectral type, and \emph{J-H}, \emph{H-K$_s$}, and \emph{J-K$_s$} colors to define pairings that effectively select spectral binary candidates, and confirmed them by comparison to both single and binary template spectra from the SpeX Prism Library.  Unusually blue L dwarfs were the main contaminant of this analysis, with four candidates classified as unusually blue but nonetheless being poorly matched to binary spectra. Exploring the separation distribution of binary systems we find suggestive evidence that spectral binaries are more closely separated than other binaries, but the confirmed sample is too small to be conclusive. We are now undertaking follow-up AO imaging and RV monitoring of these candidates to confirm them and measure orbital properties.

%~\citet{2007AJ....134.1330B} and~\citet{2002ApJ...564..421B} plus 

\section*{Acknowledgments}
The authors thank telescope operators for their assistance during our observations. DBG would like to thank the Friends of the International Center at UCSD for their generous scholarship as well as Davy Kirkpatrick and fellow graduate students Alex Mendez and David Vidmar for their helpful discussion and coding tips. This publication makes use of data from the SpeX Prism Spectral Libraries, maintained by Adam Burgasser at \url{http://www.browndwarfs.org/spexprism}; the Dwarf Archives Compendium, maintained by Chris Gelino at \url{http://DwarfArchives.org}; and the VLM Binaries Archive maintained by Nick Siegler at \url{http://vlmbinaries.org}. The authors wish to recognize and acknowledge the very significant cultural role and reverence that the summit of Mauna Kea has always had within the indigenous Hawaiian community. We are most fortunate to have the opportunity to conduct observations from this mountain.\\
\indent \emph{Facilities:} IRTF (SpeX).

%\bibliographystyle{apj}

%\bibliography{BDlibrary}

%%%%%%%%%%%%%%%%%%

%%%%%%%%%%%%%%%%%%
\begin{figure}[htbp]
\centering
\epsscale{0.7}
\plotone{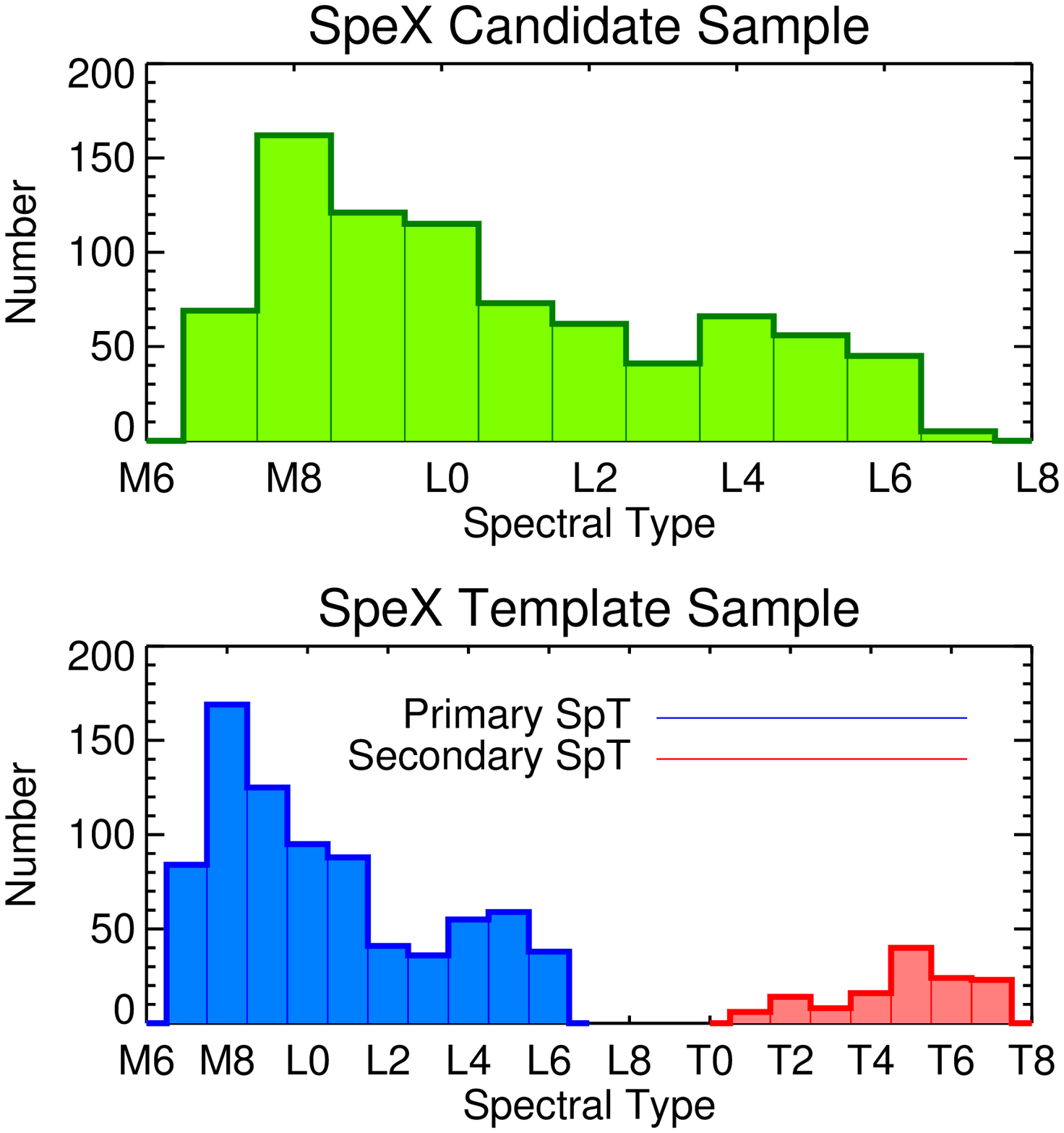}
\caption{Distribution of SpeX spectral types in the samples used for selecting candidates (top) and template fitting (bottom).}
\label{fig:sample}
\end{figure}

%%%%%%%%%%%%%%%%%%
%setcounter sets it to 1, caption adds 1 to it.
\begin{figure}
\setcounter{figure}{1}
\epsscale{0.9}
\plottwo{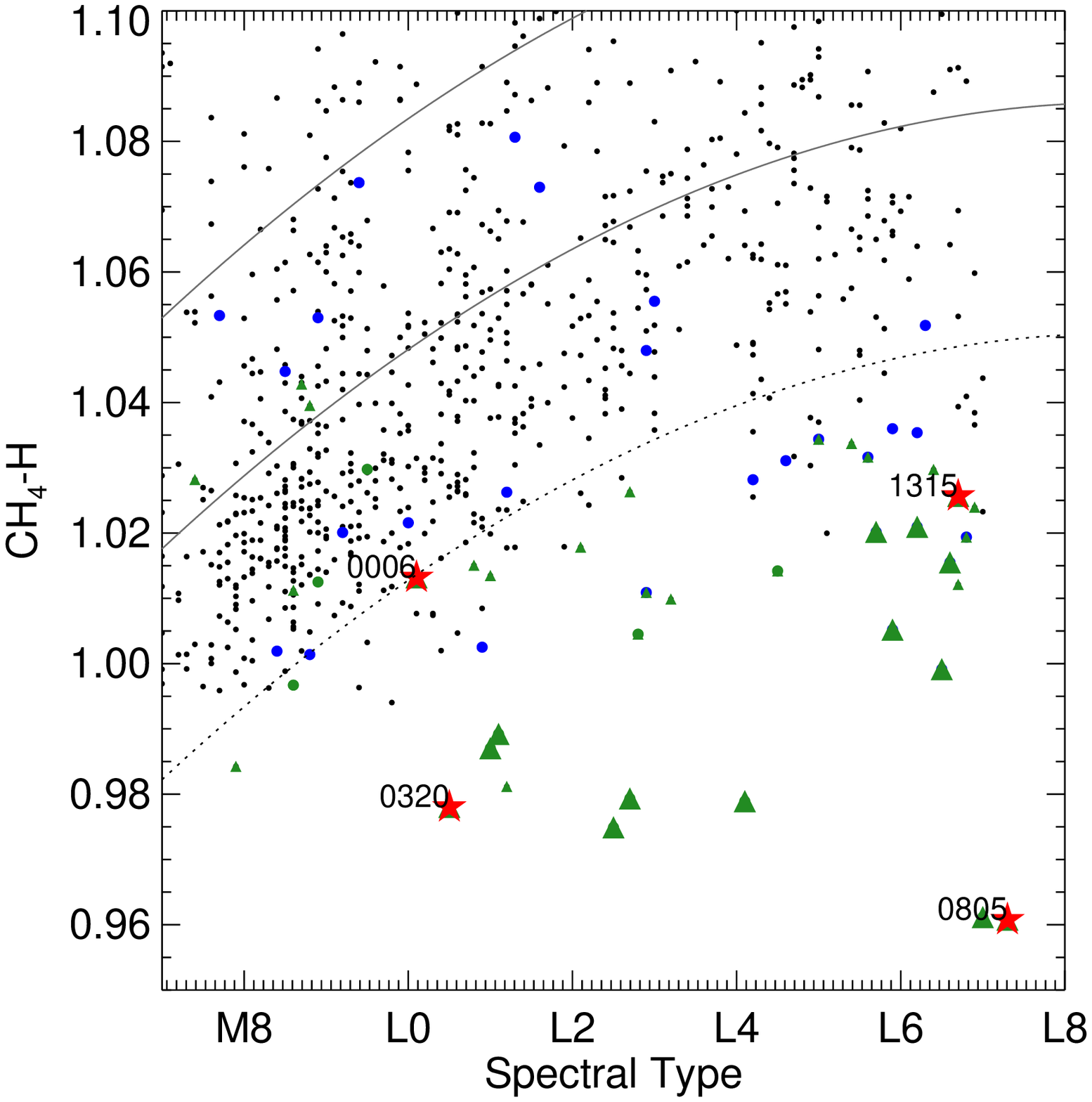}{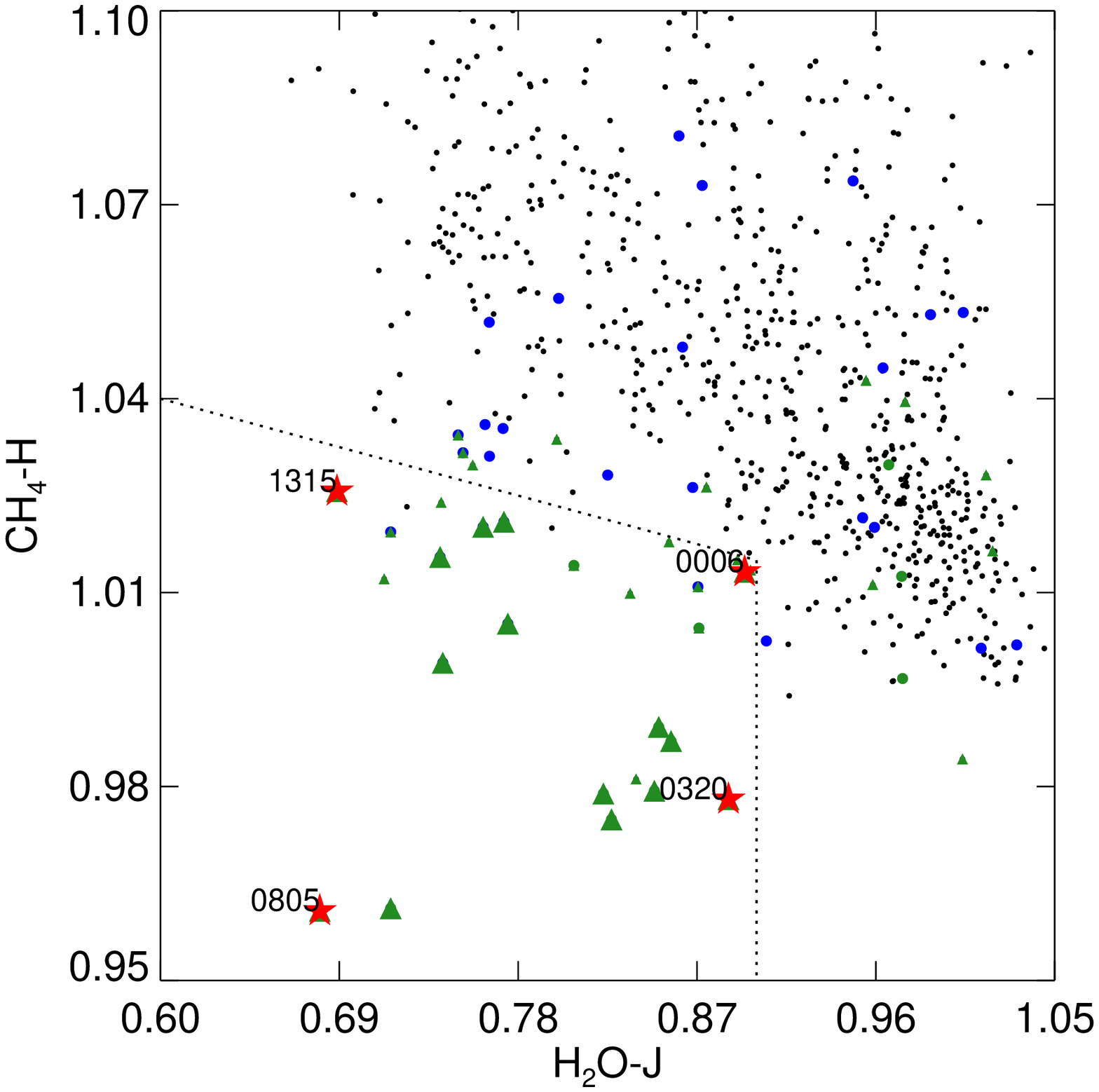}\\
\plottwo{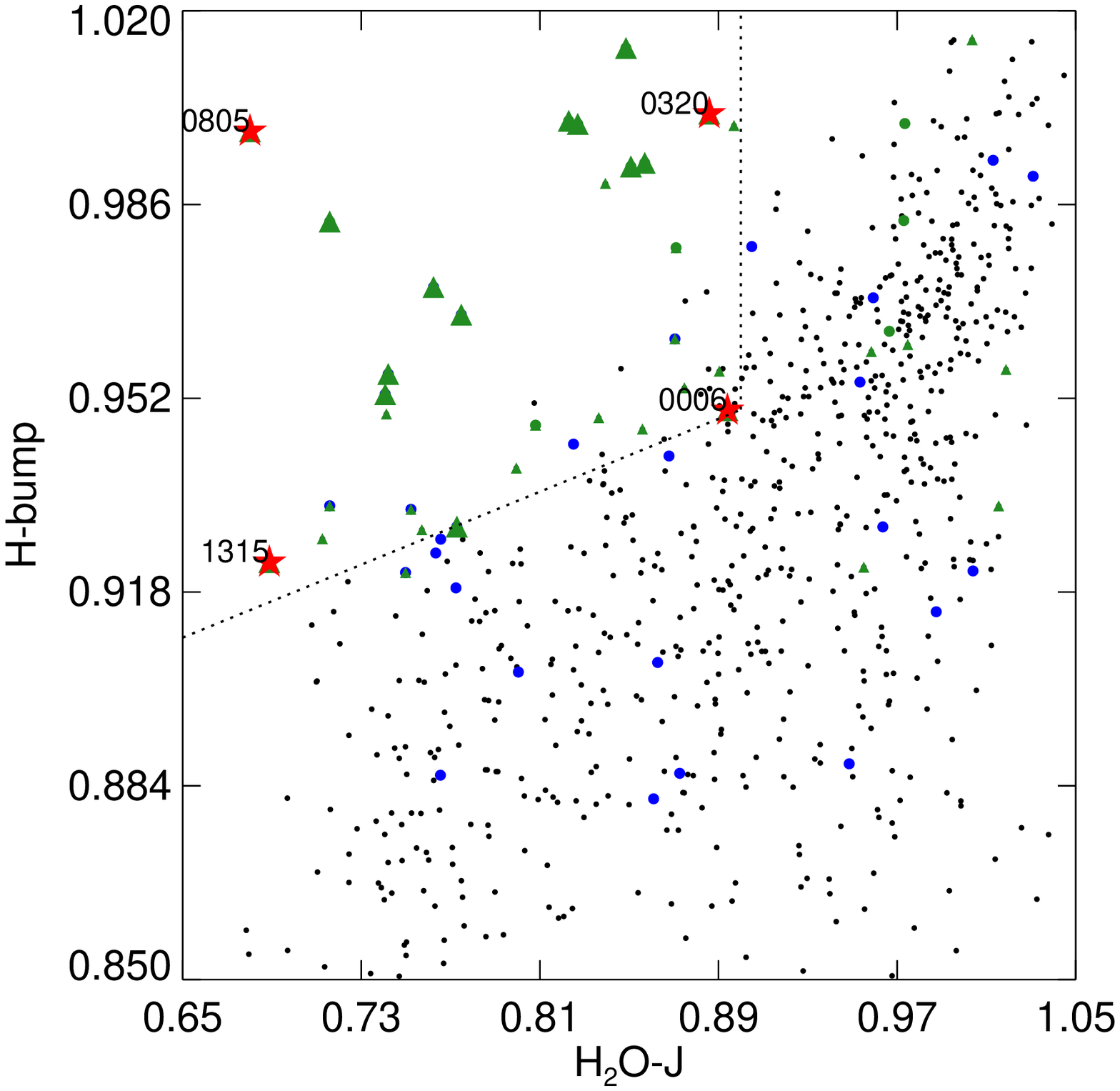}{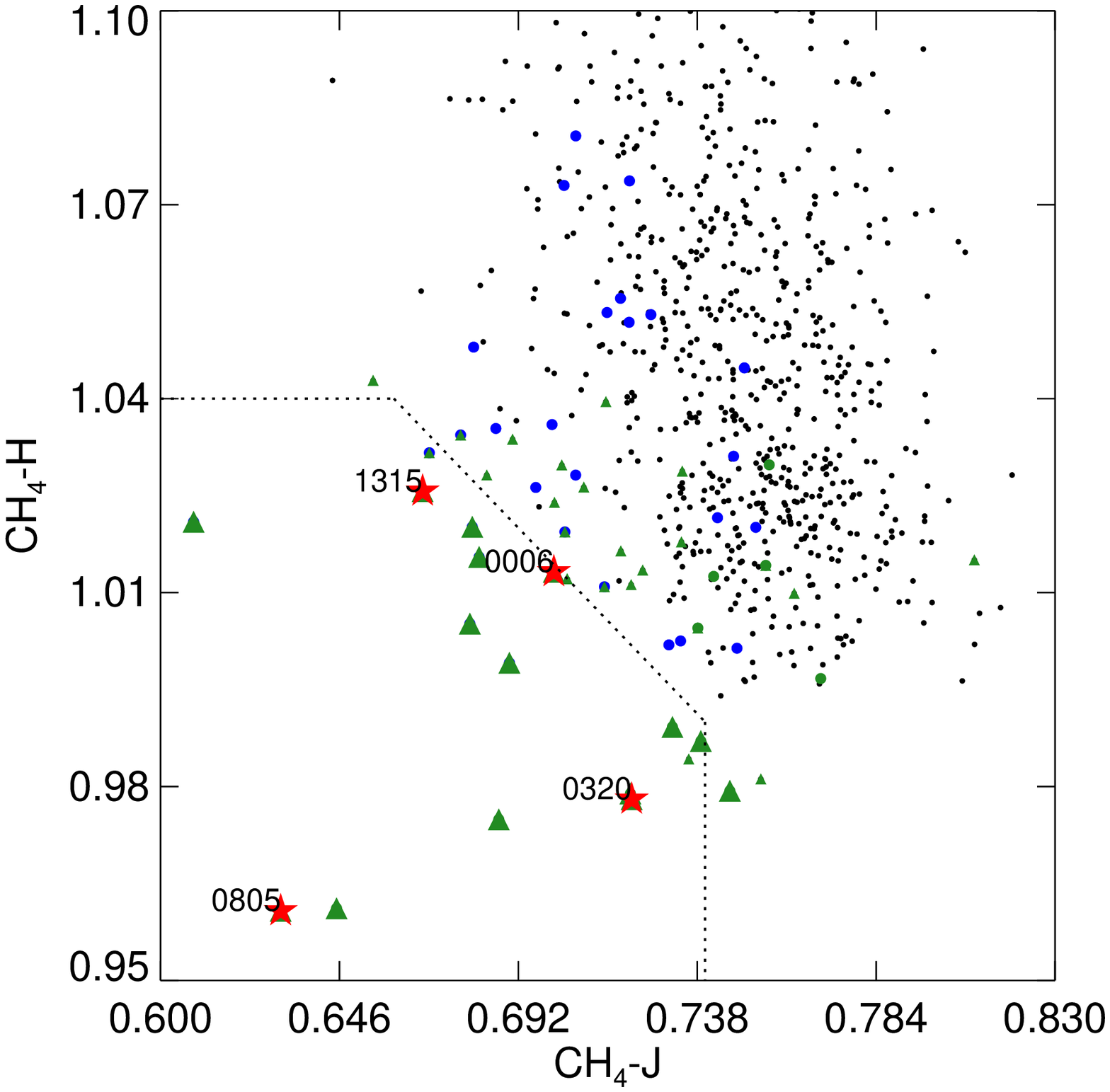}\\
\plottwo{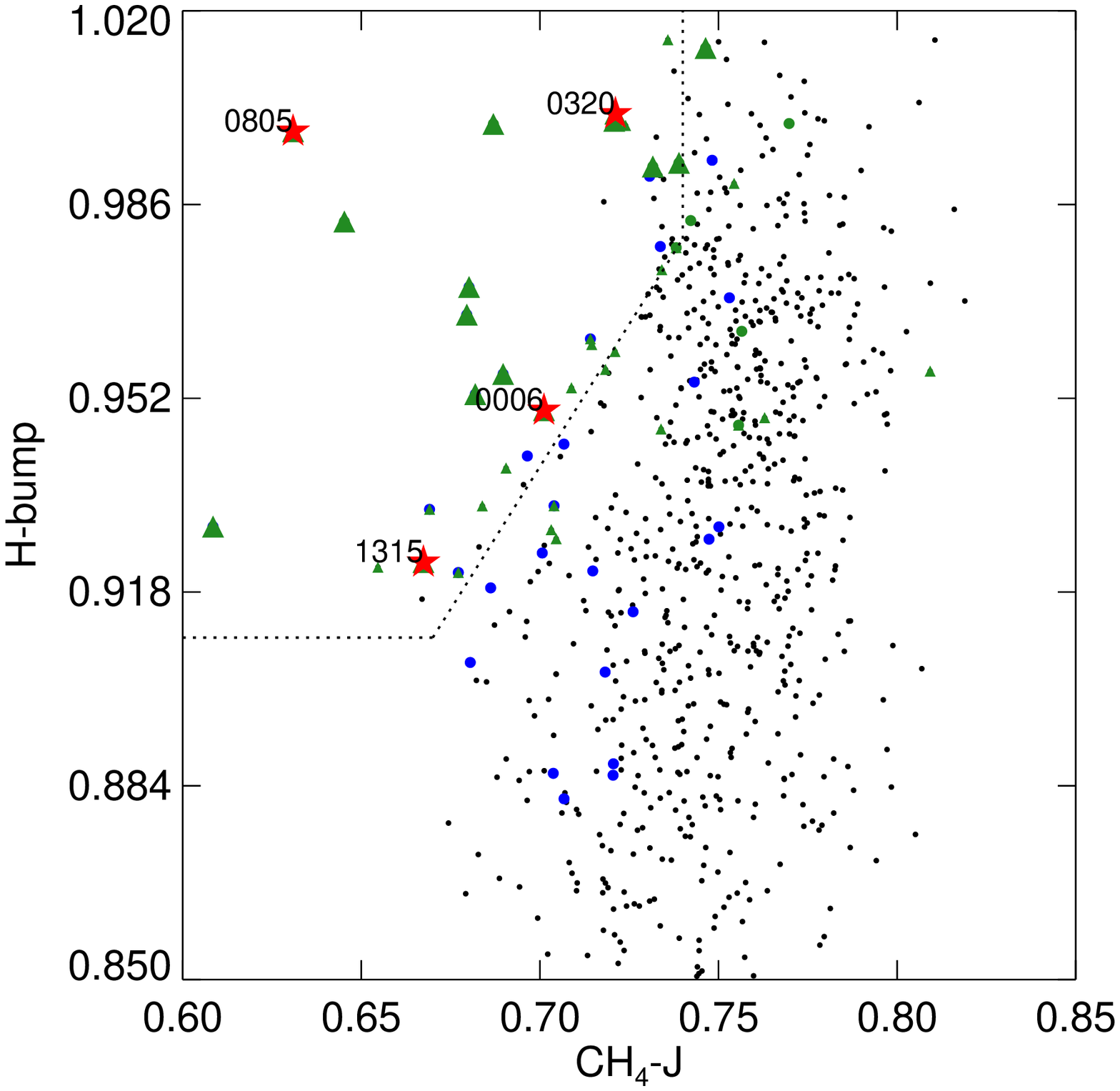}{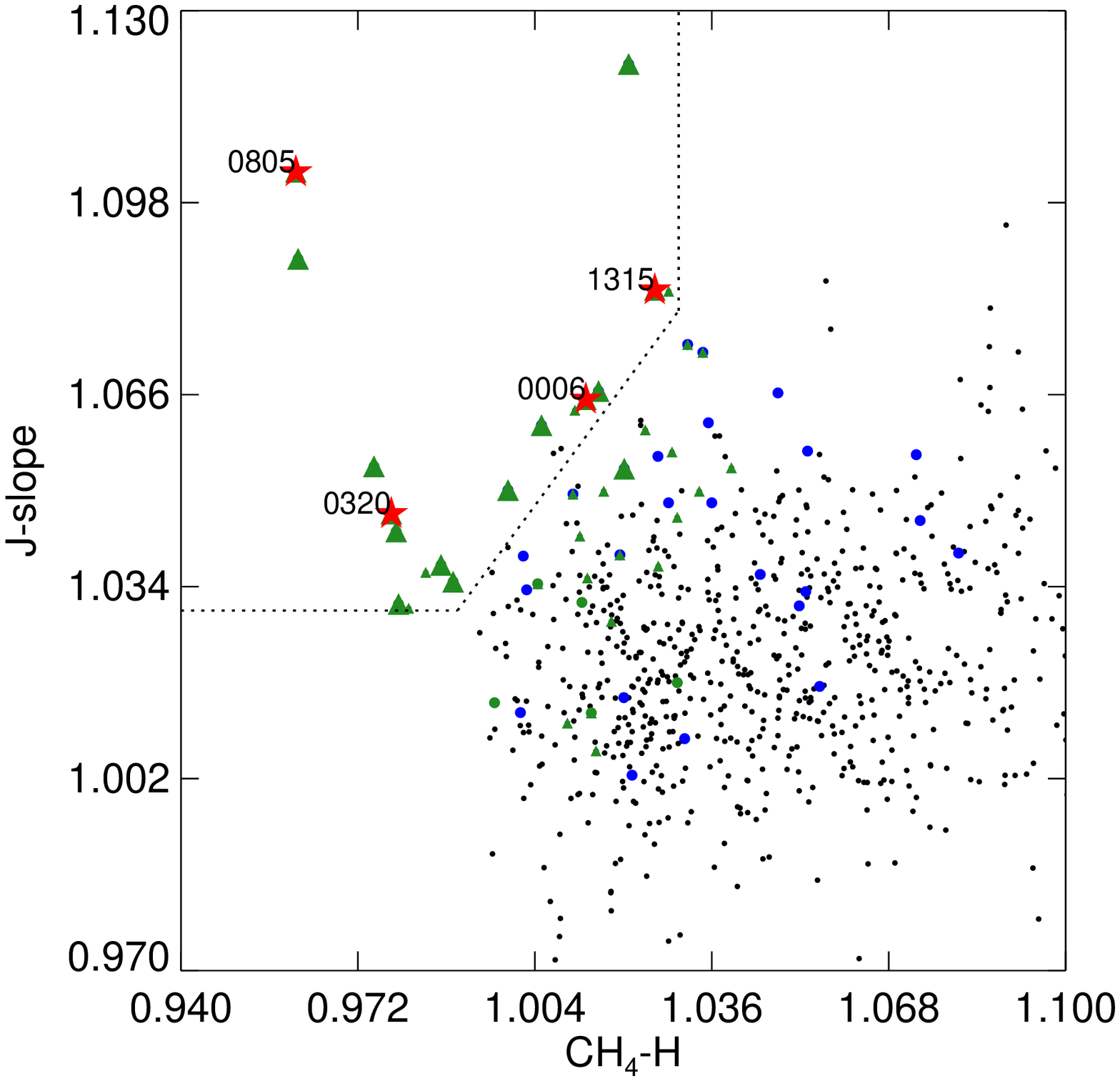}\\
\caption{Index selection of spectral binary candidates. The indices calculated from the candidate sample of SpeX spectra are shown in black. The labeled red stars represent the four binary benchmarks. Unusually blue sources are plotted as blue circles, while the large and small green triangles show the strong and weak candidates, respectively. The green circles represent the visual candidates.\label{fig:paramspace}}
\end{figure}

%%%%%%%%%%%%%%%
\begin{figure}
\epsscale{0.9}
\setcounter{figure}{1}
\plottwo{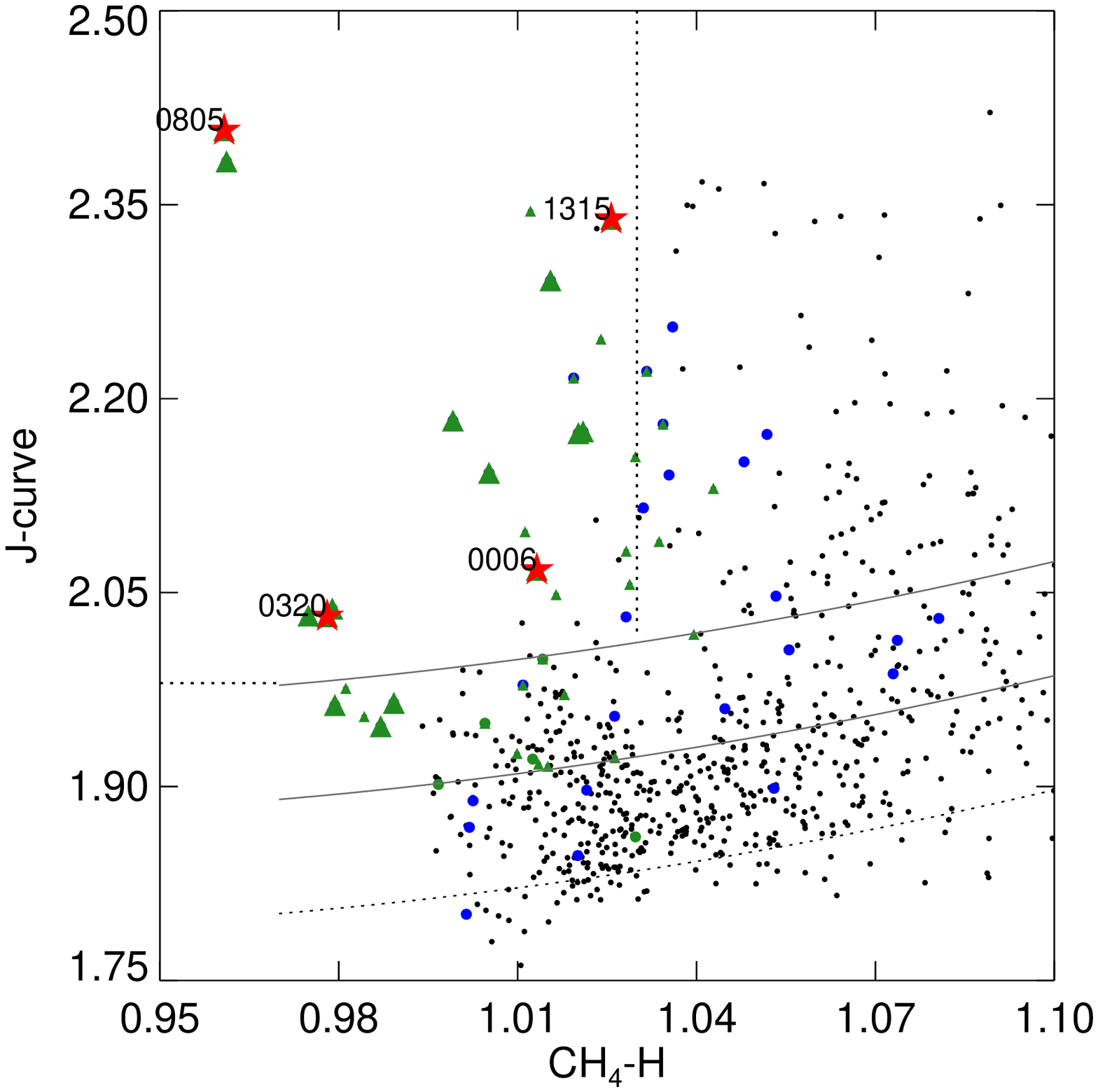}{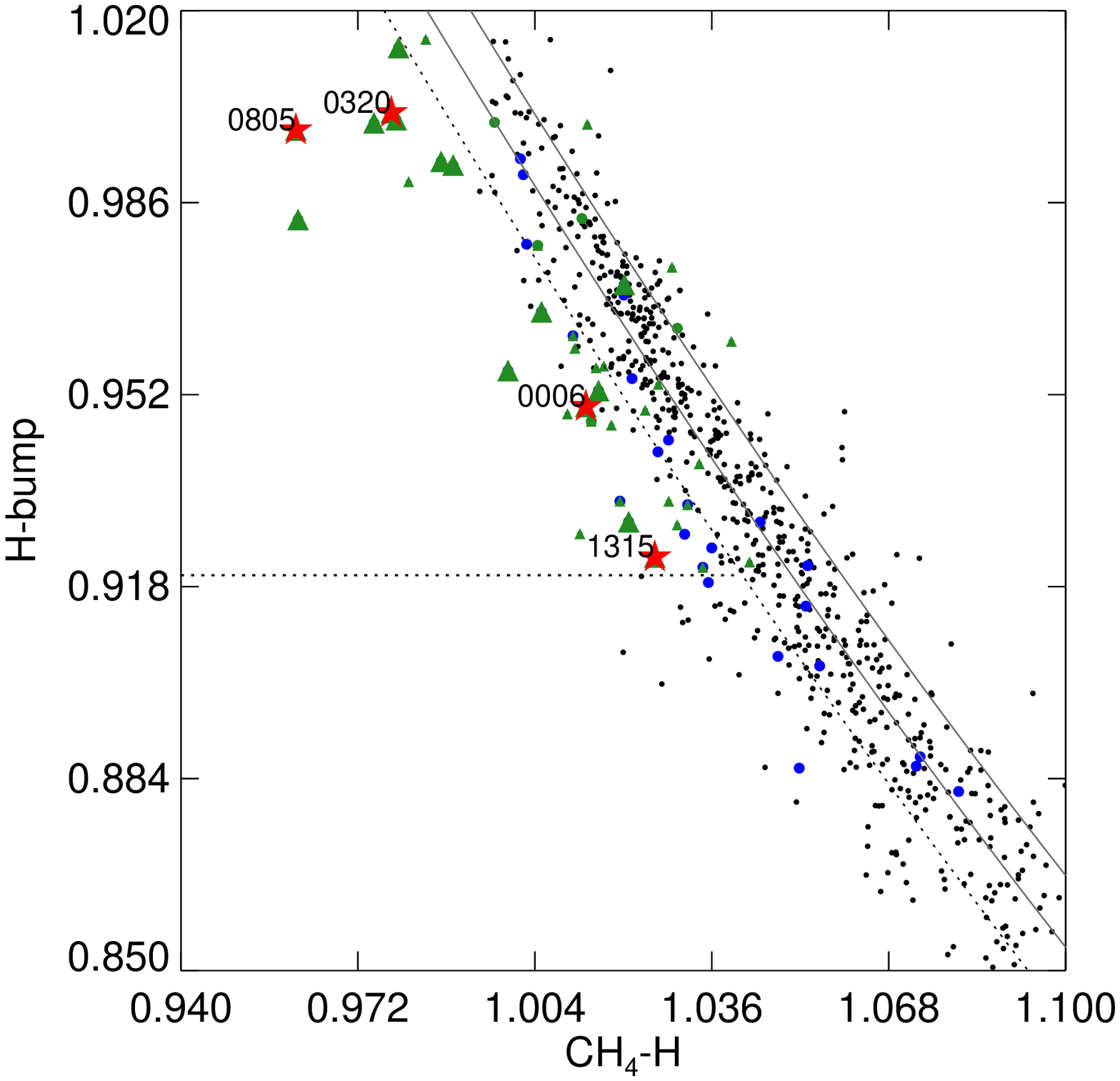}\\
\plottwo{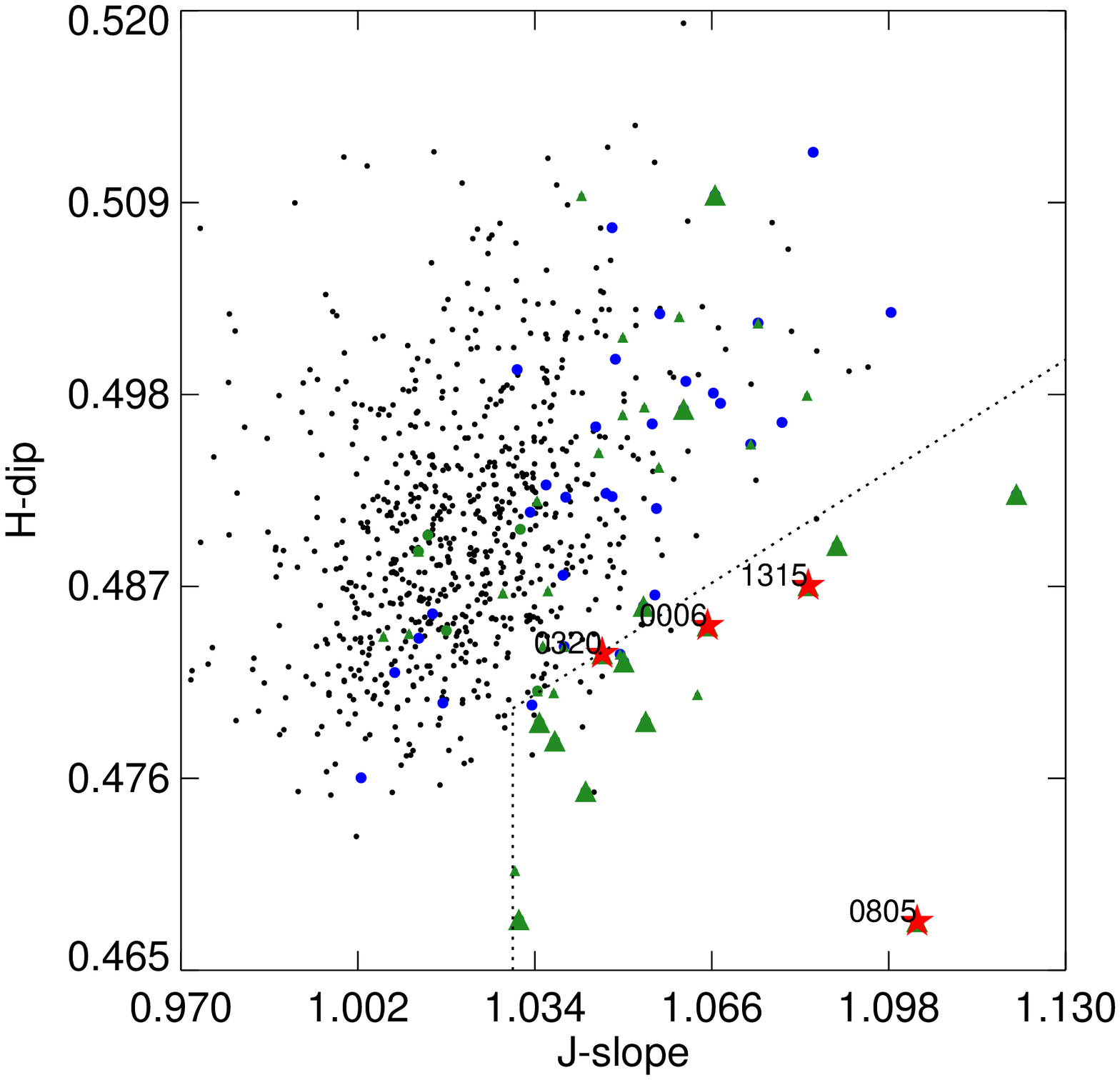}{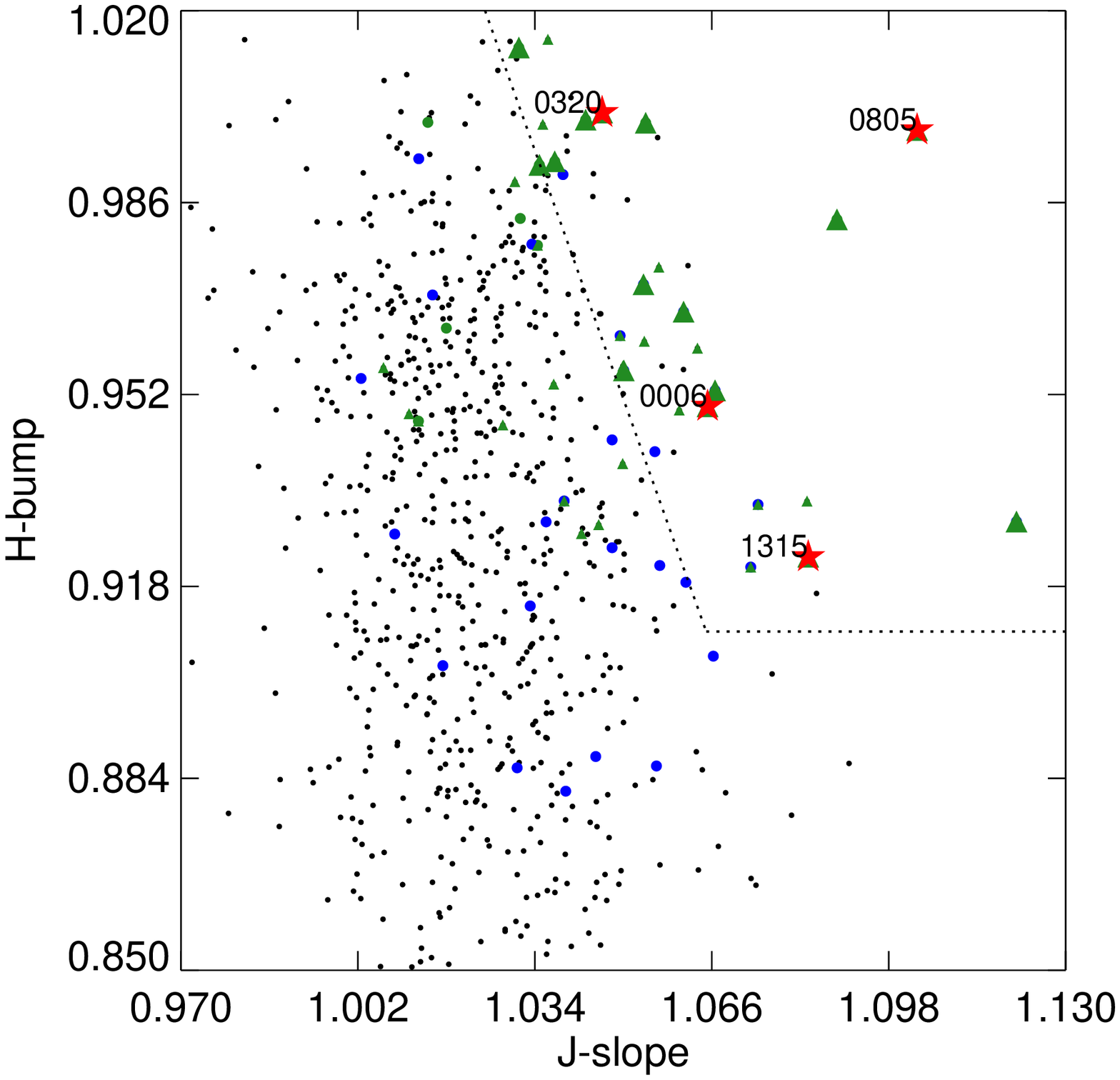}\\
\plottwo{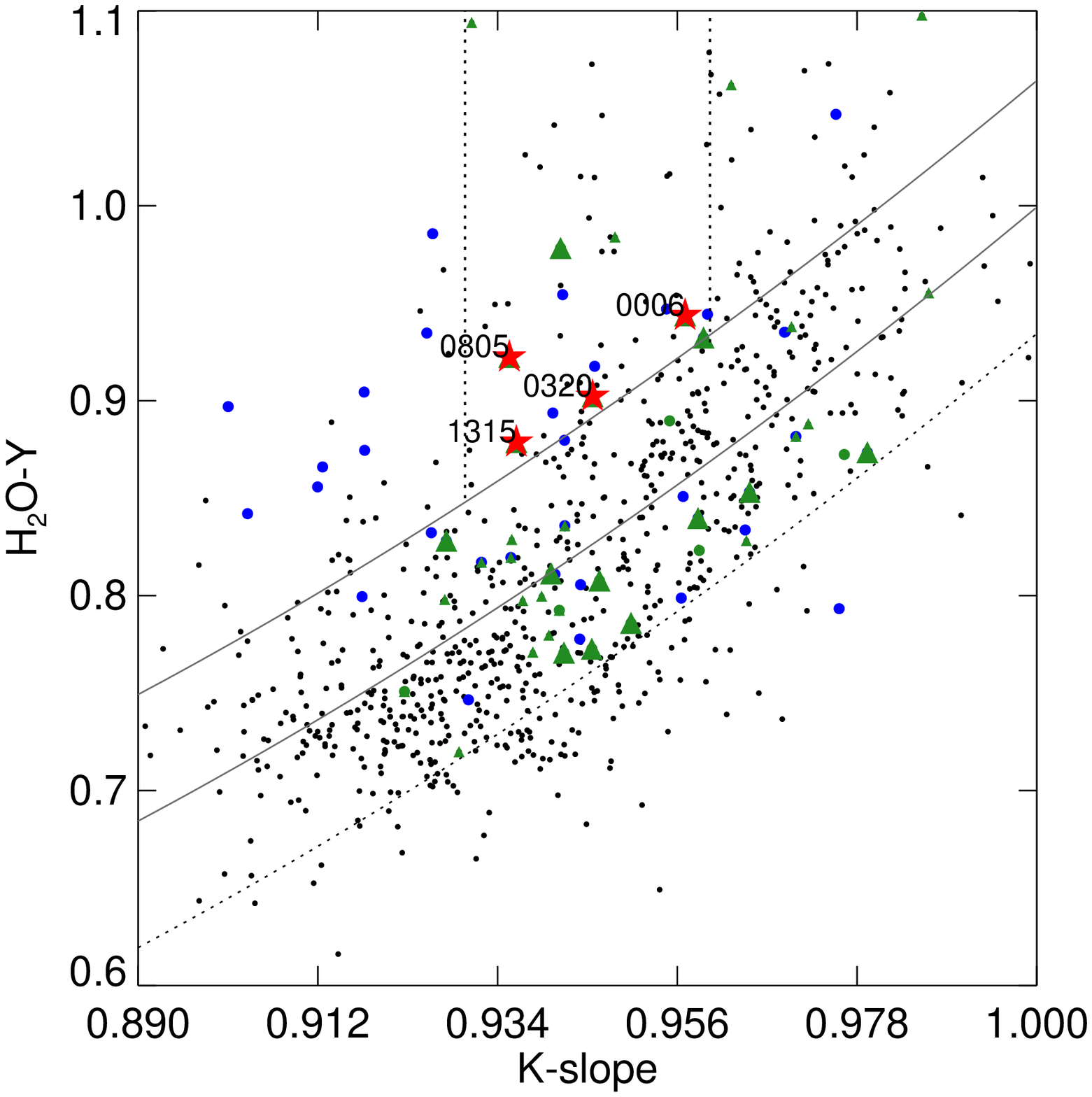}{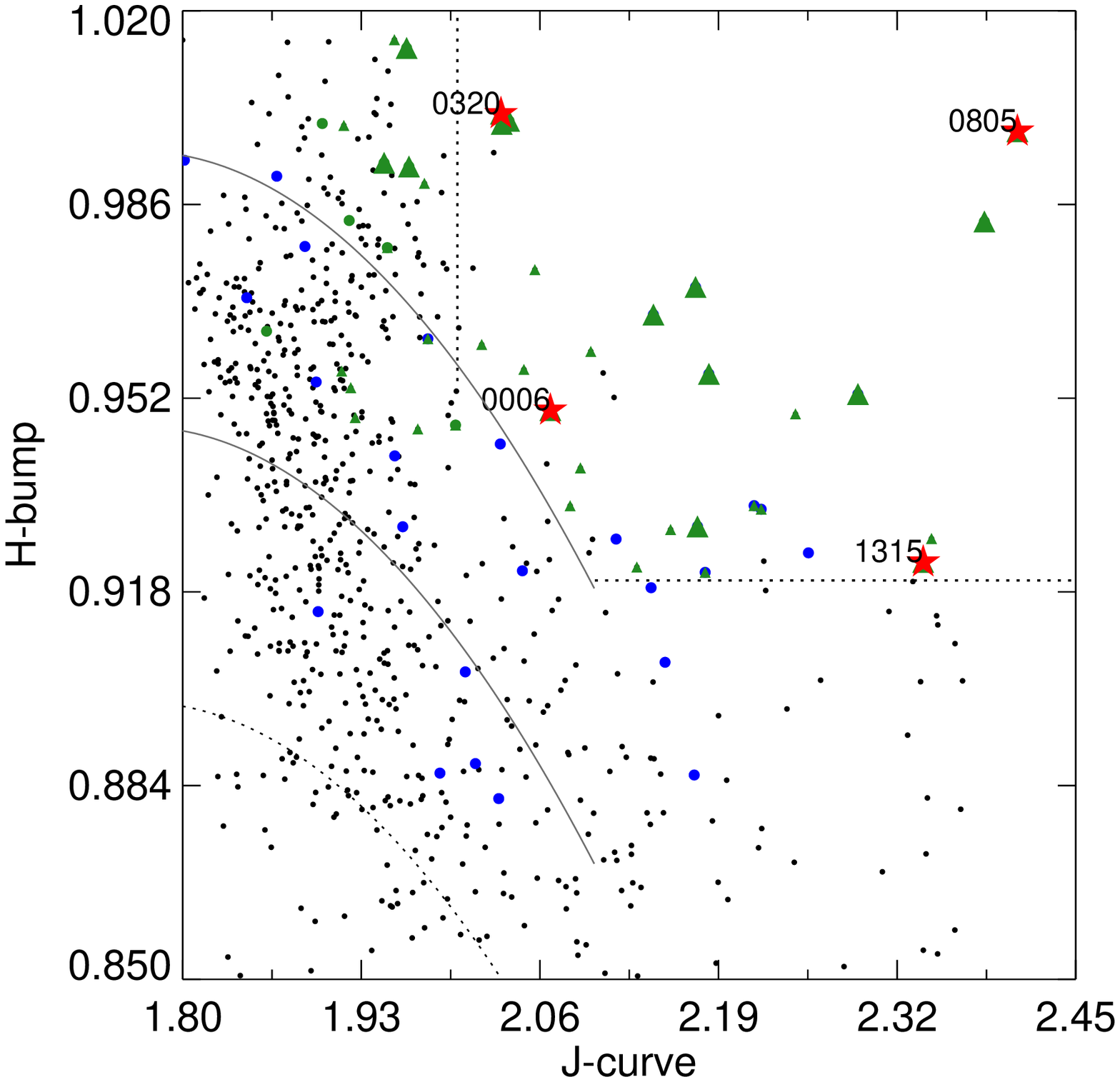}\\
\caption{Continued.\label{fig:paramspace2}}
\end{figure}

%%%%%%%%%%%%%%%
\begin{figure}\label{fig:histogram}
\setcounter{figure}{2}
\epsscale{0.8}
\plotone{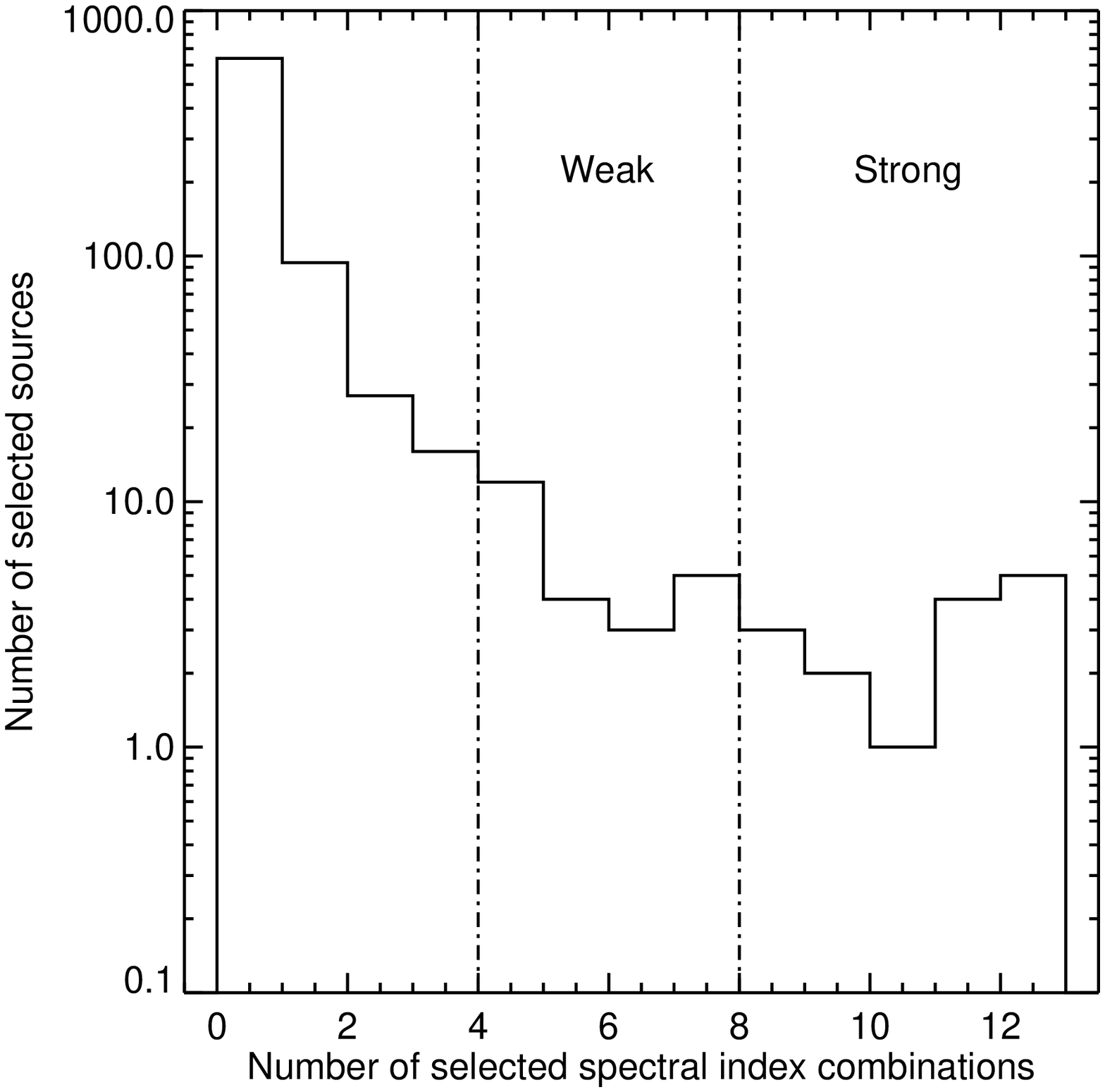}
\caption{The number of sources satisfying index combinations versus total number of combinations. Sources selected 8 or more times are considered \emph{strong} candidates. Sources selected between 4 and 8 times are considered \emph{weak} candidates.}
\end{figure}

%%%%%%%%%%%%%%%%%
\begin{figure}
\epsscale{0.95}
\setcounter{figure}{3}
\plottwo{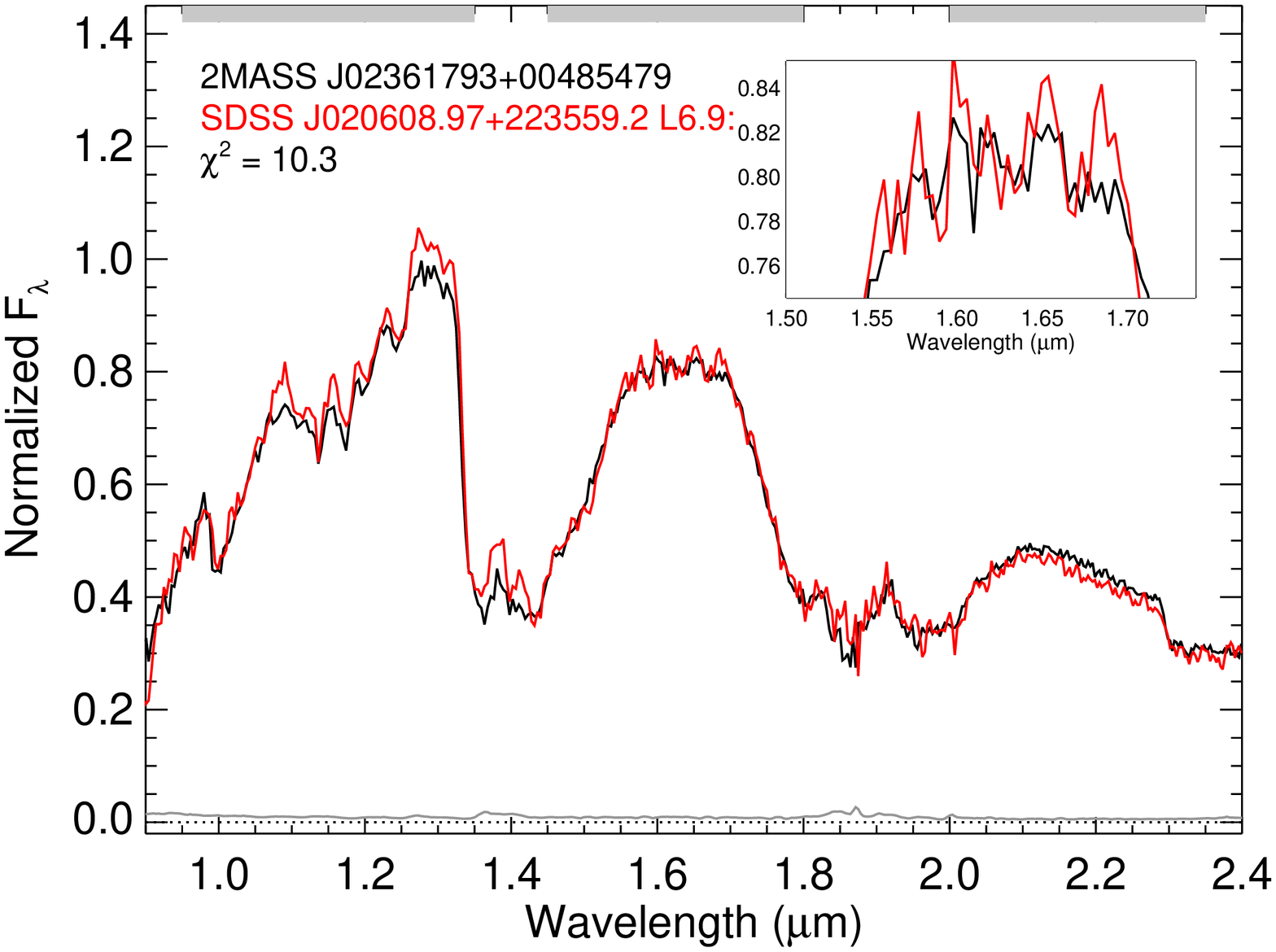}{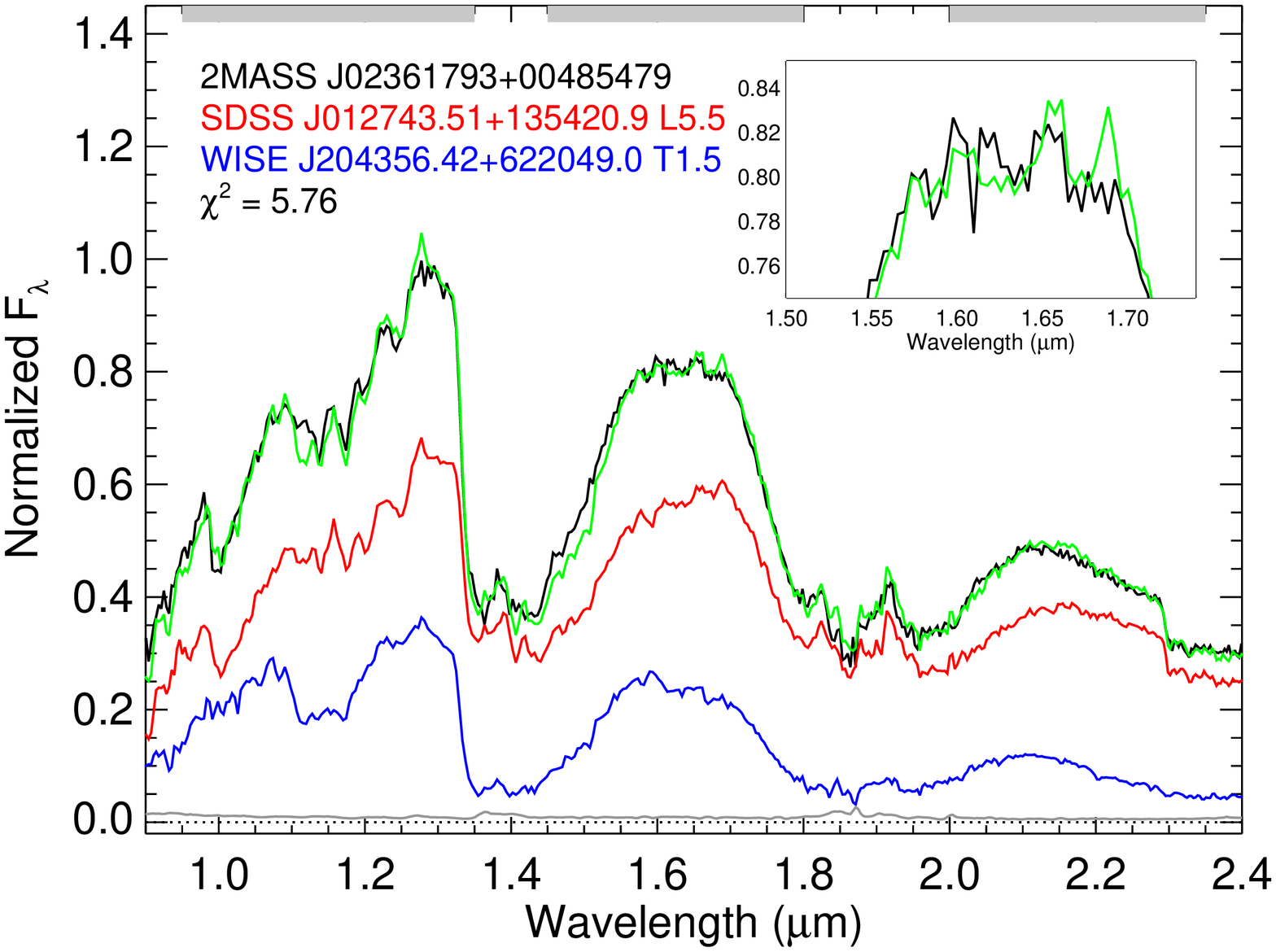}\\
\plottwo{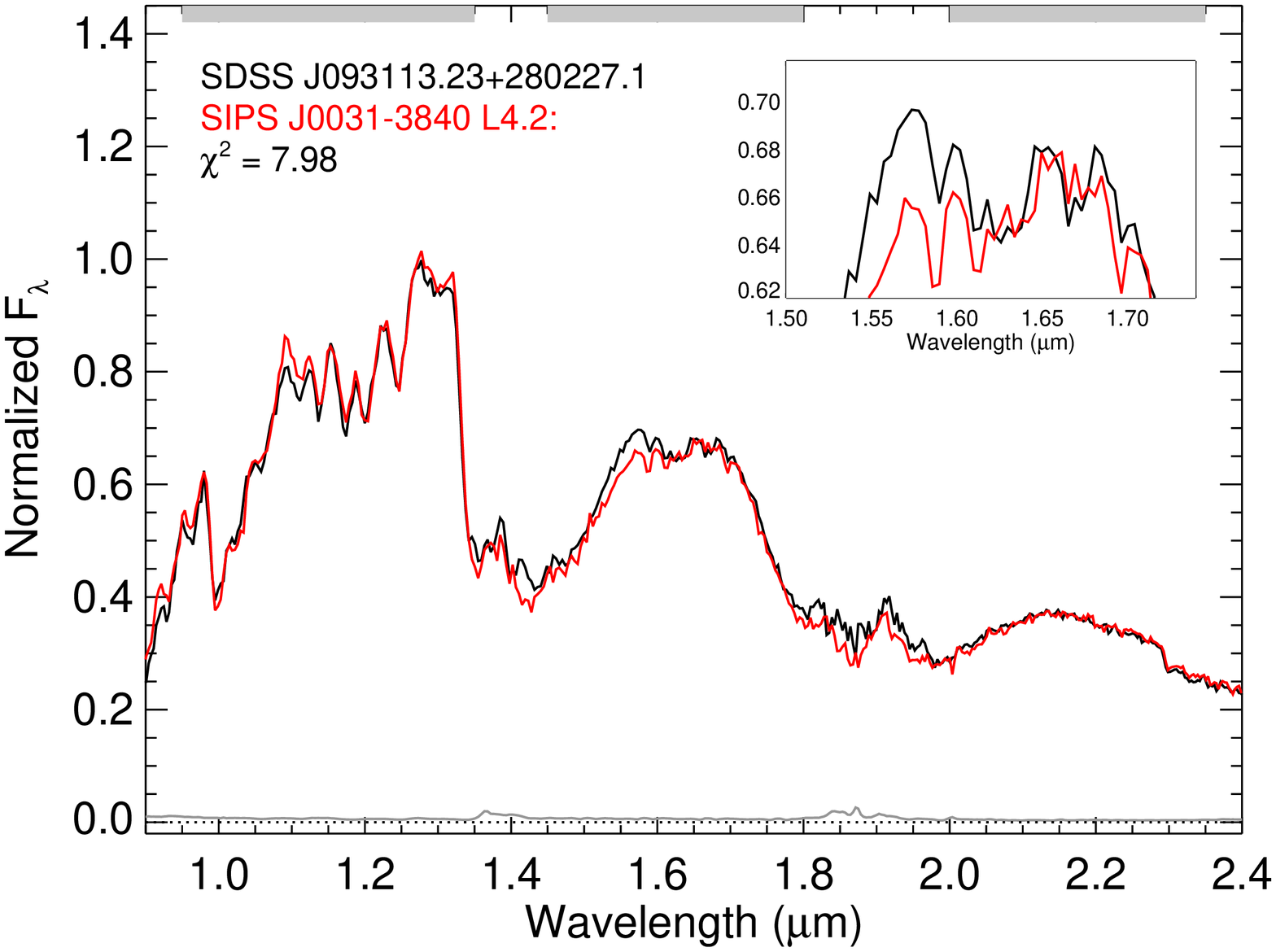}{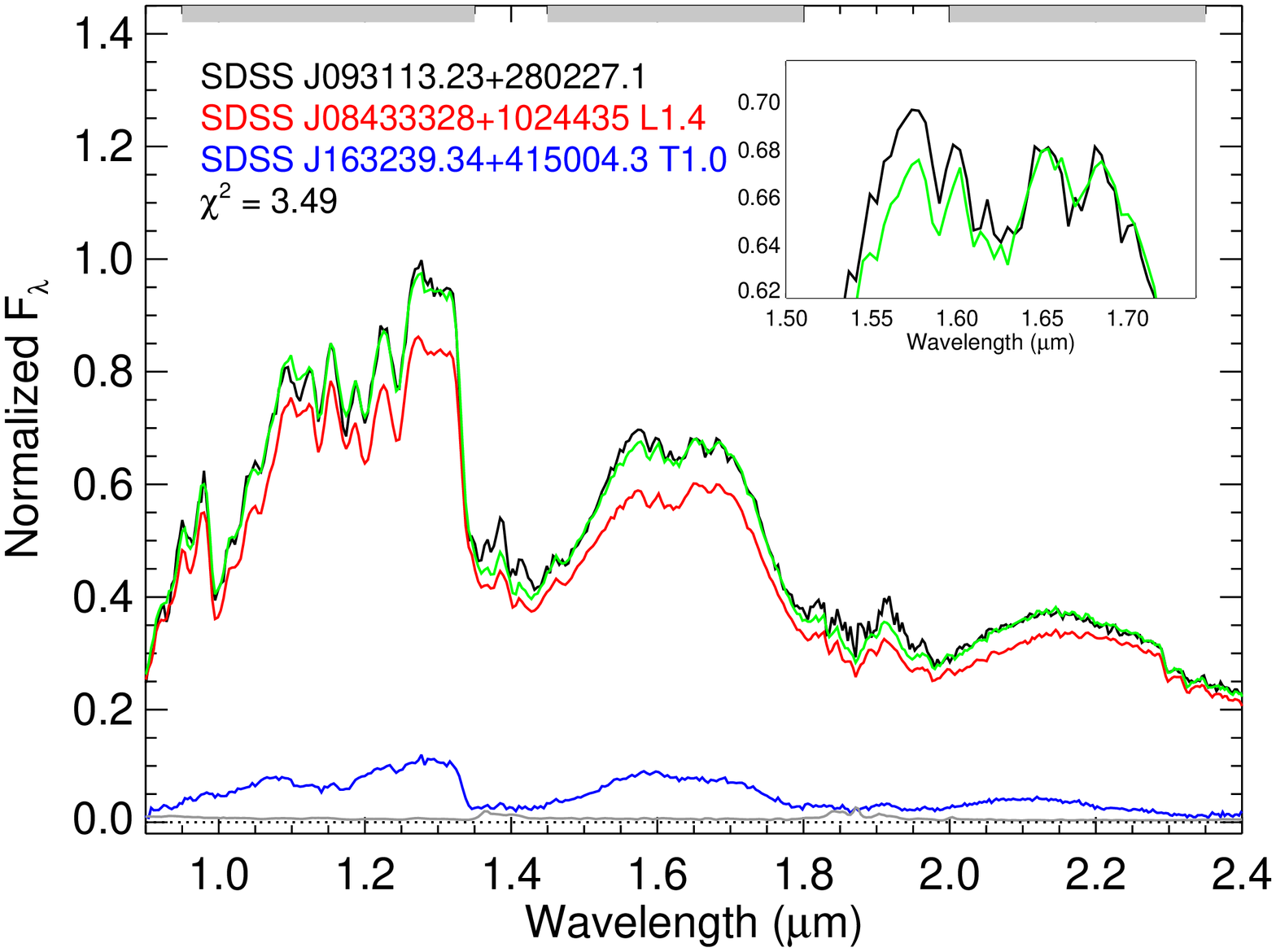}\\
\plottwo{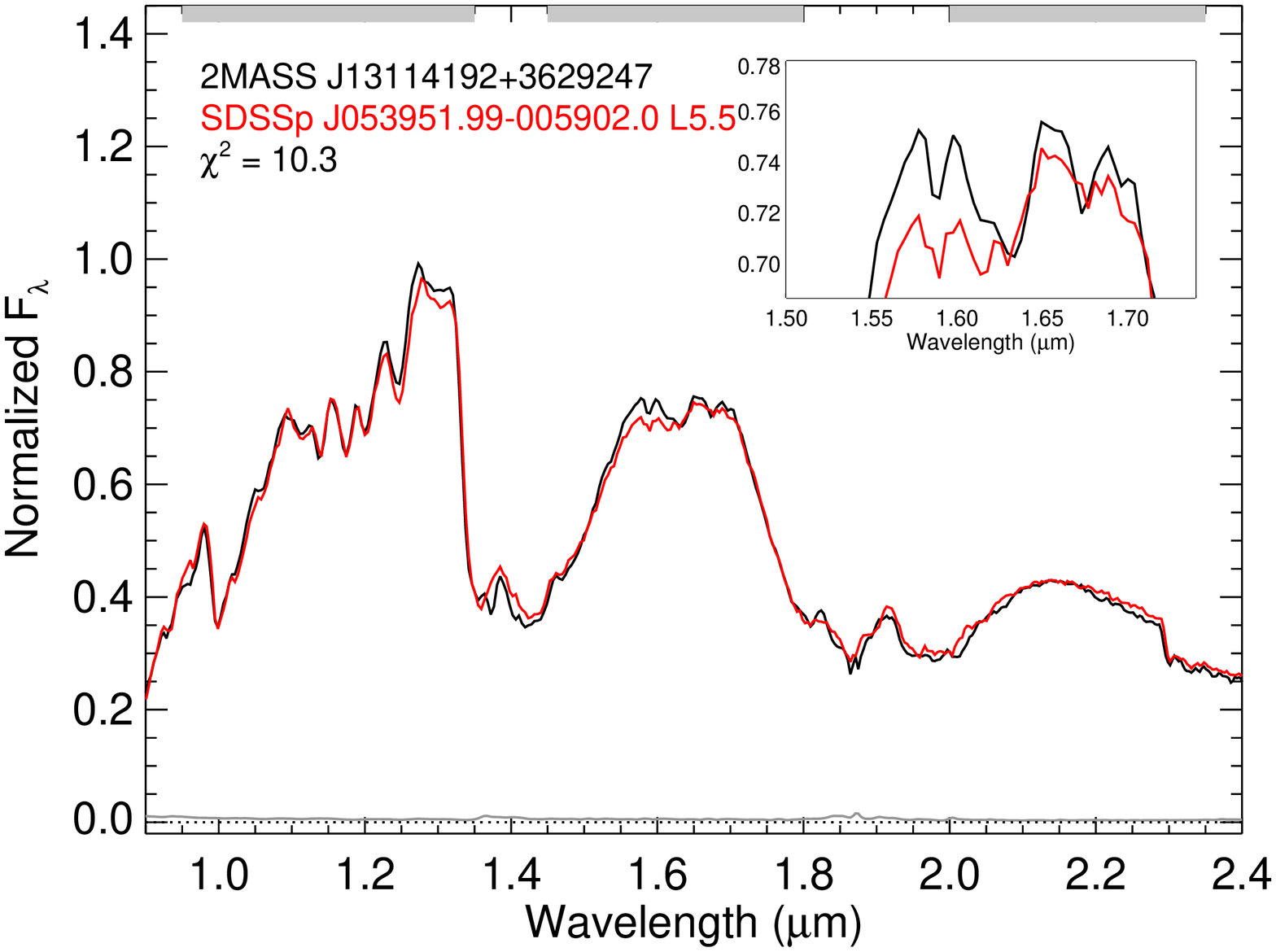}{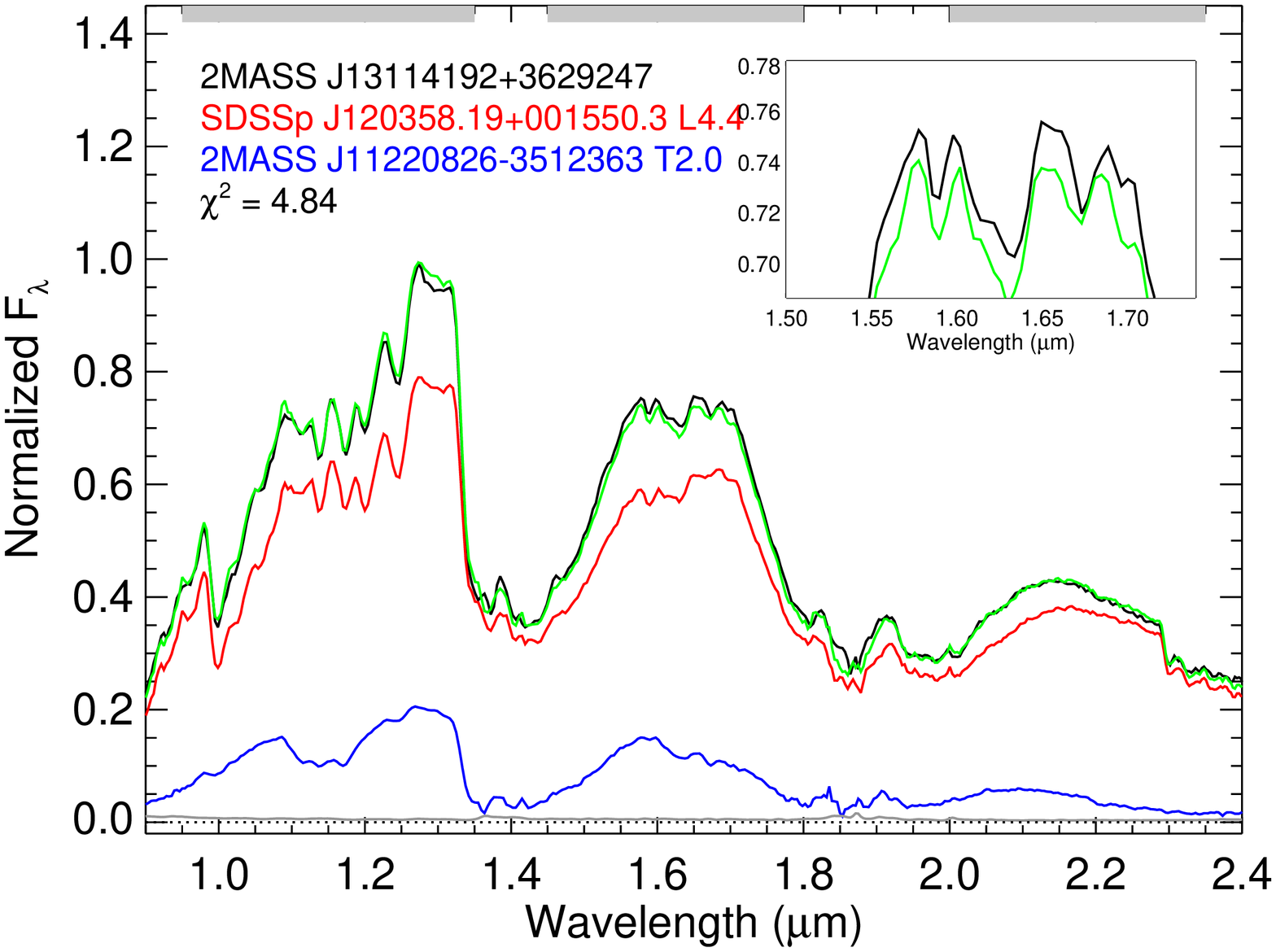}\\
\caption{Best fits to single (left) and binary (right) templates for our strong candidates. The black line shows the candidate spectrum. For the single fits, the red line is the best single template. For the binary fits, the green line is the best binary template, which is the addition of the red (primary) and blue (secondary) lines.  The gray line represents the uncertainty in the candidate spectrum. The gray horizontal bars at the top of the figures mark the parts of the spectrum being fit, while water absorption dominates the gaps. Notice the significant fitting improvement on the binary fits as compared to the single fits, particularly around the methane absorption feature centered at 1.63~$\mu$m (see inset).\label{fig:strongfit}}
\end{figure}

\begin{figure}
\epsscale{0.95}
\setcounter{figure}{3}
\plottwo{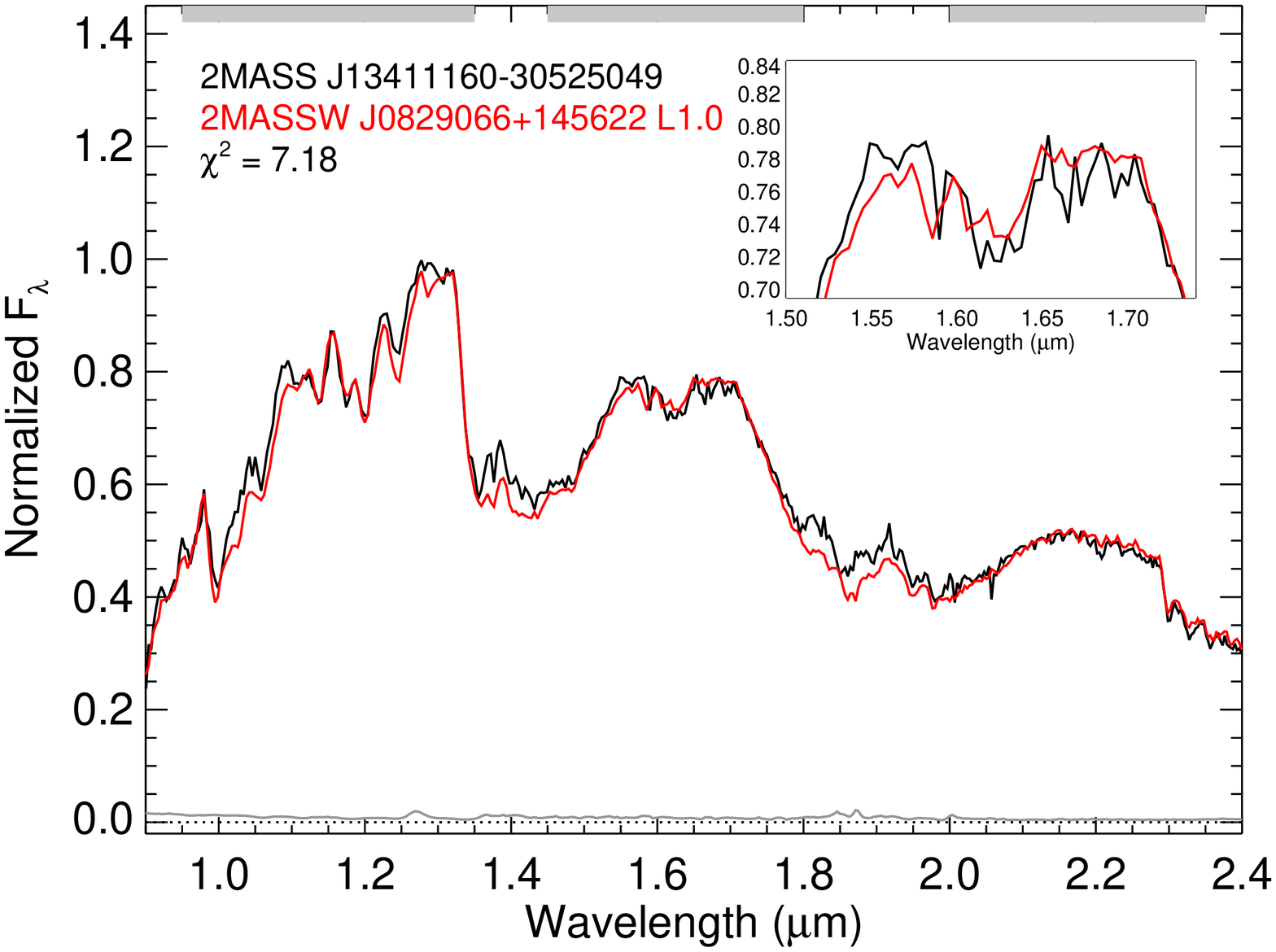}{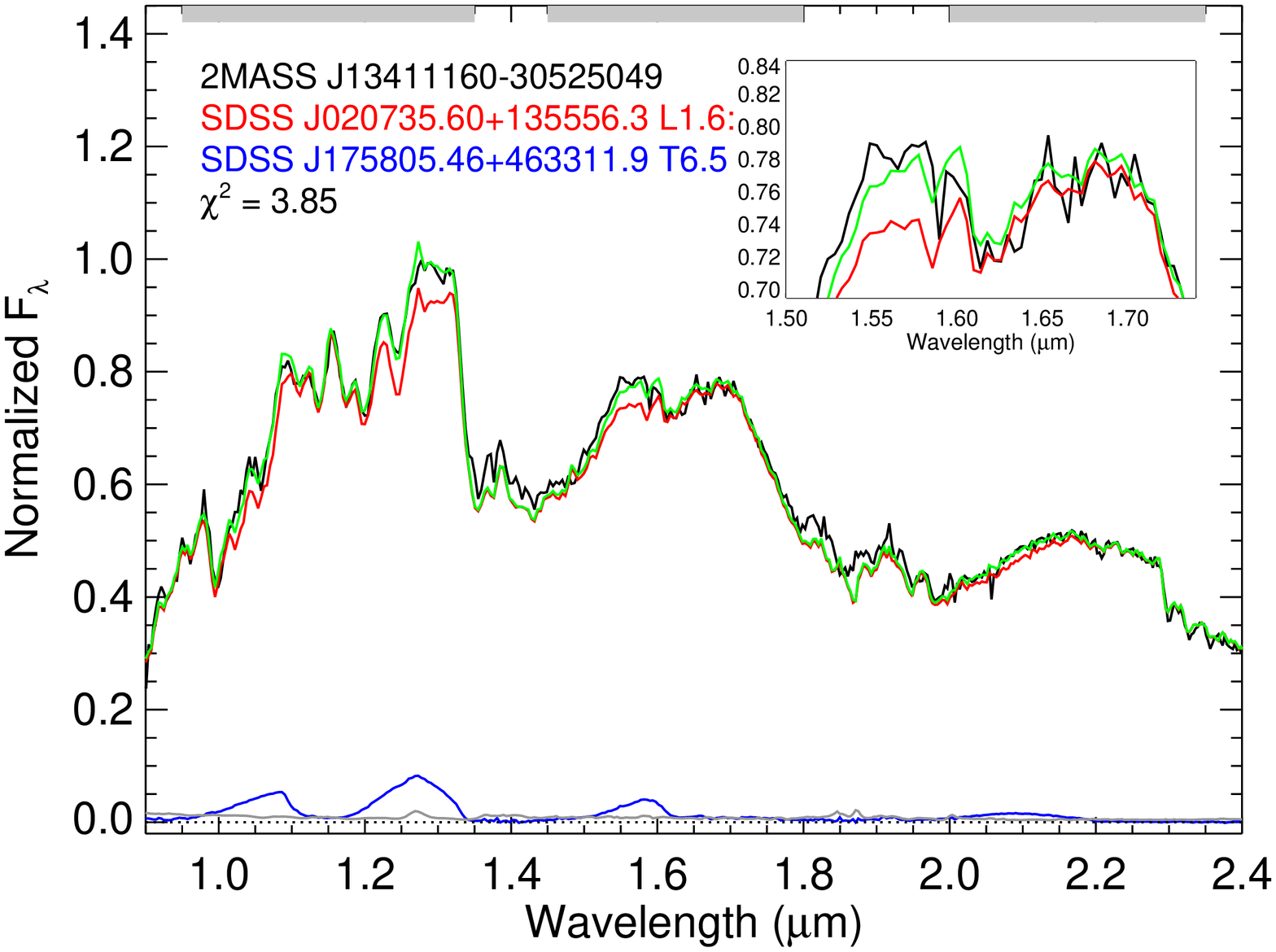}\\
\plottwo{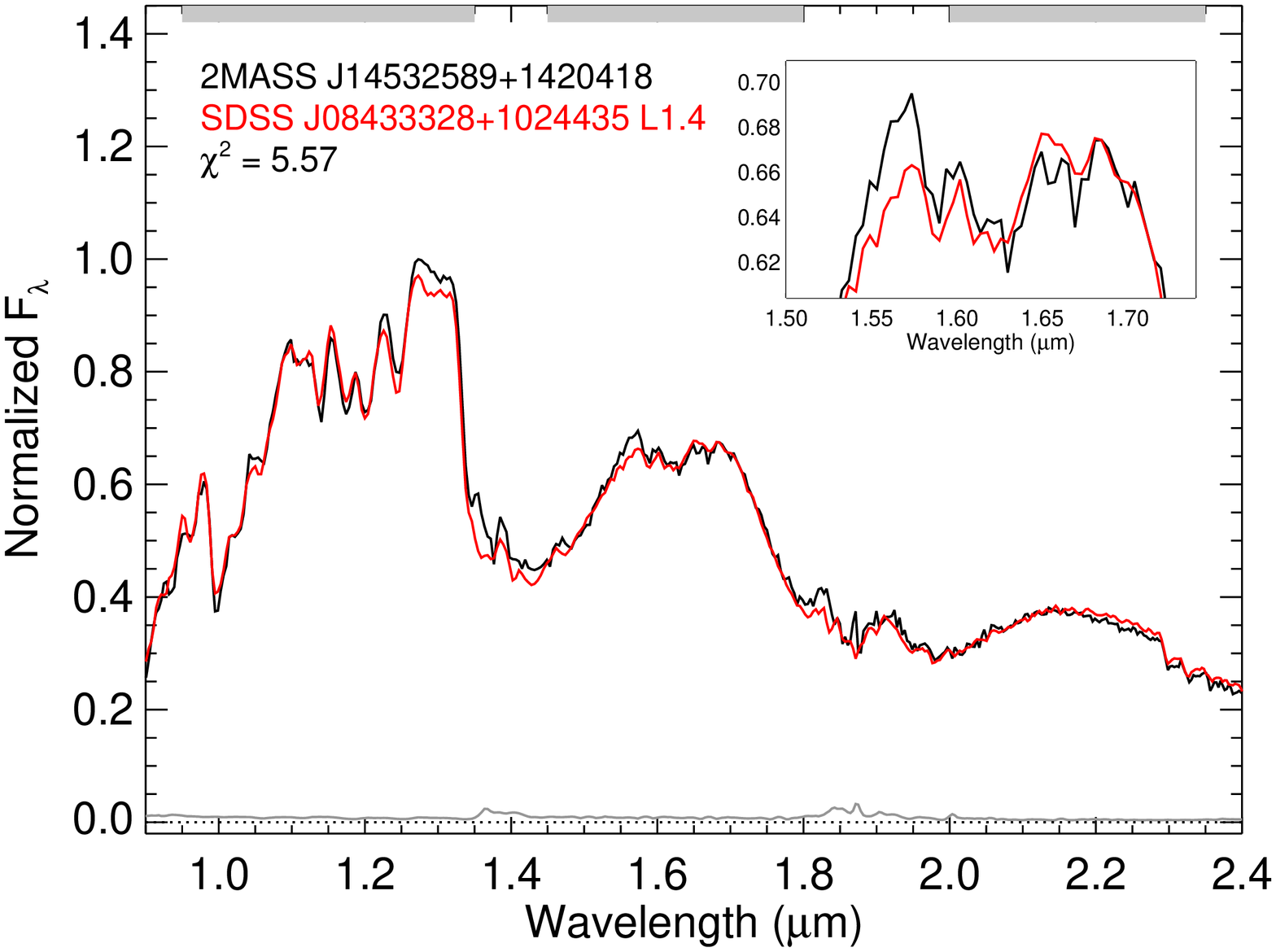}{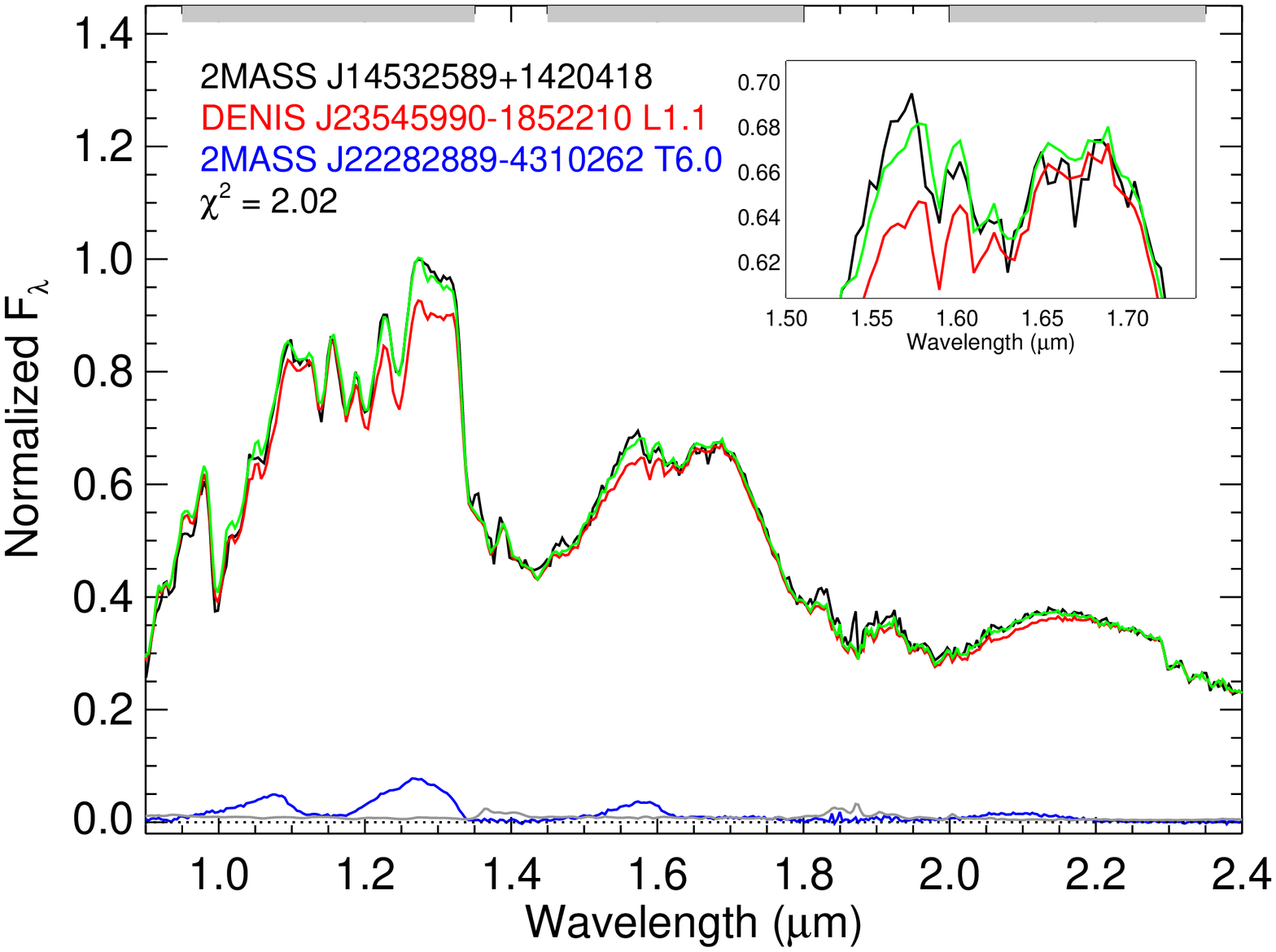}\\
\plottwo{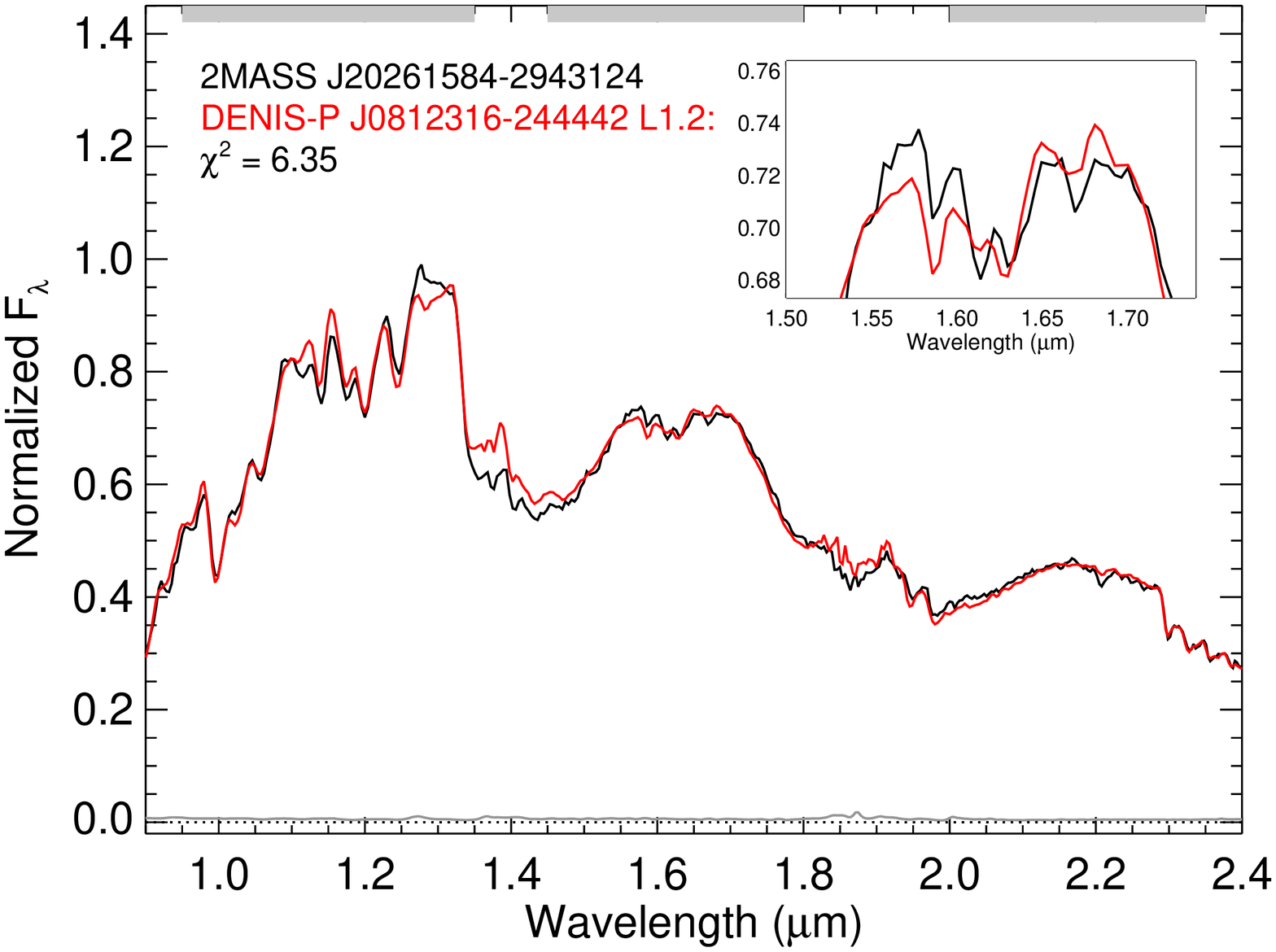}{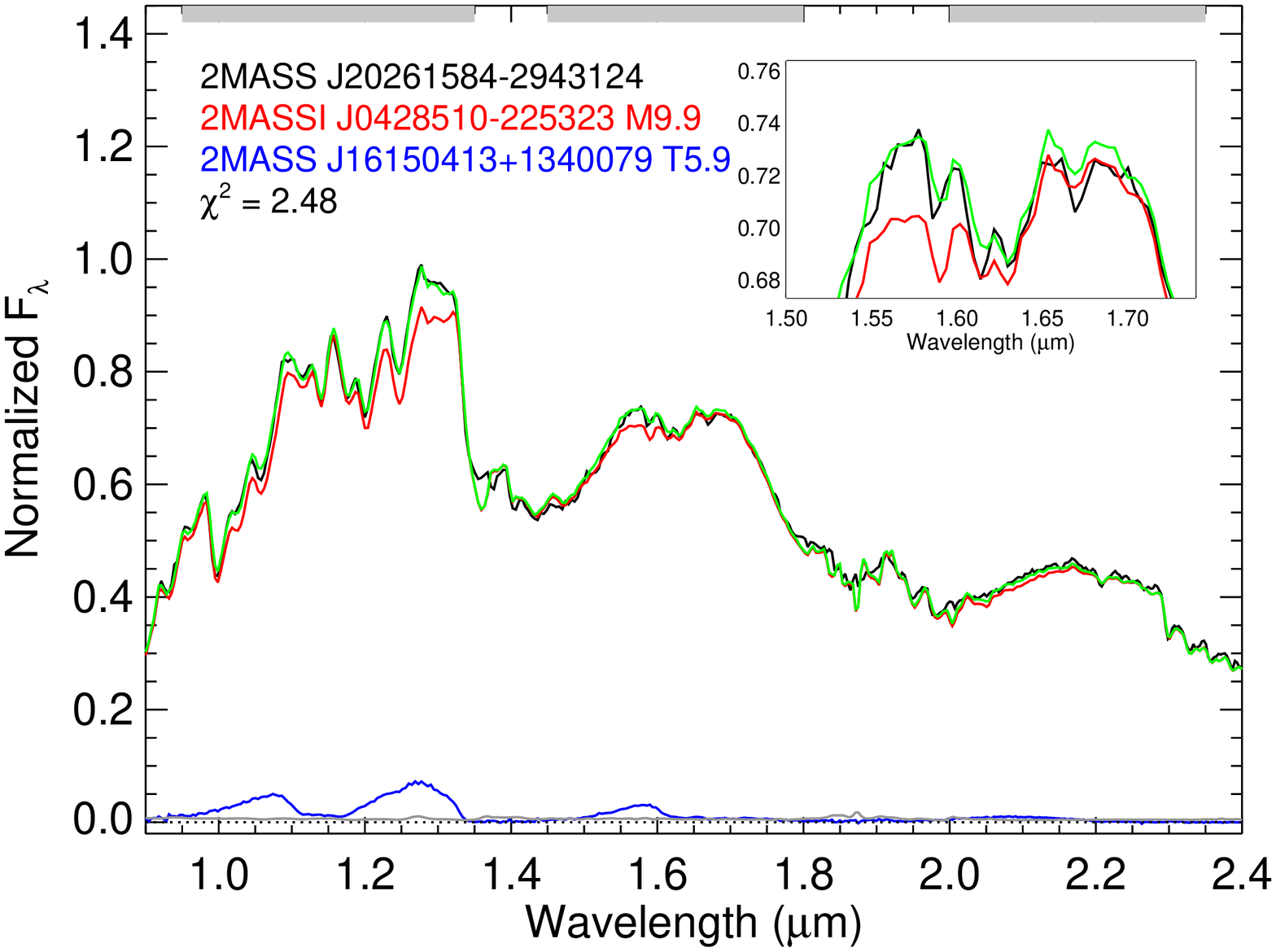}\\
\caption{Continued.\label{fig:strongfit}}
\end{figure}

%%%%%%%%%%%%%%%

\begin{figure}
\epsscale{0.95}
\setcounter{figure}{4}
\plottwo{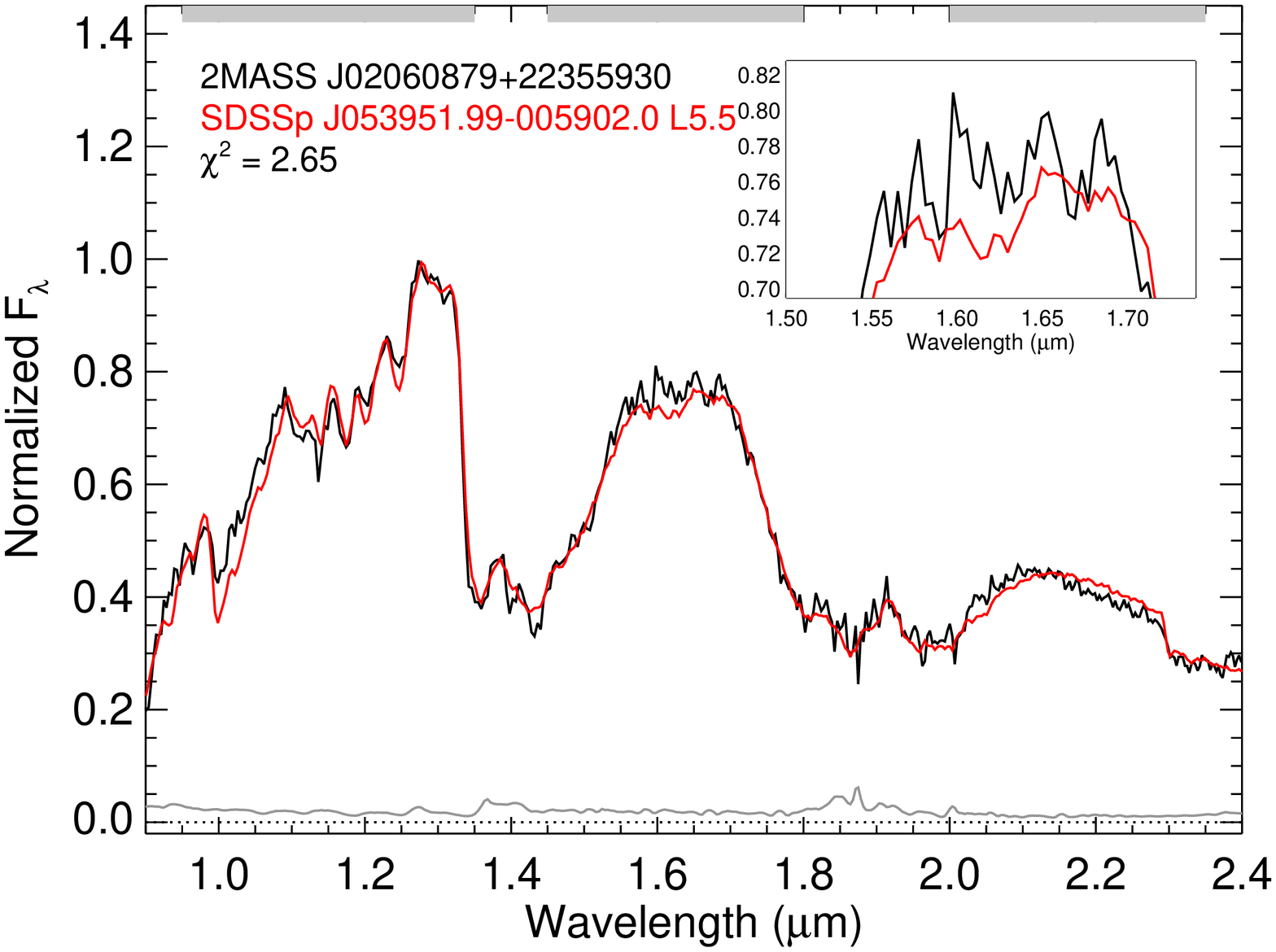}{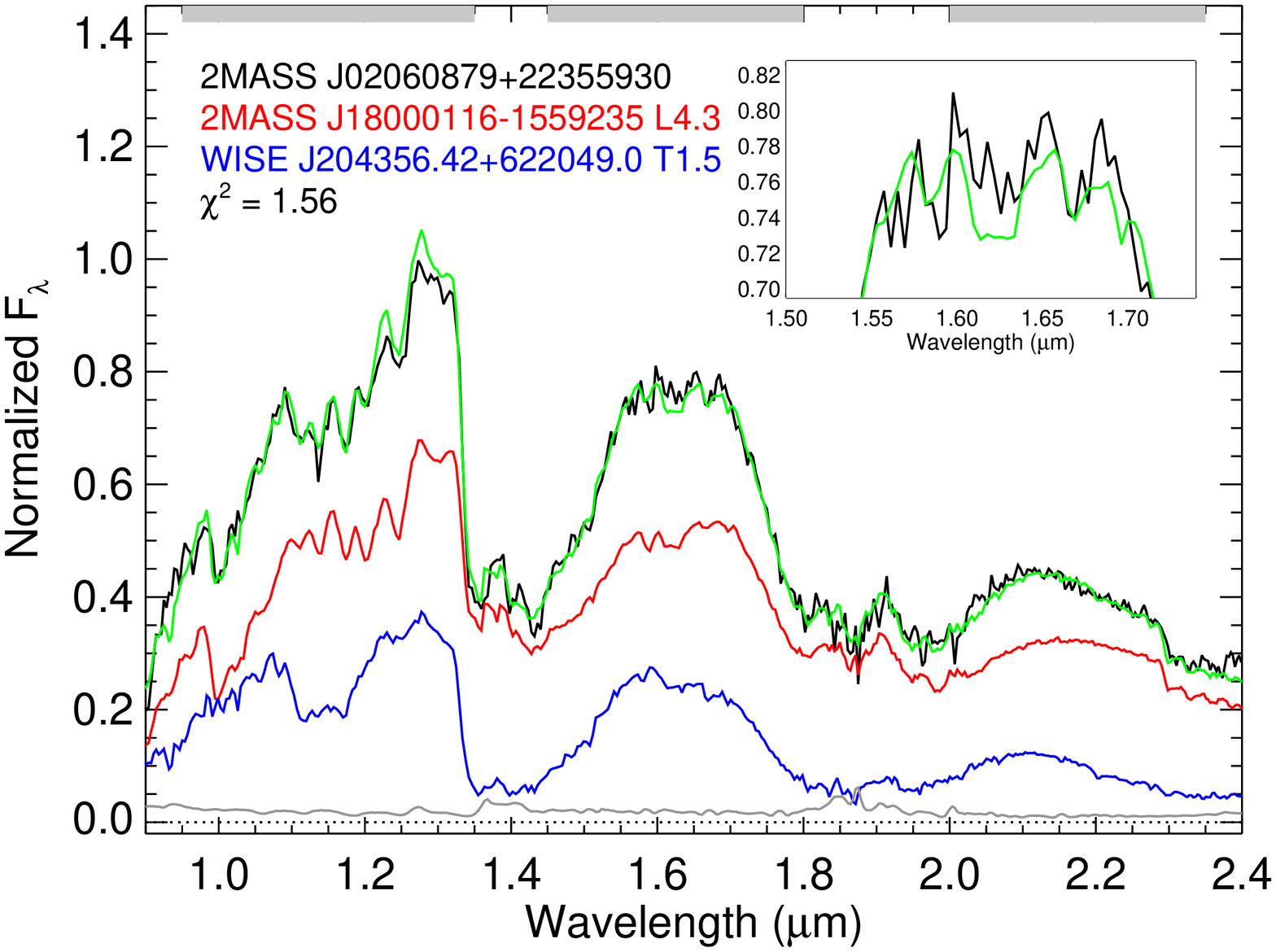}\\
\plottwo{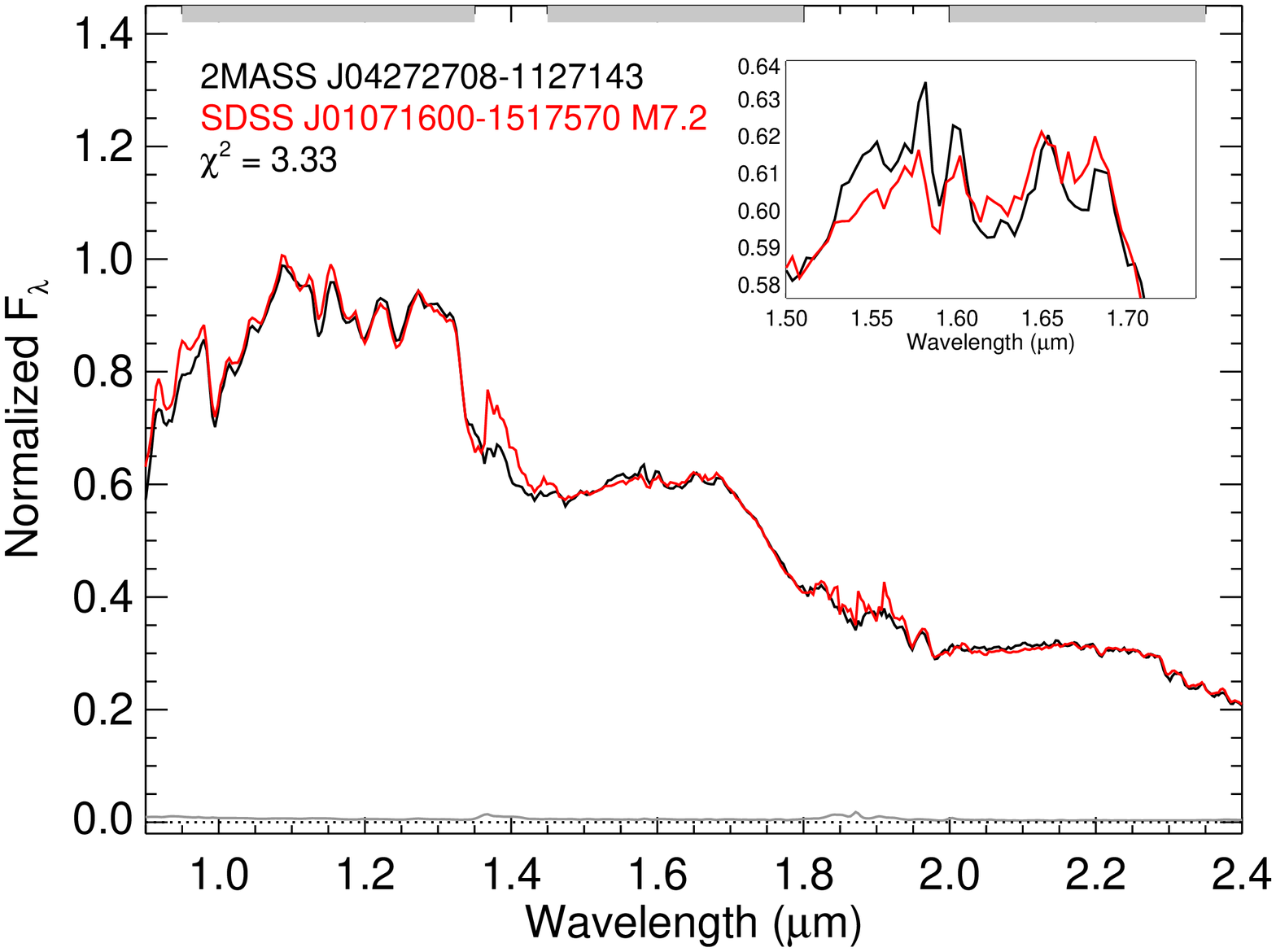}{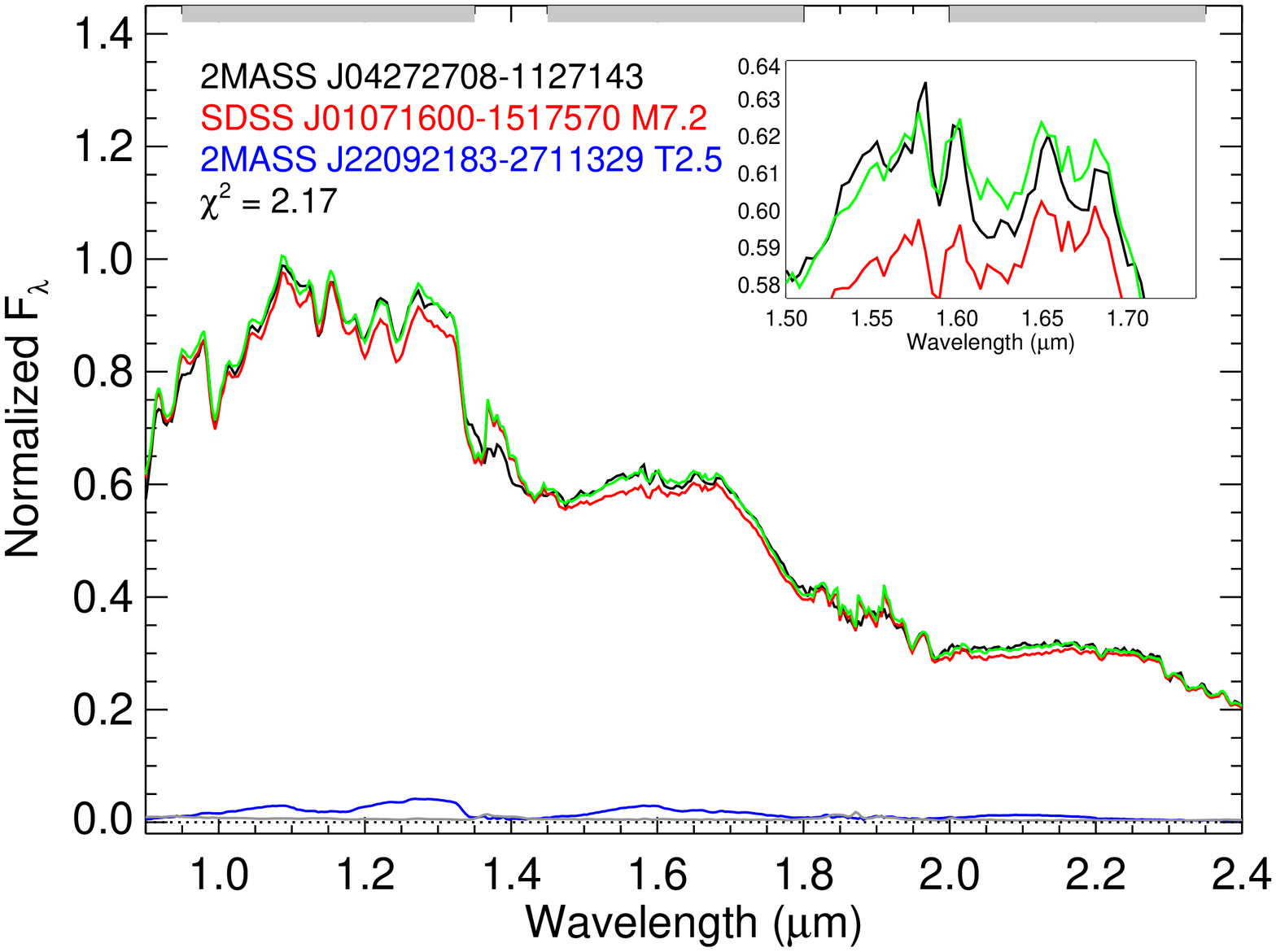}\\
\plottwo{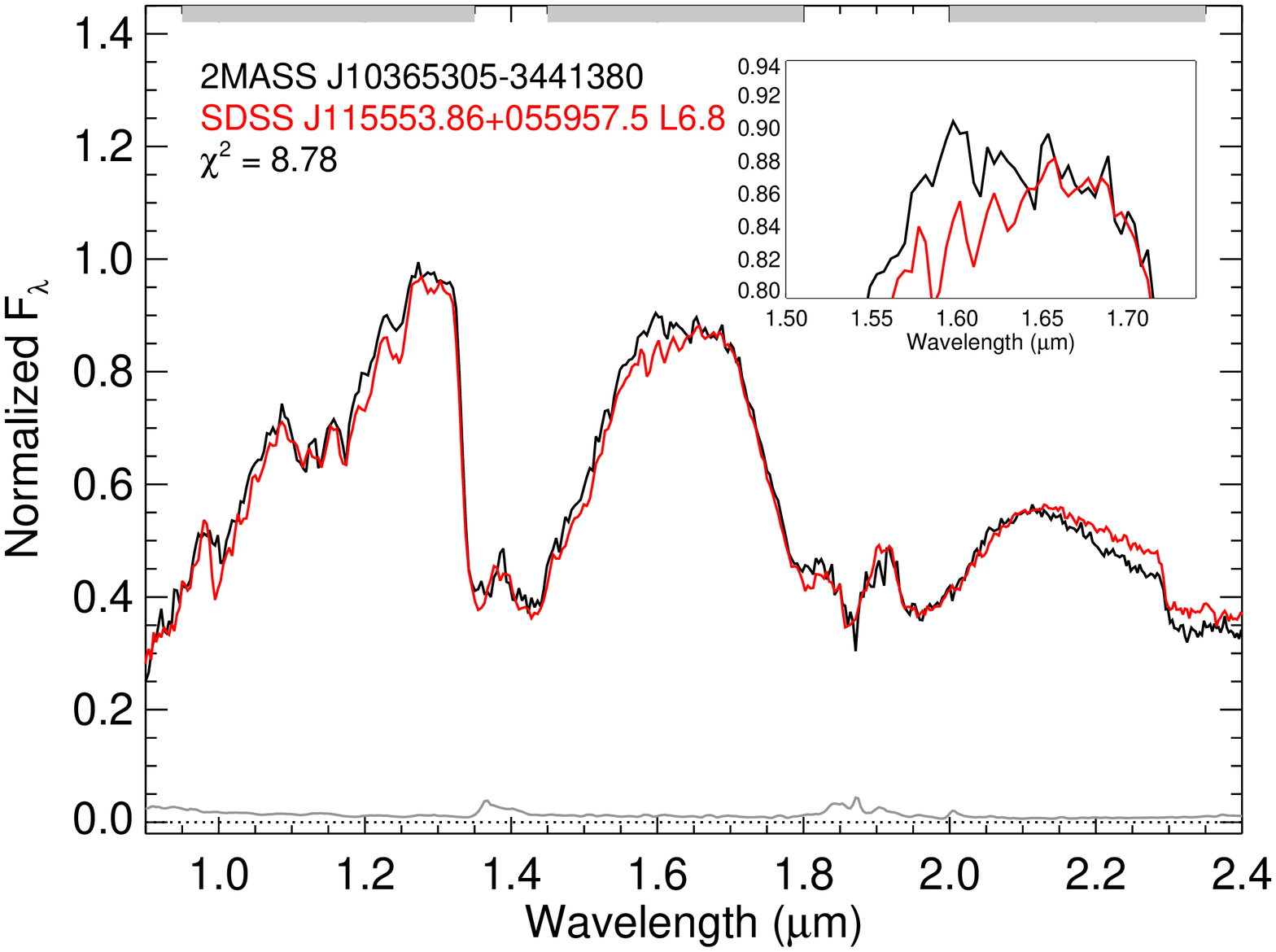}{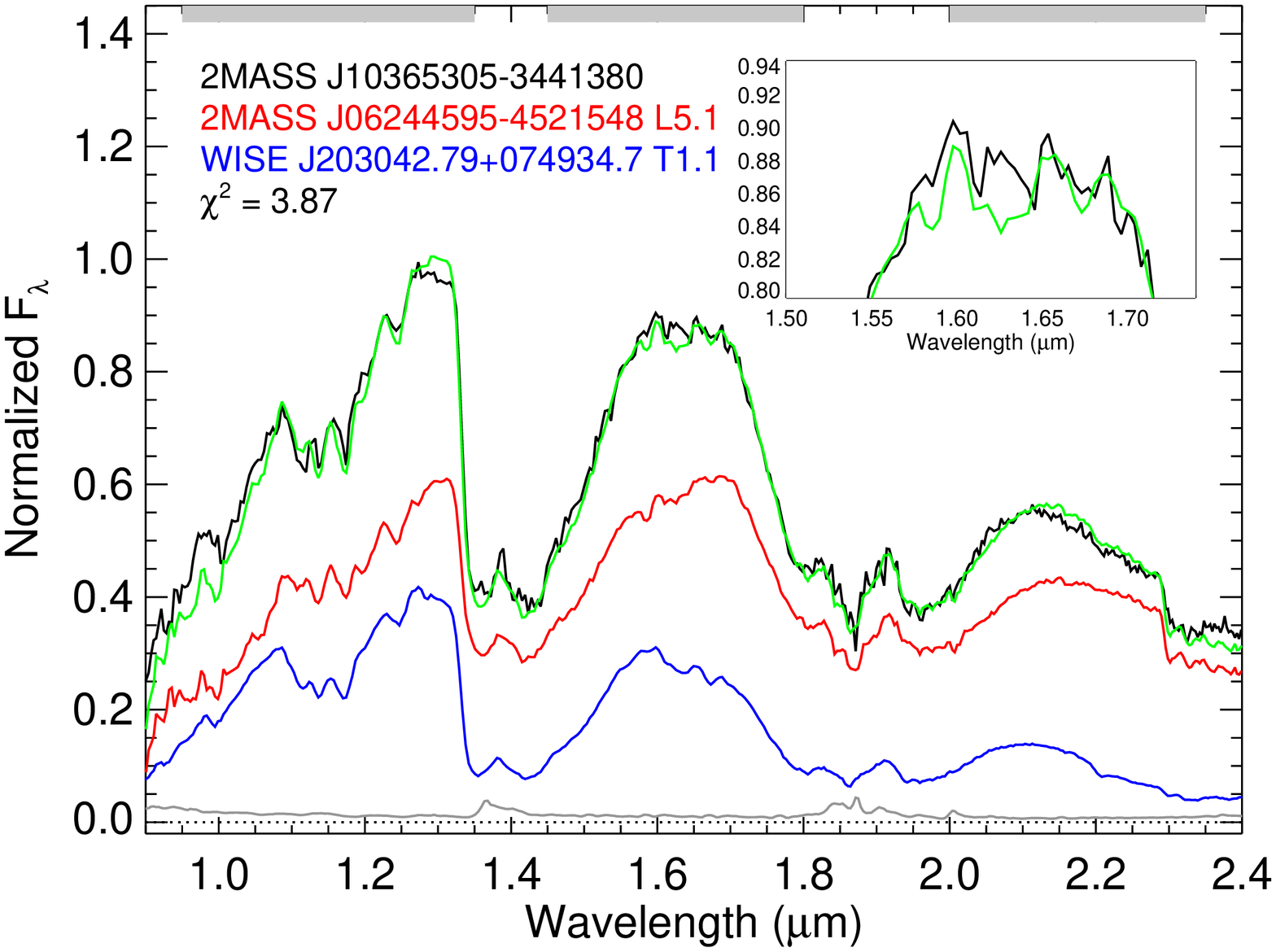}\\
\caption{Best fits to single (left) and binary (right) templates for our weak candidates. Same color code as for Figure~\ref{fig:strongfit}.\label{fig:weakfit}}
\end{figure}

\begin{figure}
\epsscale{0.95}
\setcounter{figure}{4}
\plottwo{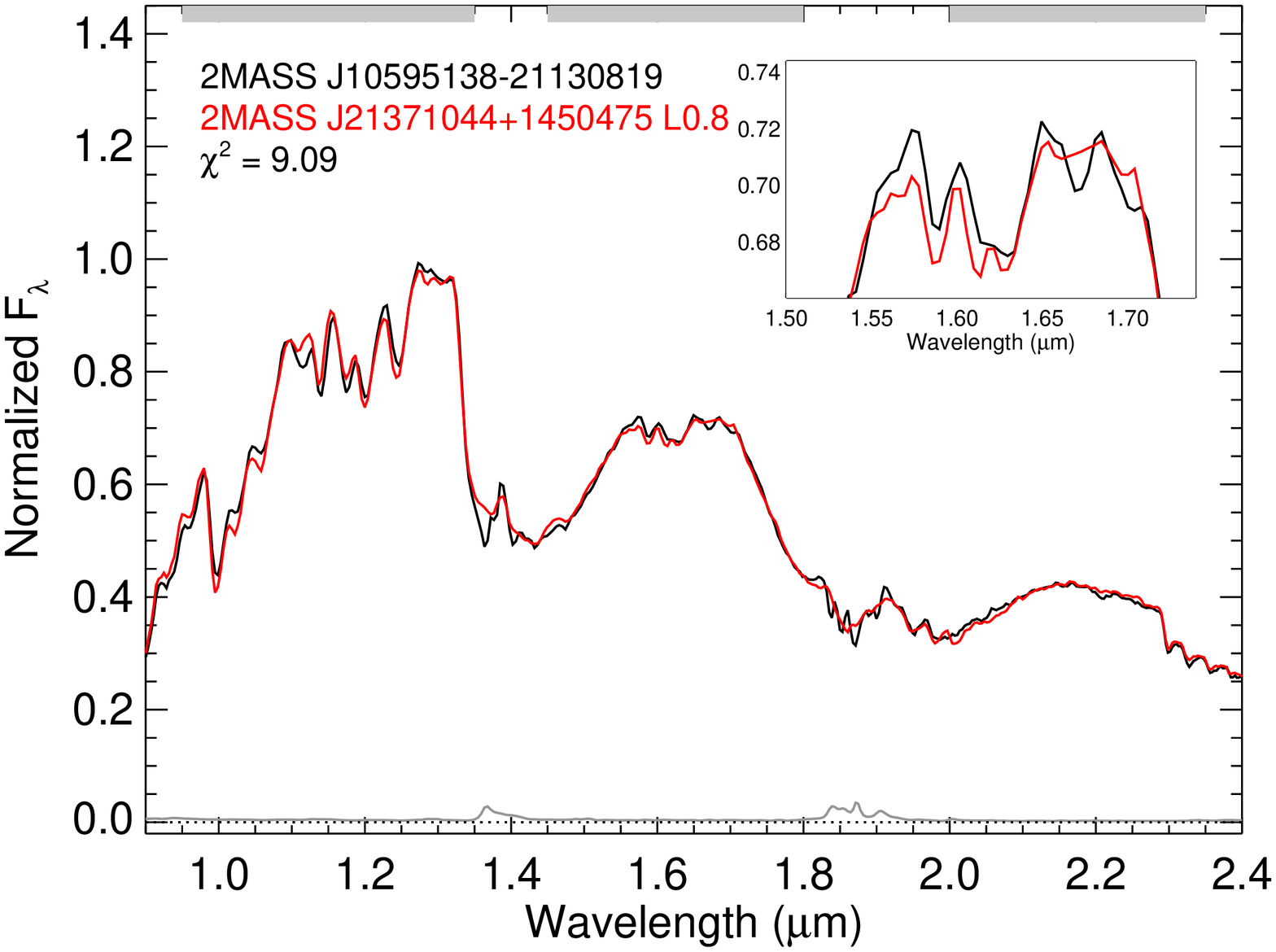}{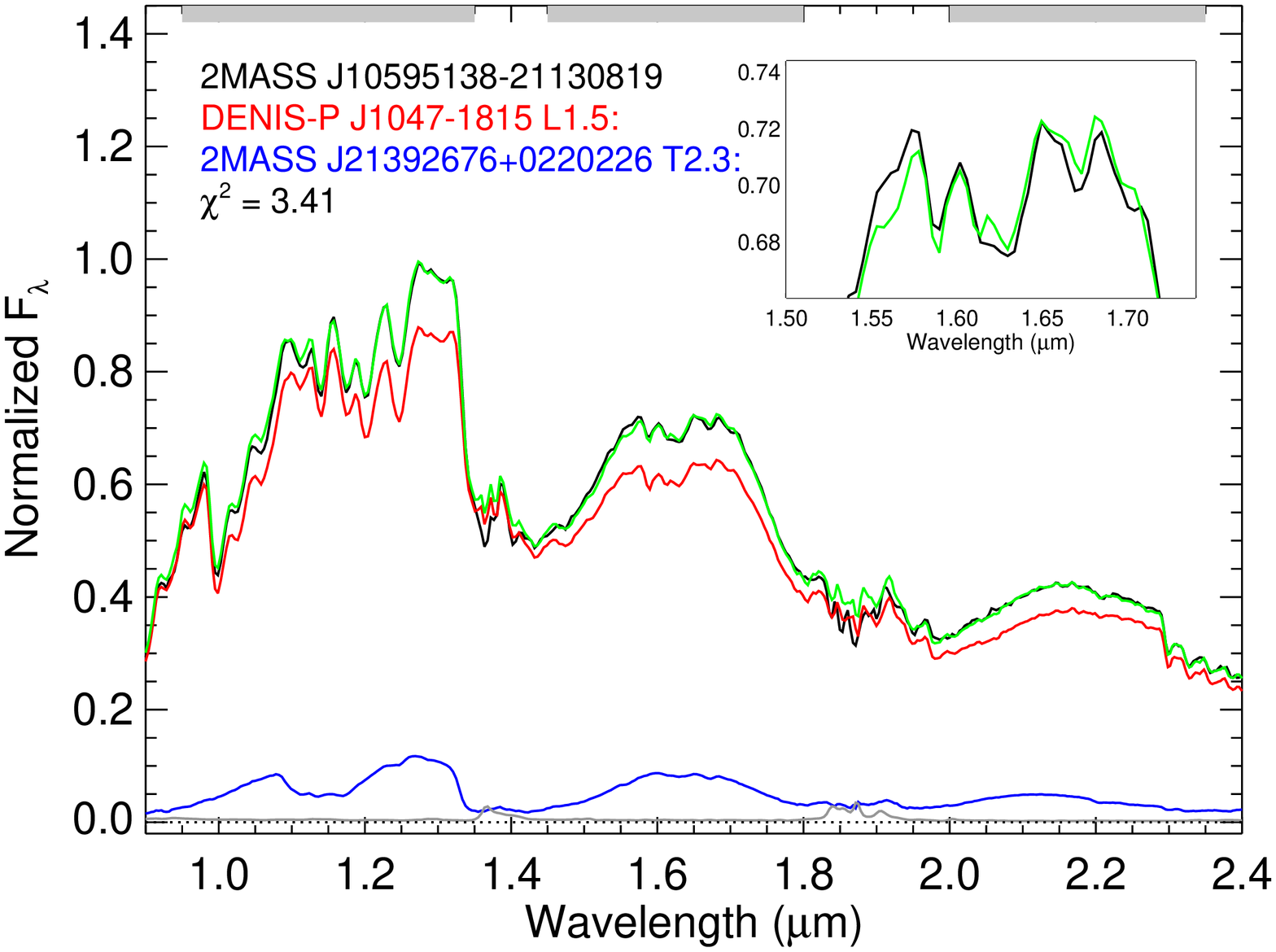}\\
\plottwo{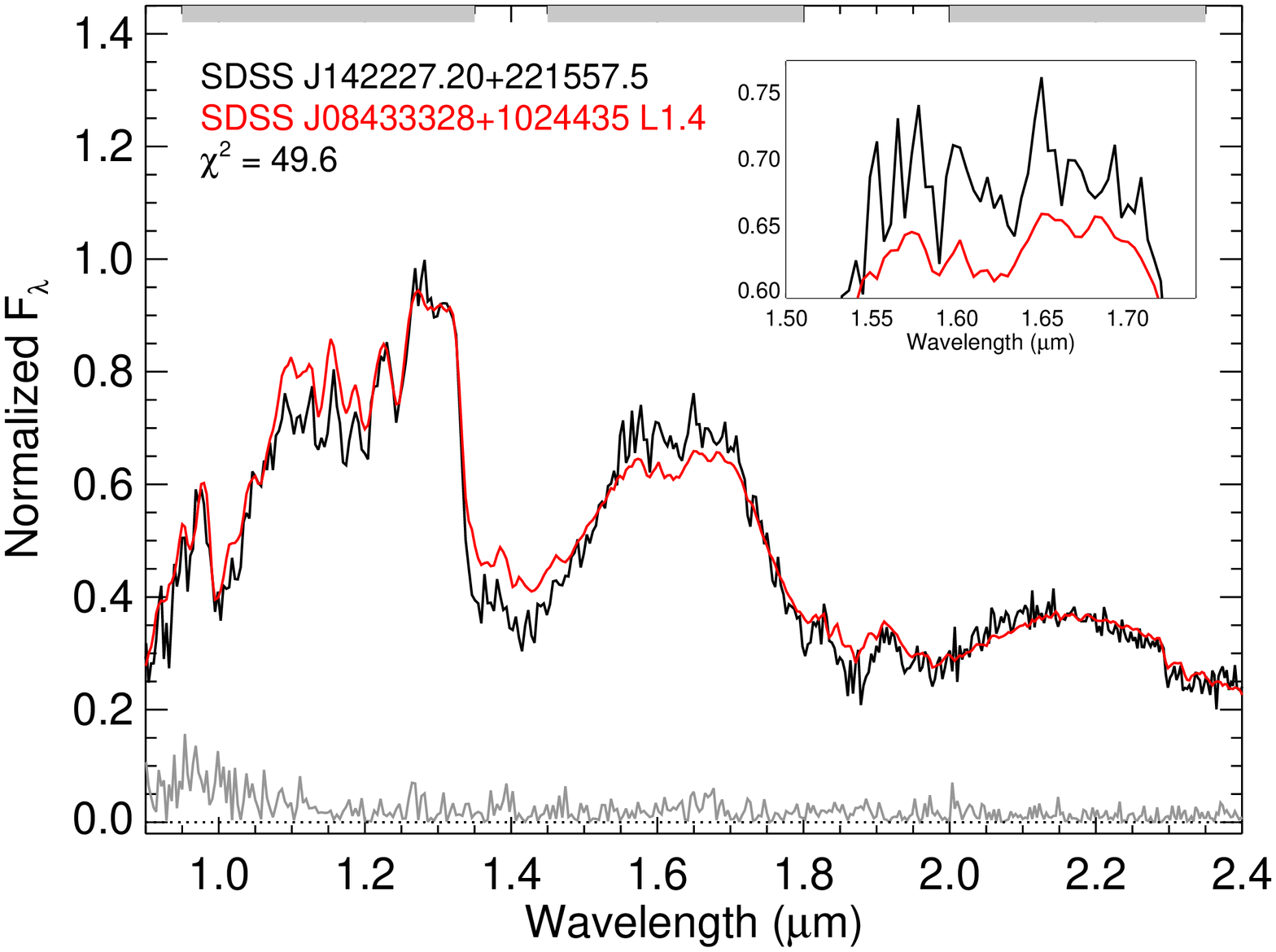}{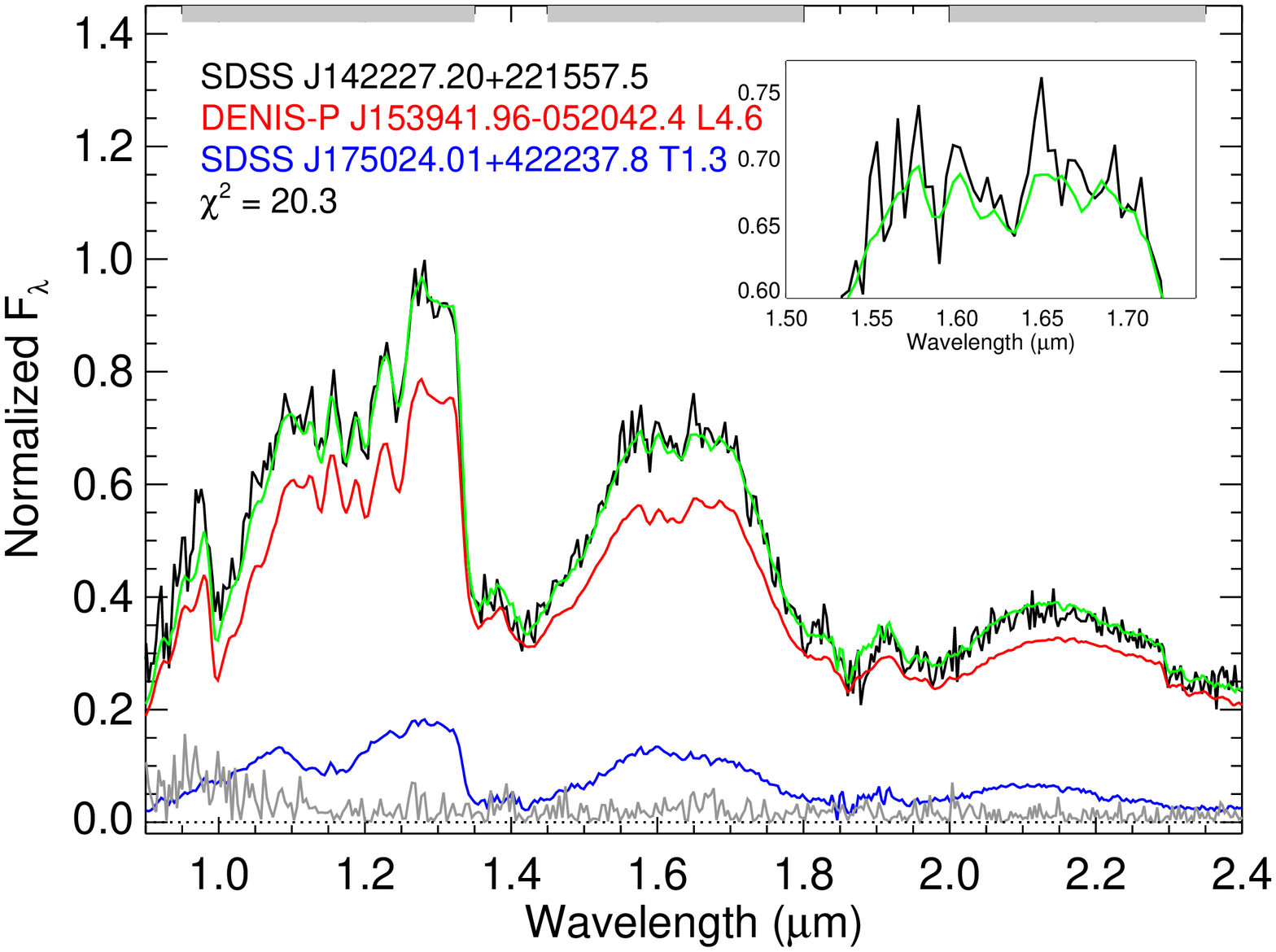}\\
\plottwo{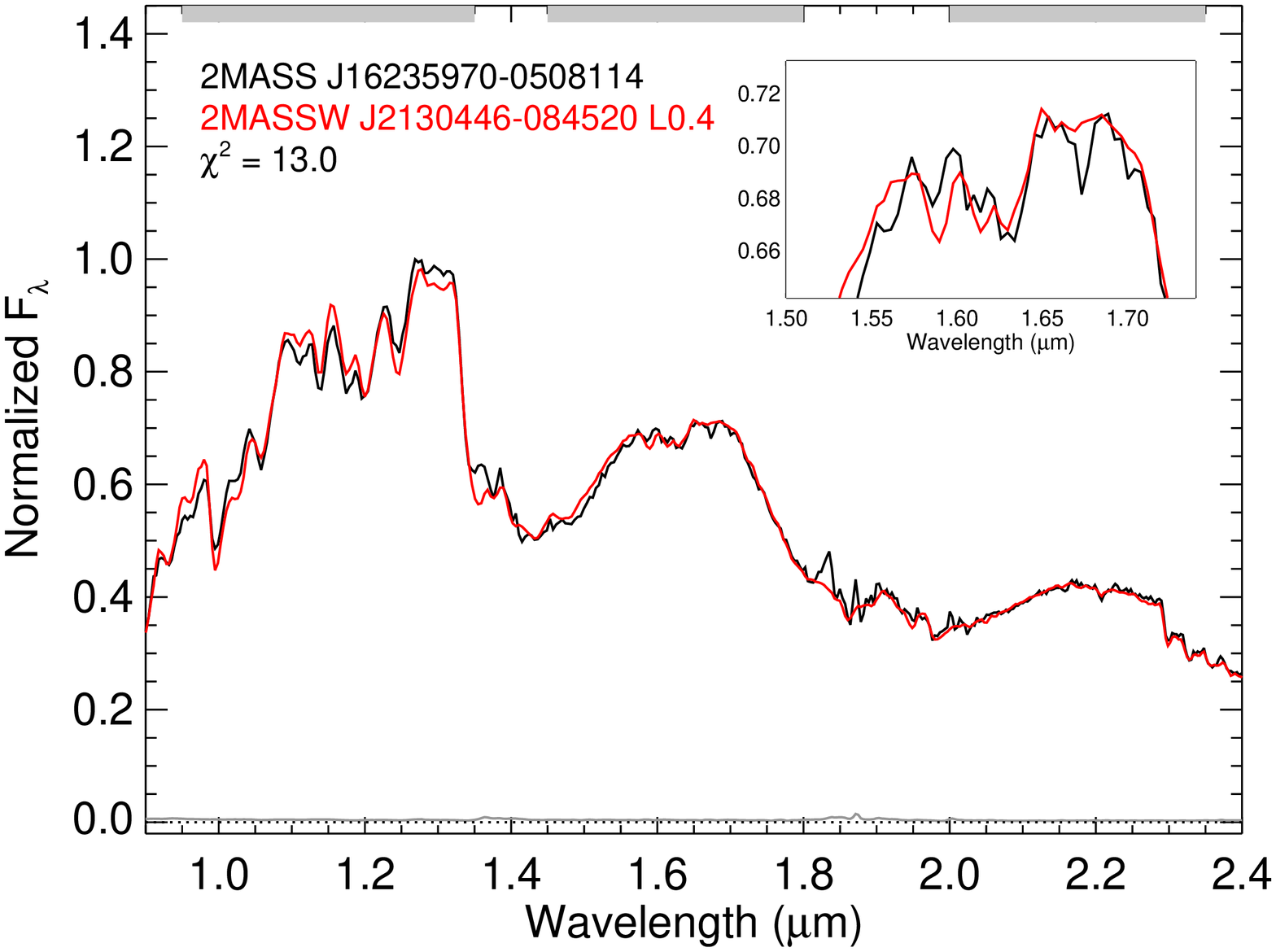}{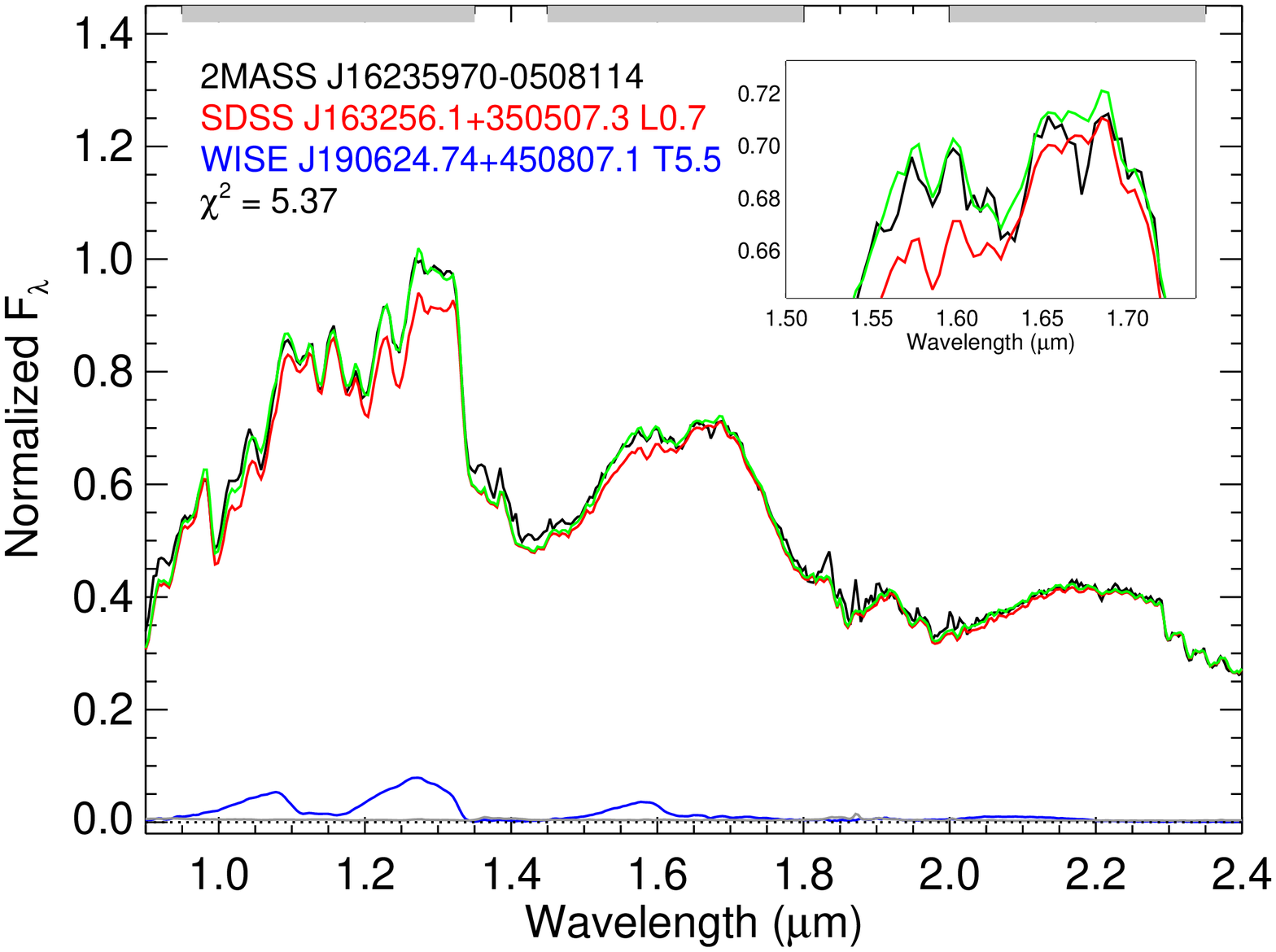}\\
\caption{Continued.\label{fig:weakfit}}
\end{figure}

\begin{figure}
\epsscale{0.95}
\setcounter{figure}{4}
\plottwo{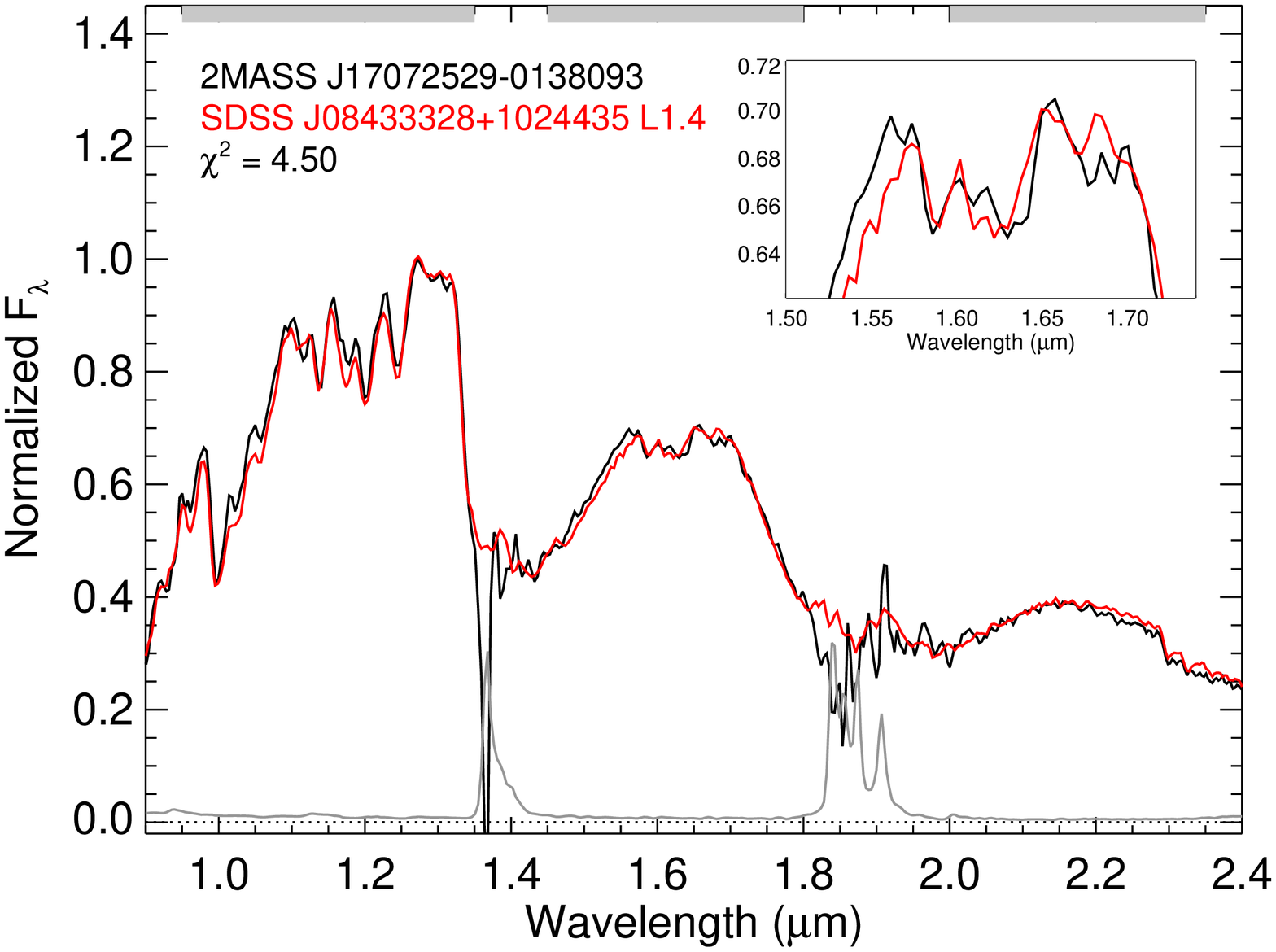}{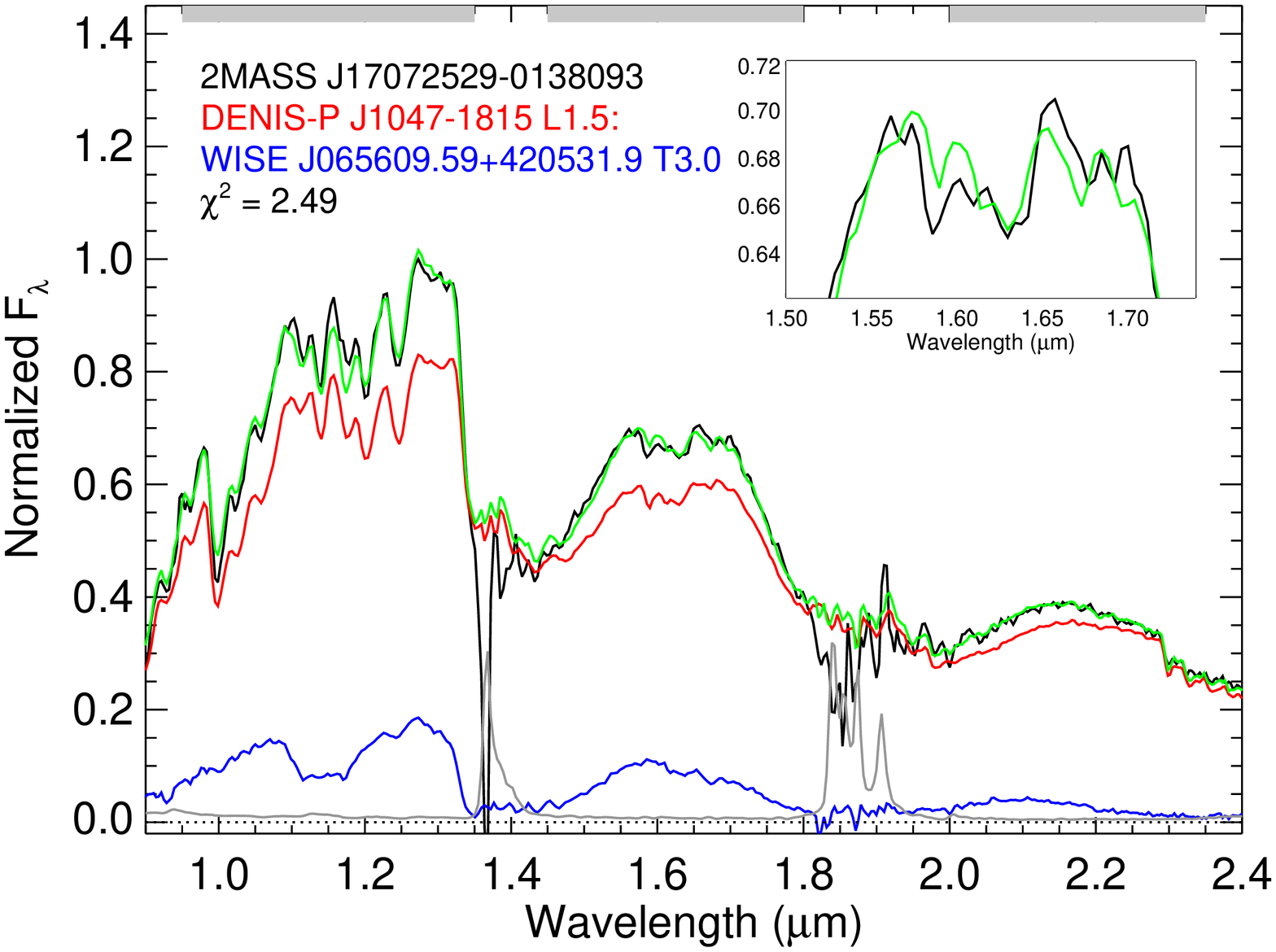}\\
\caption{Continued.\label{fig:weakfit}}
\end{figure}

%%%%%%%%%%%

\begin{figure}
\epsscale{0.95}
\setcounter{figure}{5}
\plottwo{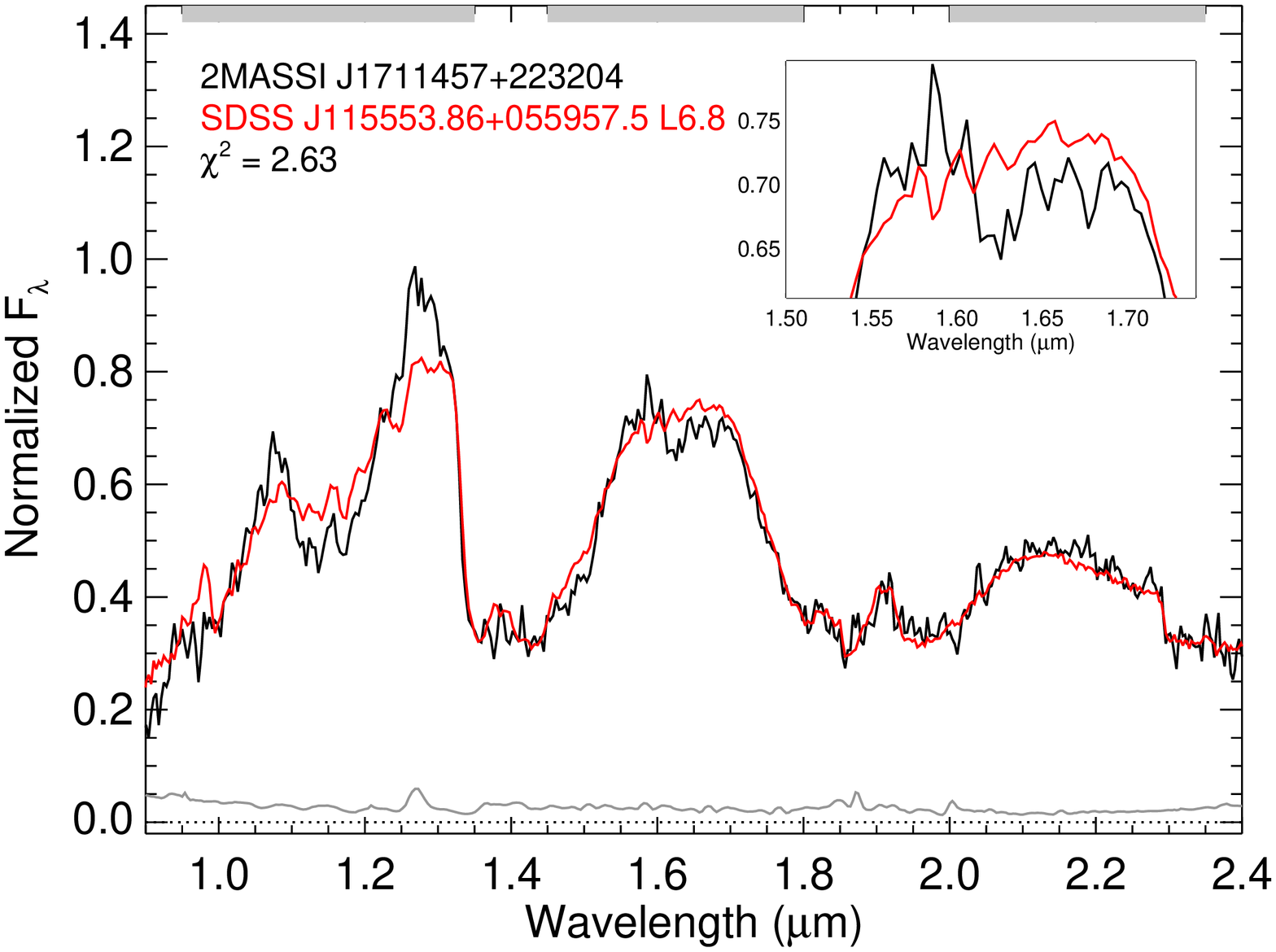}{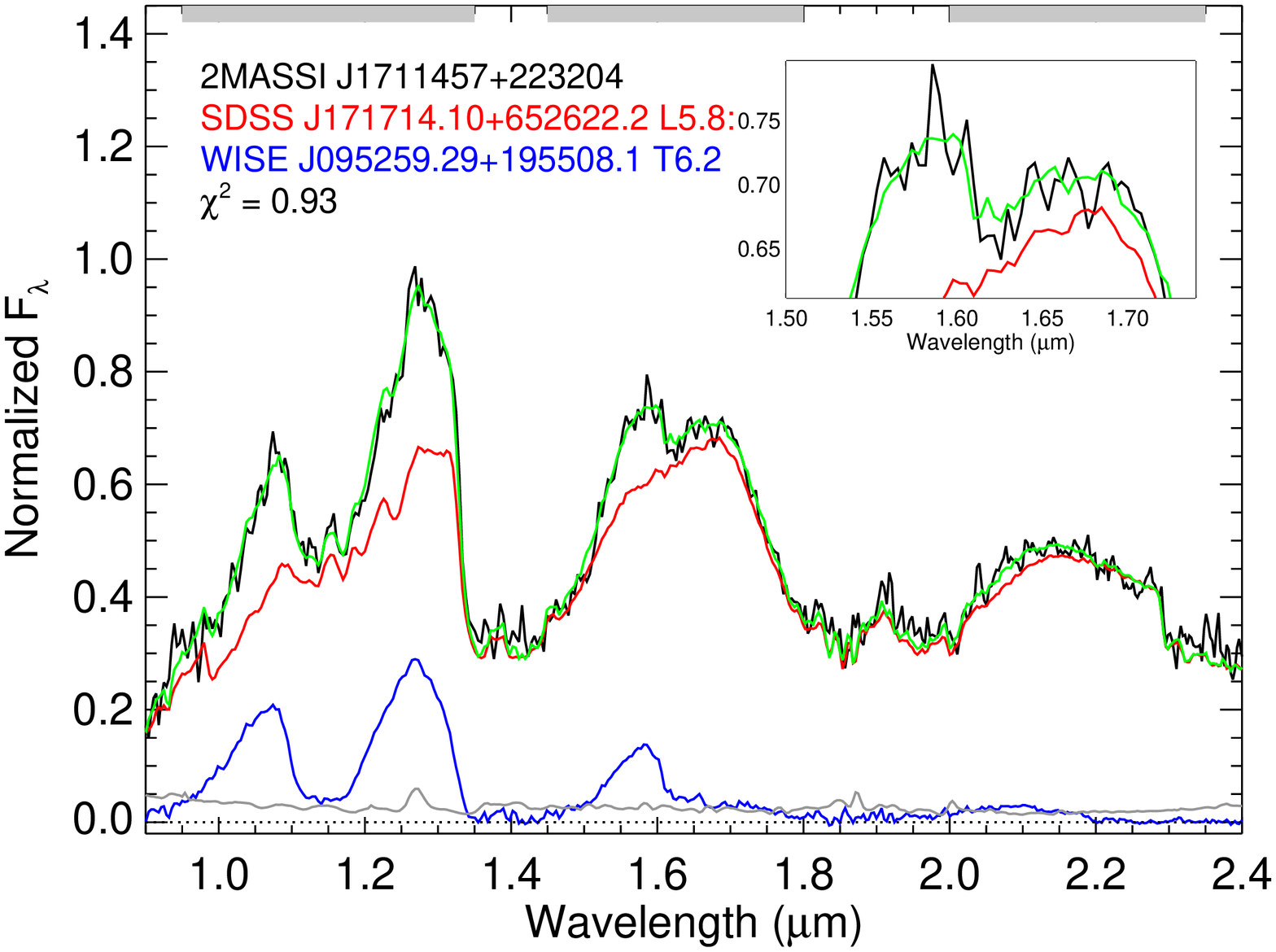}\\
\caption{Best fits to single (left) and binary (right) templates for the only visual candidate not selected by indices. Same color code as for Figure~\ref{fig:strongfit}.\label{fig:visualfit}}
\end{figure}

%%%%%%%%%%%

\begin{figure}
\epsscale{0.95}
\setcounter{figure}{6}
\plottwo{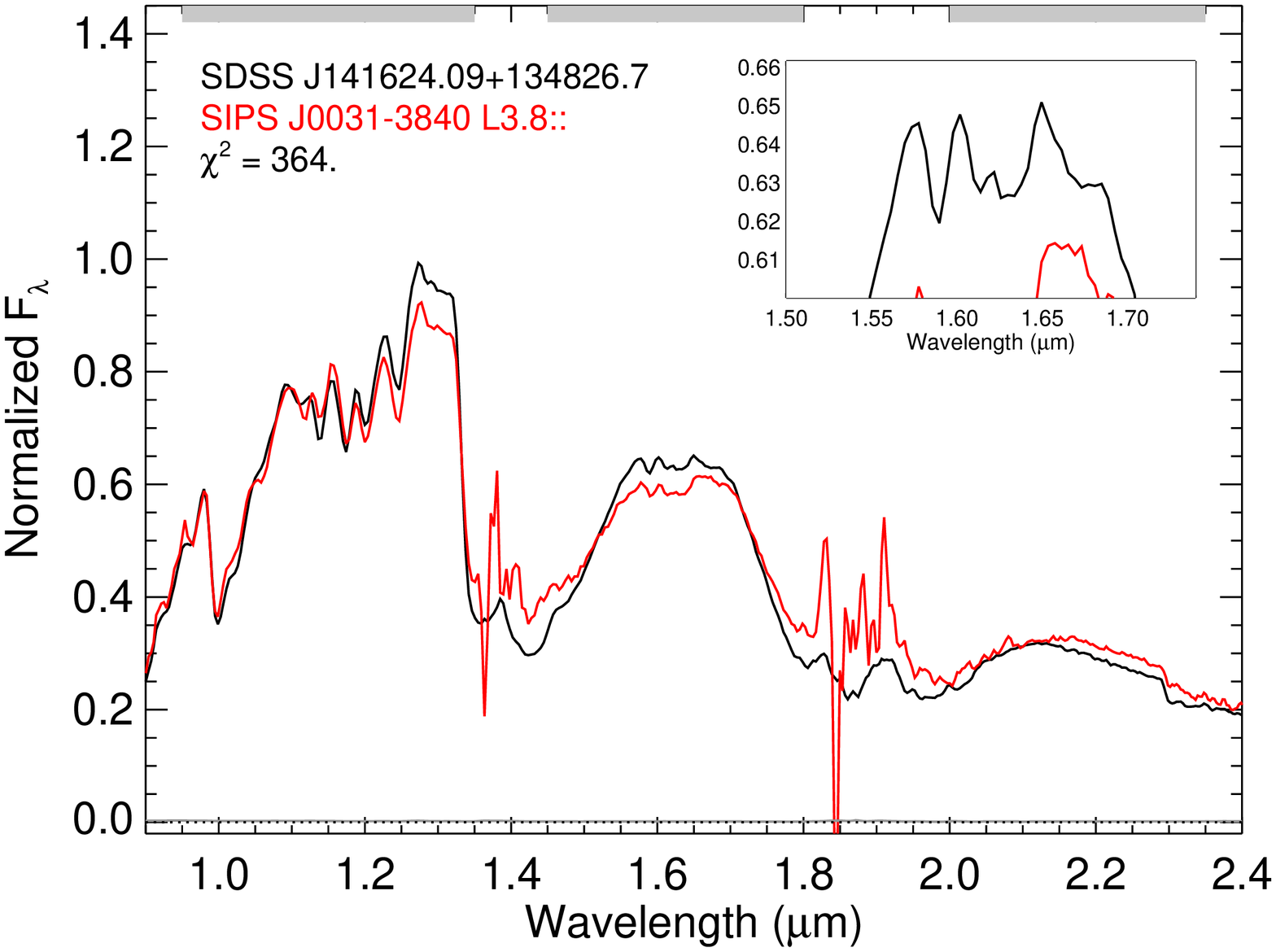}{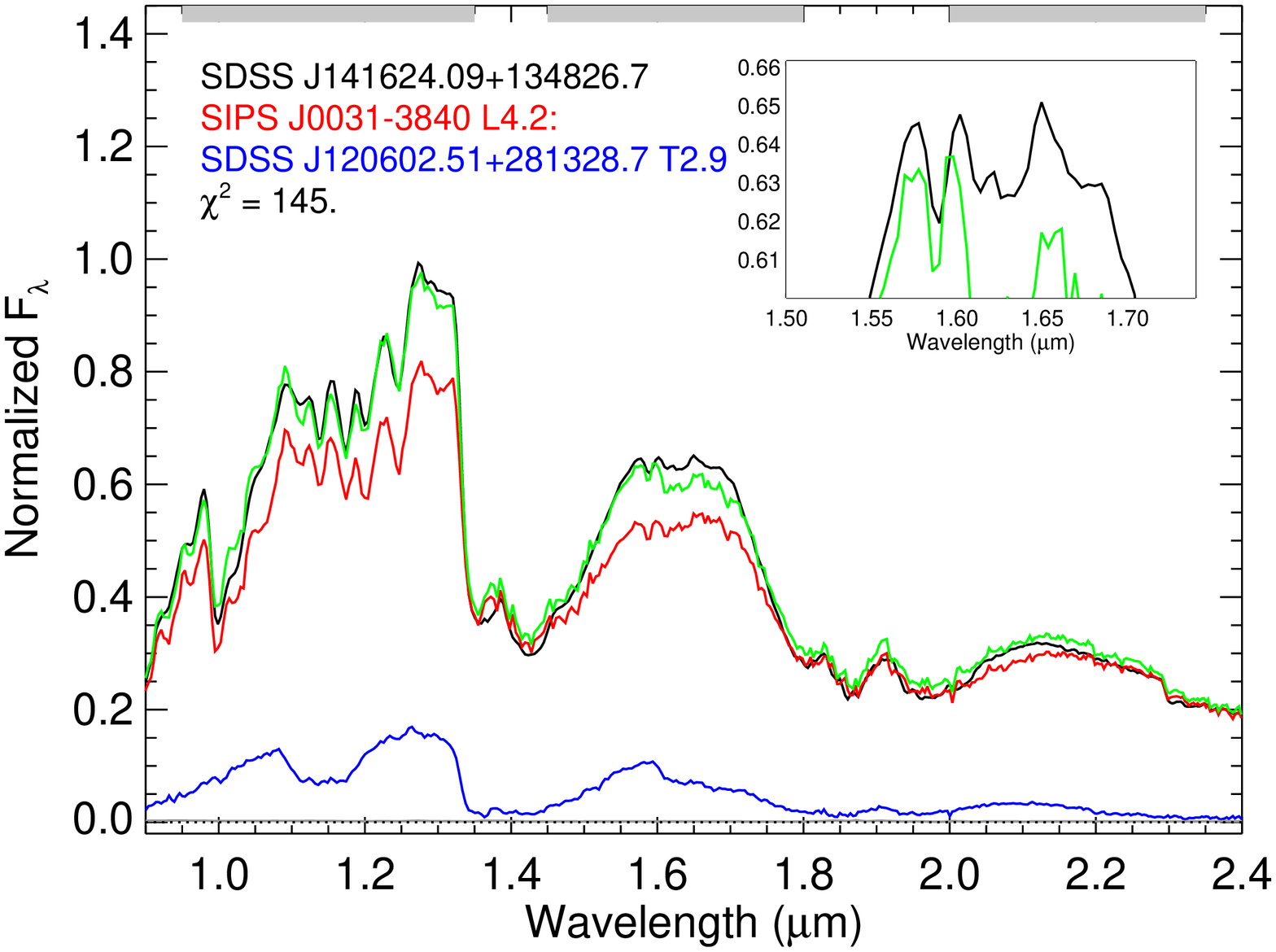}\\
\caption{Example of best fits to the blue L dwarf SDSS J141624.09+134826.7.\label{fig:blueLexample}}
\end{figure}

%%%%%%%%%%%%%%%%
\begin{figure}
\epsscale{0.8}
\setcounter{figure}{7}
\plotone{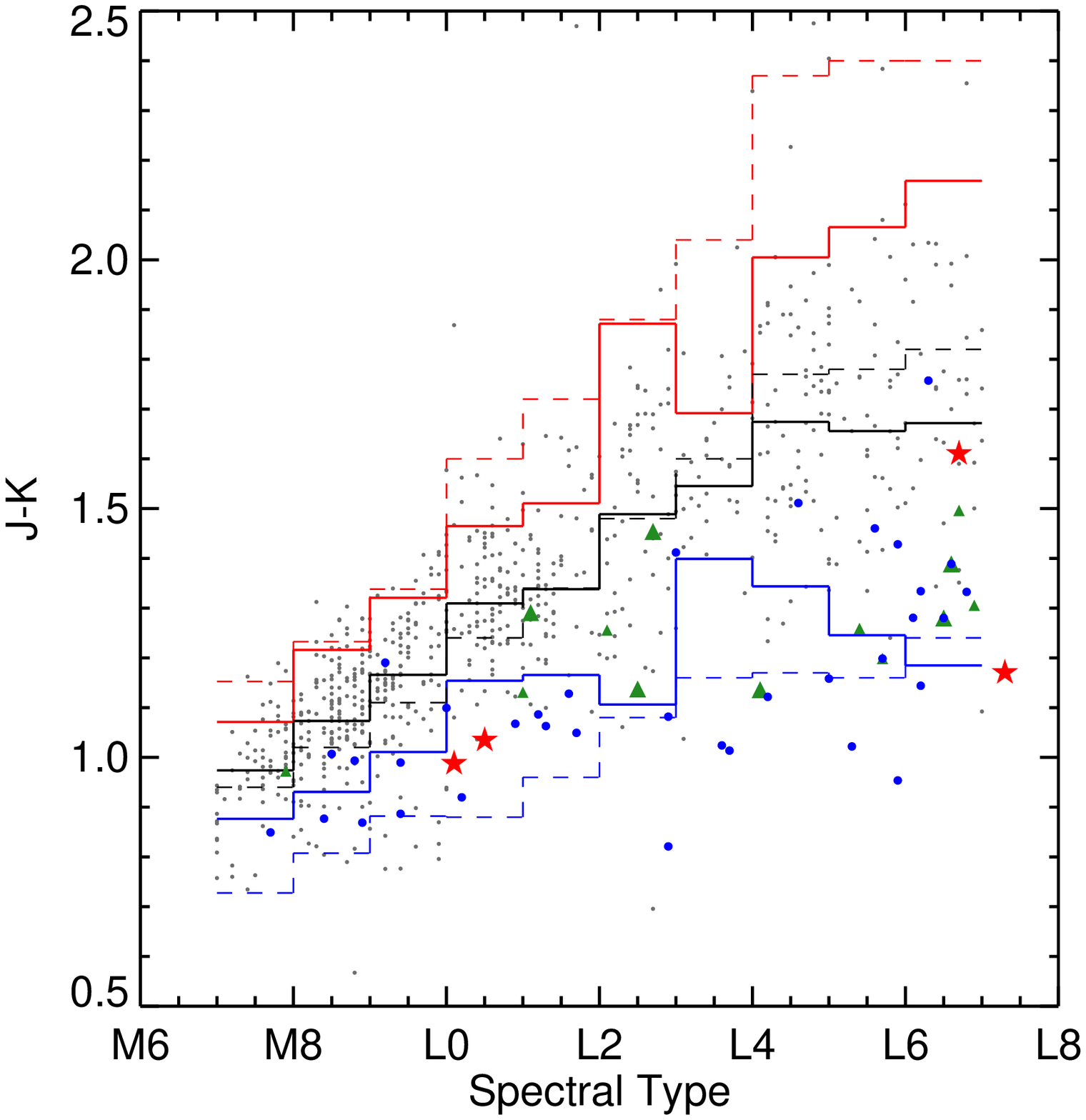}
\caption{Comparison of spectrophotometric J-K$_s$ colors of the ``candidate'' sample as a function of spectral type. The solid black line shows the median J-K colors  from the sample, while the dashed black line represents the median J-K colors as calculated by~\citet{2011AJ....141...97W} and~\citet{2010AJ....139.1808S} from samples of M and L dwarfs. The +2$\sigma$ and -2$\sigma$ boundaries are indicated in red and blue, respectively. The dashed lines indicate the +2$\sigma$ and -2$\sigma$ boundaries from~\citet{2011AJ....141...97W} and~\citet{2010AJ....139.1808S} . Outliers to these regions indicate unusually red and blue dwarfs as described by~\citet{2009AJ....137....1F}. Red stars indicate confirmed M/L+T binaries, while large and small green triangles are strong and weak binary candidates as selected by spectral indices. Blue circles represent unusually blue sources as listed in the literature.\label{fig:J-K}}
\end{figure}

%%%%%%%%%%%%%
\begin{figure}
\setcounter{figure}{8}
\label{fig:separation}
\epsscale{0.8}
\plotone{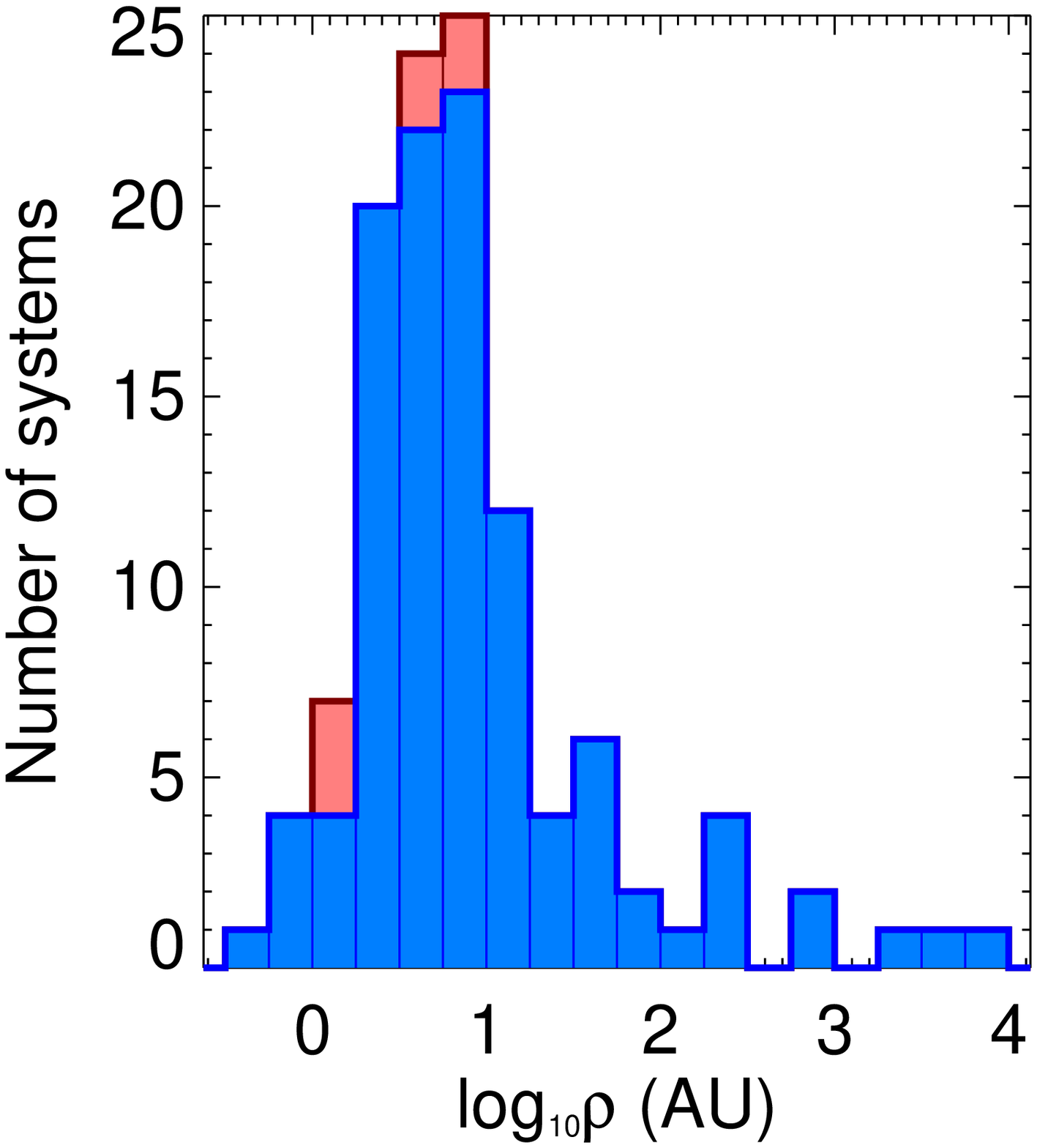}
\caption{Projected separation ($\rho$) distribution of 122 confirmed brown dwarf and VLM binary systems from the Very Low Mass Binaries Archive.  Spectral binaries are shown in red. Binary systems with only upper limits in separation have been excluded.}
\end{figure}

%%%%%%%%%%%%%%%%%%%

%comes from SBcompilation.txt
\begin{landscape}
\begin{deluxetable}{lccccccccccc}
\tabletypesize{\footnotesize}
\tabletypesize{\scriptsize}
\tablecolumns{11} 
\tablewidth{0pt}
 \tablecaption{Compilation of all confirmed and candidate spectral binaries discovered to date.\label{tab:SB}}
 \tablehead{
  & & \multicolumn{2}{c}{Spectral Type}\\
\cline{2-5}
 \colhead{Name} &
 \colhead{Optical} &
 \colhead{NIR} & 
 \colhead{Primary}& 
 \colhead{Secondary}& 
 \colhead{J mag}& 
 \colhead{J-K$_s$}& 
 \colhead{$\Delta$ J}& 
 \colhead{Separation\tablenotemark{a}} & 
 \colhead{Confirmation\tablenotemark{b}} &
 \colhead{Ref.}\\ 
  \colhead{}&
 \colhead{} &
 \colhead{} & 
 \colhead{} & 
 \colhead{} & 
 \colhead{} & 
 \colhead{} & 
 \colhead{} & 
 \colhead{(AU)} & 
 \colhead{method} & 
 \colhead{SB; Conf.}}
 \startdata 
SDSS J000649.16-085246.3\tablenotemark{*,\dagger} & M9 & \nodata  & M8.5$\pm$0.5 & T5$\pm$1 & 14.14$\pm$0.04 & 1.01$\pm$0.05 & 3.15$\pm$0.31 & 0.29$\pm$0.01 & RV  & 2; 2\\
ULAS J004757.41+154641.4 & \nodata  & T2.0$\pm$2.0 & L8.0 & T7.0 & 17.83$\pm$0.05 & 1.41$\pm$0.07 & \nodata & $<65$ & & 22\\
2MASS J00521232+0012172 & L5 & \nodata & L4 & T3 & 16.36$\pm$0.11 & 0.90$\pm$0.19 & \nodata & $<49$ & & 24\\%d=49.24
SDSS J011912.22+240331.6 & \nodata  & T2 & T0$\pm$0.7 & T4$\pm$0.4 & 17.02$\pm$0.18 & $<-0.02$ & -0.42$\pm$0.19 & $<43$ & & 21\\%d=43.15
ULAS J020529.62+142114.0 & \nodata & T1.0$\pm$0.5 & T1.0 & T3.0 & 17.99$\pm$0.04 & 1.06$\pm$0.07 & -- & $<71$ & & 22\\%d=70.96
2MASS J02060879+22355930 & \nodata & L5.5 & L5.1$\pm$0.5 & T3.2$\pm$2.3 & 16.56$\pm$0.11 & 1.39$\pm$0.17 & 1.61$\pm$0.89 & \nodata & & 26\\%no distance available
2MASS J02361794+0048548 & L6.5 & \nodata & L5.0$\pm$0.6 & T1.9$\pm$1.1 & 16.10$\pm$0.08 & 1.43$\pm$0.12 & 1.05$\pm$0.48 & $<48$ & & 26\\%d=29\pm3 pc
SDSS J024749.90-163112.6 & \nodata & T2.0$\pm$1.5 & T0$\pm$0.2   & T7$\pm$0.3 & 17.19$\pm$0.18 & 1.57$\pm$0.27 & 0.68$\pm$0.10 & $<36$ & & 21\\
\enddata
\tablenotetext{*}{Used as M+T binary benchmark.}
\tablenotetext{\dagger}{Spectral binaries with confirmed separations. See figure~\ref{fig:separation}.}
%\tablenotetext{}{Upper limits based on 2MASS (1.5") or SDSS (1") discovery and reported distance.}
\tablenotetext{a}{Upper limits based on SDSS angular resolution (1" - \citealt{2000AJ....120.1579Y}) and distance reported or estimated from absolute magnitudes~\citep{2008ApJ...685.1183L}.}
\tablenotetext{b}{RV = Radial Velocity, DI = Direct Imaging, AV = Astrometric Variability, OL = Overluminous.}
\tablenotetext{c}{Estimated from F110W filter~\citep{2006ApJS..166..585B}.}
%\tablenotetext{a}{HST/NICMOS} 0518, 0926
%\tablenotetext{b}{Keck II/LGS AO} 2026, 1315 
%\tablenotetext{c}{VLT/NACO} 2052
\tablerefs{(1)~\citet{2008ApJ...678L.125B}; (2)~\citet{2012ApJ...757..110B}; (3)~\citet{2011ApJ...739...49B}; (4)~\citet{2010ApJ...710.1142B}; (5)~\citet{2008ApJ...681..579B}; (6)~\citet{2007AJ....134.1330B};   (7)~\citet{2006ApJS..166..585B}; (8)~\citet{2011ApJ...743..141C}; (9)~\citet{2006AJ....131.2722C}; (10)~\citet{2004ApJ...604L..61C}; (11)~\citet{2012ApJS..201...19D}; (12)~\citet{2012ApJ...752...56F}; (13)~\citet{2010AJ....140..110G}; (14)~\citet{2008ApJ...685.1183L}; (15)~\citet{2013arXiv1310.2191M}; (16)~\citet{2008ApJ...676.1281M}; (17)~\citet{2011A&A...525A.123S};(21)~\citet{2010ApJ...710.1142B}; (22)~\citet{2013MNRAS.430.1171D}; (23)~\citet{2011ApJ...732...56G}; (24)~\citet{2008ApJ...676.1281M}; (25)~\citet{2008arXiv0811.0556S}; (26) This paper; (27)~\citet{2011ApJS..197...19K}; (28)~\citet{2010ApJS..190..100K}.}
\end{deluxetable}
\end{landscape}

%%%%%%%%%%%%%%%%%%
% 530 sources
\begin{deluxetable}{lllllccccl}
\tablecaption{New SpeX Observations of M7-L6 Dwarfs\label{tab:newobs}}
\tabletypesize{\scriptsize}
\tablewidth{0pt}
\rotate
\tablehead{
 & & \multicolumn{3}{c}{Spectral Type} & \multicolumn{2}{c}{2MASS} \\
\cline{3-5} \cline{6-7}
\colhead{Source} &
\colhead{Designation} &
\colhead{Opt} &
\colhead{NIR} &
\colhead{SpeX\tablenotemark{a}} &
\colhead{$J$} &
\colhead{$J-K_s$} &
\colhead{Date} &
\colhead{$\lambda/\Delta\lambda$} &
\colhead{Ref\tablenotemark{b}} \\
}
\startdata
2MASS J0000286-124515 & J00002867-1245153 & M8.5 & \nodata & M9.2 & 13.20 & 1.23 & 2013 Aug 14 & 120 & 44; 44 \\
LEHPM 1-162 & J00054768-2157176 & M8.5 & M8 & M8.7 & 13.27 & 1.07 & 2006 Sep 03 & 120 & 32; 75 \\
2MASSI J0006205-172051 & J00062050-1720506 & L2.5 & \nodata & L3.0 & 15.66 & 1.65 & 2008 Sep 08 & 120 & 25; 25 \\
SDSS J000632.60+140606.4 & J00063260+1406064 & L1 & \nodata & L0.4 & 15.85 & 0.79 & 2009 Jun 30 & 120 & 48; 48 \\
SDSS J000646.81+151225.8 & J00064681+1512258 & L4:: & \nodata & L4.3 & 16.22 & 1.41 & 2013 Sep 03 & 120 & 62; 62 \\
2MASS J0007078-245804 & J00070787-2458042 & M7 & \nodata & M7.7 & 13.12 & 1.05 & 2013 Oct 23 & 120 & 44; 44 \\
2MASS J00100009-2031122 & J00100009-2031122 & L0 & \nodata & M8.8: & 14.13 & 1.25 & 2008 Jul 14 & 120 & 44; 44 \\
2MASSI J0013578-223520 & J00135779-2235200 & L4 & \nodata & L5.7 & 15.78 & 1.74 & 2009 Nov 04 & 120 & 25; 25 \\
\enddata
\end{deluxetable}

%%%%%%%%%%%%%%%%%%
\begin{deluxetable}{ccccc}
\tabletypesize{\footnotesize}
\tablecolumns{4} 
\tablewidth{0pt}
 \tablecaption{Spectral indices\tablenotemark{*}
\label{tab:index}}
 \tablehead{
 \colhead{Spectral Index} & \colhead{Numerator Range ($\mu$m)} &\colhead{Denominator Range ($\mu$m)} & \colhead{Feature} & \colhead{Ref.}} 
 \startdata 
  H$_2$O-$J$ & 1.14-1.165 & 1.26-1.285 & 1.15~$\mu$m H$_2$O & 1\\
  CH$_4$-$J$ & 1.315-1.335 & 1.26-1.285 & 1.32~$\mu$m CH$_4$ & 1\\  
  H$_2$O-$H$ & 1.48-1.52 & 1.56-1.60 & 1.40~$\mu$m H$_2$O & 1\\
  CH$_4$-$H$ & 1.635-1.675 & 1.56-1.60 & 1.65~$\mu$m CH$_4$ & 1\\
  H$_2$O-$K$ & 1.975-1.995 & 2.08-2.10 & 1.90~$\mu$m H$_2$O & 1\\
  CH$_4$-$K$ & 2.215-2.255 & 2.08-2.12 & 2.20~$\mu$m CH$_4$ & 1\\
  $K/J$ & 2.06-2.10 & 1.25-1.29 & $J-K$ color & 1\\
  $H$-dip & 1.61-1.64 & 1.56-1.59~+~1.66-1.69\tablenotemark{a} & 1.63~$\mu$m FeH/CH$_4$ & 2\\
  $K$-slope & 2.06-2.10 & 2.10-2.14 & $K$-band shape/CIA H$_2$ & 3\\ 
  $J$-slope & 1.27-1.30 & 1.30-1.33 & 1.28~$\mu$m flux peak shape & 4\\
  $J$-curve & 1.04-1.07~+~1.26-1.29\tablenotemark{b} & 1.14-1.17 & Curvature across $J$-band & 4\\
  $H$-bump & 1.54-1.57 & 1.66-1.69 & Slope across $H$-band peak & 4\\
  H$_2$O-$Y$ & 1.04-1.07 & 1.14-1.17 & 1.15~$\mu$m H$_2$O & 4\\
 \enddata
\tablenotetext{*}{Indices were calculated by integrating flux between the specified wavelength ranges.}
\tablenotetext{a}{Denominator is average of these two wavelength ranges.}
\tablenotetext{b}{Numerator is average of these two wavelength ranges.}
\tablerefs{(1)~\citet{2006ApJ...637.1067B}; (2)~\citet{2010ApJ...710.1142B}; (3)~\citet{2002ApJ...564..421B}; (4)This paper.}
\end{deluxetable}
\clearpage

%%%%%%%%%%%%%%%%%
\begin{deluxetable}{cl}
\tabletypesize{\footnotesize}
\tablecolumns{2} 
\tablewidth{0pt}
 \tablecaption{Delimiters for selection regions in parameter spaces.\label{tab:selectreg}}
 \tablehead{
 \colhead{x vs. y} \vspace{-0.2cm}& \colhead{Limits}\\ \vspace{-0.2cm}}
 \startdata
 SpT vs. CH$_4$-\emph{H} & Best fit curve: $y = -4.3\times10^{-4}x^{2} + 0.0253x + 0.7178$, $\sigma$ = 0.0354.\\
 & Select points below the $-1\sigma$ curve.\\
 H$_2$O-\emph{J} vs. CH$_4$-\emph{H} & Intersection of: $y = -0.08x+1.09$ and $x = 0.90$.\\
 & Select points on lower left corner.\\
 H$_2$O-\emph{J} vs. \emph{H}-bump & Intersection of: $y = 0.16x+0.806$ and $x = 0.90$.\\
 & Select points on upper left corner.\\
 CH$_4$-\emph{J} vs. CH$_4$-\emph{H} & Intersection of: $y= -0.56x + 1.41$ and $y=1.04$.\\
 &  Select points on lower left corner.\\
\enddata
\end{deluxetable}

%%%%%%%%%%%%
\begin{deluxetable}{lccrcccl}
\tabletypesize{\footnotesize}
\tabletypesize{\scriptsize}
\tablewidth{0pt}
 \tablecaption{Binary candidates resulting from spectral fitting.\label{tab:cand}}
 \tablehead{
 \colhead{Source} \vspace{-0.2cm} & 
 \colhead{Primary SpT~\tablenotemark{a}} &
 \colhead{Secondary SpT~\tablenotemark{a}} & 
 \colhead{Confidence\tablenotemark{b}} & 
 \colhead{$\Delta$ \emph{J}} & 
 \colhead{$\Delta$ \emph{K}} &
 \colhead{SB Ref.} &
 \colhead{Comments~\tablenotemark{c}} \\\vspace{-0.2cm}}
 \startdata 
 \cutinhead{Strong Candidates}
2MASS J02361794+0048548 & L5.0$\pm$0.6 & T1.9$\pm$1.1 & $97\%$ & 1.05$\pm$0.48 & 1.80$\pm$0.55 & 1 & \nodata\\
SDSS J093113.23+280227.1 & L1.4$\pm$0.1 & T2.6$\pm$0.9 & $>99\%$ & 2.22$\pm$0.23 & 2.74$\pm$0.25 & 1 & V, B\\
2MASS J13114227+3629235 & L4.8$\pm$0.6 & T4.1$\pm$2.7 & $>99\%$ & 2.19$\pm$1.02 & 3.14$\pm$1.33 & 5 & V\\
2MASS J13411160-30525049 & L1.2$\pm$0.3 & T6.3$\pm$1.0 & $98\%$ & 3.28$\pm$0.53 & 4.82$\pm$0.63 & 1 & V\\
2MASS J14532589+1420418 & L1.1$\pm$0.0 & T6.0$\pm$1.1 & $>99\%$ & 3.27$\pm$0.46 & 4.42$\pm$0.63 & 1 & V, (B)\\
2MASS J20261584-2943124 & L0.1$\pm$0.5 & T5.8$\pm$1.0 & $>99\%$ & 3.42$\pm$0.40 & 4.82$\pm$0.57 & 4 & V\\ 
 \cutinhead{Weak Candidates}
2MASS J02060879+22355930 & L5.1$\pm$0.5 & T3.2$\pm$2.3 & $95\%$ & 1.61$\pm$0.89 & 2.36$\pm$1.25 & 1 & \nodata \\
DENIS-P J04272708-1127143 & M7.4$\pm$0.2 & T5.1$\pm$1.5 & $92\%$ & 4.13$\pm$0.62 & 4.98$\pm$0.84 & 1 & \nodata \\
\enddata
\tablenotetext{a}{Uncertainties include systematics from spectral classification of~\citet{2007ApJ...659..655B}.}
\tablenotetext{b}{Confidence that the source fits the binary template better than the single template based on a one-sided F-test. See Section~\ref{sec:specfit}.}
\tablenotetext{c}{B = Unusually blue L dwarf, (B) = From this paper; V = Also a visual candidate.}
\tablenotetext{d}{See Section~\ref{sec:blue}.}
\tablerefs{(1) This paper; (2)~\citet{2003AJ....126.1526B}; (3)~\citet{2011ApJ...732...56G}; (4)~\citet{2010AJ....140..110G}; (5)~\citet{2011ApJS..197...19K}; (6)~\citet{2010ApJ...710.1142B}.}
\end{deluxetable}
\clearpage

\end{document}